\documentclass[final,3p]{elsarticle}

\usepackage[T1]{fontenc}
\usepackage[british]{babel}
\usepackage[utf8]{inputenc}
\usepackage{caption}
\usepackage{subcaption}
\usepackage{wrapfig}
\usepackage{makeidx}
\usepackage{multicol}
\usepackage{footmisc}
\usepackage{amsmath}
\usepackage{amsthm}
\usepackage{amssymb}
\usepackage{amsbsy}
\usepackage{tikz}
\usepackage{color}
\usepackage{url}
\usepackage[firstpage]{draftwatermark}
\usepackage{graphicx}
\usepackage{cleveref}
\usepackage{todonotes}
\usepackage{booktabs}   % Nicer rules in tables
\usepackage{pgfplots}
\usepackage{centernot}
\usepackage{xspace}
\usepackage{nomencl} % for nomenclature
\usepackage{stmaryrd} % for big-O notation
\usepackage{xfrac} % for slanted fractions, e.g. 1/2 \sfrac
\usepackage{units} % for fysiske enheter
\usepackage{multirow} % to have multirow figures in tables
\usepackage{epstopdf}

\crefname{chapter}{Chapter}{Chapters}
\crefname{section}{Section}{Sections}
\crefname{appendix}{Appendix}{Appendices}
\crefname{subsection}{Section}{Sections}
\crefname{subsubsection}{Section}{Sections}
\crefname{equation}{Equation}{Equations}
\crefname{figure}{Figure}{Figures}
\crefname{table}{Table}{Tables}
\crefname{subfigure}{Figure}{Figures}
\crefname{listing}{Listing}{Listings}

\graphicspath{{figures/}}

\usepackage{tgcursor}  %Courier font for code listings, called TeX Gyre Cursor
\usepackage[activate={true,nocompatibility},final,kerning=true,tracking=true,spacing=true]{microtype}
\microtypecontext{spacing=nonfrench}

\usetikzlibrary{calc} % for coordinate calculation

\newcommand{\nomunit}[1]{\renewcommand{\nomentryend}{\hspace*{\fill}#1}}
\makenomenclature

\usetikzlibrary{plotmarks,shapes,arrows}
\tikzstyle{decision} = [diamond, draw, fill=blue!20,
    text width=4.5em, text badly centered, node distance=3cm, inner sep=0pt]
\tikzstyle{block} = [rectangle, draw, fill=blue!20,
    text width=13em, text centered, rounded corners, minimum height=4em]
\tikzstyle{line} = [draw, -latex']
\tikzstyle{cloud} = [draw, ellipse,fill=red!20, node distance=3cm,
    minimum height=2em]
\pgfplotsset{compat=1.6}

\newcommand{\eg}{e.g.\xspace}
\newcommand{\ie}{i.e.\xspace}
\newcommand{\etal}{et al.\xspace}
\newcommand{\NPT}{$NpT\;$}
\newcommand{\NVT}{$NVT\;$}

\renewcommand{\v}[1]{\ensuremath{\mathbf{#1}}} % for vectors
\newcommand{\gv}[1]{\ensuremath{\mbox{\boldmath$ #1 $}}}  % for vectors of Greek letters
\newcommand{\pd}[2]{\frac{\partial #1}{\partial #2}} % for partial derivatives
\newcommand{\pdd}[2]{\frac{\partial^2 #1}{\partial #2^2}} % for double partial derivatives
\newcommand{\grad}[1]{\gv{\nabla} #1} % for gradient
\renewcommand{\div}[1]{\gv{\nabla} \cdot #1} % for divergence
\let\baraccent=\= % rename builtin command \= to \baraccent
\renewcommand{\=}[1]{\stackrel{#1}{=}} % for putting numbers above =

\journal{Journal of Computational Physics}

\begin{document}

\begin{frontmatter}

\title{A multiscale method for simulating fluid interfaces covered with
large molecules such as asphaltenes}
\author[ntnu]{Åsmund Ervik\corref{cor1}}
\ead{asmunder@pvv.org}
\cortext[cor1]{Corresponding author}
\author[ntnumat]{Morten Olsen Lysgaard}
\ead{morten@lysgaard.no}
\author[bath]{Carmelo Herdes}
\ead{c.e.herdes.moreno@bath.ac.uk}
\author[imperial]{Guadalupe Jim{\'e}nez-Serratos}
\ead{m.jimenez-serratos@imperial.ac.uk}
\author[imperial]{Erich~A.~Müller}
\ead{e.muller@imperial.ac.uk}
\author[sintef]{Svend Tollak Munkejord}
\ead{svend.t.munkejord@sintef.no}
\author[ntnu]{Bernhard Müller}
\ead{bernhard.muller@ntnu.no}
\address[ntnu]{Department of Energy and Process Engineering, NTNU, NO-7491 Trondheim, Norway}
\address[ntnumat]{Department of Mathematics, NTNU, NO-7491 Trondheim, Norway}
\address[imperial]{Department of Chemical Engineering, Imperial College London, London SW7 2AZ, United Kingdom}
\address[bath]{Department of Chemical Engineering, University of Bath, Claverton
Down, Bath, Somerset BA2 7AY, United Kingdom}
\address[sintef]{SINTEF Energy Research, P.O. Box 4761 Sluppen, NO-7465 Trondheim, Norway}

\begin{abstract}
The interface between two liquids is fully described by the interfacial tension
only for very pure liquids. In most cases the system also contains surfactant
molecules which modify the interfacial tension according to their concentration
at the interface. This has been widely studied over the years, and interesting
phenomena arise, \eg the Marangoni effect. An even more complicated situation
arises for complex fluids like crude oil, where large molecules such as
asphaltenes migrate to the interface and give rise to further phenomena not seen
in surfactant-contaminated systems. An example of this is the ``crumpling drop''
experiments, where the interface of a drop being deflated becomes non-smooth at
some point. In this paper we report on the development of a multiscale method
for simulating such complex liquid-liquid systems. We consider simulations
where water drops covered with asphaltenes are deflated, and reproduce the
crumpling observed in experiments. The method on the nanoscale is based on using coarse-grained
molecular dynamics simulations of the interface, with an accurate model for
the asphaltene molecules. This enables the calculation of interfacial
properties. These properties are then used in the macroscale simulation, which
is performed with a two-phase incompressible flow solver using a novel hybrid
level-set/ghost-fluid/immersed-boundary method for taking the complex interface
behaviour into account. We validate both the
nano- and macroscale methods. Results are presented from nano- and
macroscale simulations which showcase some of the interesting behaviour caused
by asphaltenes affecting the interface. The molecular simulations
presented here are the first in the literature to obtain the correct
interfacial orientation of asphaltenes. Results from the macroscale
simulations present a new physical explanation of the crumpled drop
phenomenon, while highlighting shortcomings in previous hypotheses.

\end{abstract}


\end{frontmatter}
\section{Introduction}
Interfacial tension is a remarkable phenomenon, in that the tumultuous interactions
of fluid molecules of different types and shapes give rise to macroscopic
interfaces being not only smooth and stable, but indeed well-described by
a single material constant, viz. \emph{the interfacial tension} $\gamma$. This holds
true for an impressive number of fluid molecules that may be polar or non-polar
and may have different topology and size. It is by
adding a third phase to the system, this phase being interfacially active either
by virtue of amphiphilicity or through being poorly soluble in either fluid,
that more complicated dynamics arise. The simplest case is that of amphiphilic
surfactant molecules, where \citet{marangoni1865,marangoni1870} first described
the effects of nonuniform interfacial tension, and Gibbs in his seminal treatise
\citep{gibbs1878} was the first to consider the effective elasticity of the
interface imparted by the surfactants. The first mention of the deliberate use
of surfactants to alter the interfacial properties of liquids is, however, much
earlier. In book two of Pliny the Elder's encyclopedic \emph{Naturalis Historia}
\cite{pliny1906,pliny1949} (77 AD) it is
mentioned that divers would release small oil drops from their mouth, to smooth
the surface of the sea and thus increase the amount of light transmitted down to them.

Even though surfactants give rise to richer dynamics, their effects can be
modelled using simple equations dependent on macroscopic parameters, using the
approaches developed by \citet{gibbs1878}, \citet{pockels1893}, \citet{szyszkowski1908},
\citet{langmuir1918} and \citet{frumkin1925}  (among others); see
\citet{levich1969} (or the book by \citet{levich1962}) for a good
introduction. Apart from studies using these approaches for systems with low
surfactant concentration, there is also a rich field of study into the various phases and
phase transitions in the three-component systems when the number of surfactant
molecules becomes comparable to the number of fluid molecules. However, we will not
discuss this in further detail here. In the
end, the effect of surfactants at low concentration is an interfacial tension which may vary along the interface,
according to the variation in interfacial concentration of the surfactant. The
relation between concentration and interfacial tension is given by \eg
a Langmuir equation of state. 

In more recent years, as experimental techniques have increased in
sophistication and soft matter has come to be a field of its own, attention has
also turned to larger molecules at fluid interfaces. These larger molecules may
originate from biological systems, being \eg proteins or lipids, or they may
originate from complex fluids such as crude oil, for instance asphaltenes.
These molecules cause effects beyond those seen in surfactant systems, \eg the
crumpling upon deflation of red blood cells \cite{fai2013} or asphaltene-covered
drops \cite{yeung1999,pauchard2014}. It follows that something more than a (possibly varying)
interfacial tension appears in these systems. We shall focus our attention here on
asphaltenes, and will use this term rather than ``large molecules'' for the
remainder of the paper, but the method remains general. 

Let us then consider the asphaltenes. Having earned a reputation as the
``cholesterol of crude oil'', this component causes many problems in petroleum
extraction, processing and refinement. As the name suggests, asphaltenes are
similar in appearance to asphalt, and are found in large quantities in bitumen,
but are present to some degree in most crude oils. The first reported
discussion of asphaltenes is by \citet{boussingault1836} who coined the term in
1836 (see \citet[Chapter 1.1]{hoepfner2013} for a historical review)
but there is still today disagreement about their properties. Indeed, there was
debate in the literature \cite{herod2008} up until recently on what their
average molecular weight is. Furthermore, the precise definition of what constitutes an
asphaltene is still not agreed upon. A commonly used operational definition is
that it is the part of crude oil which is soluble in toluene but insoluble in
$n$-heptane, as codified \eg in the ASTM D 6560 -- 00 standard. (From here on,
we will denote $n$-heptane simply by heptane; branched alkanes are not
considered in this work.) But this definition in terms of solubilities is more
a description of how to isolate asphaltenes in the laboratory, as opposed to a
definition of what they are.

The case is not clearer from the
molecular perspective, as advanced experimental techniques such as neutron
scattering or high-resolution mass spectrometry \cite{klein2006} have shown that there
are thousands of different empirical formulae in
a given asphaltene sample. Samples from different oil fields around the world
have different asphaltene compositions.  Furthermore, since asphaltenes have of
the order of 50 carbon atoms, even for a single empirical formula the
number of isomers is in the trillions. One may compare asphaltene molecules to
snowflakes; no two are exactly the same. It has been argued that asphaltenes are
among the most complex materials ever studied \cite{hoepfner2013}.

The effect of asphaltenes on liquid-liquid interfaces is also complicated. One
such effect is that they make it hard to separate emulsions of water and crude
oil. See \eg
\citet{jones1978,gafonova2001,sjoblom2001,kokal2005,kilpatrick2012} for
reviews on water-in-crude oil emulsion stability. Emulsion stability has been
directly linked to the properties that the asphaltenes impart on the
interface \cite{dicharry2006,bi2015}. The asphaltenes give rise to
interesting phenomena such as the previously mentioned crumpled drops reported
by \citet{yeung1999,pauchard2014}. 

In the present paper we report on our development of a multiscale method that
enables the simulation of liquid-liquid interfaces covered with
asphaltenes (or other large molecules). This multiscale method is loosely
coupled, with equilibrium simulations at the nanoscale providing parameters for
dynamic simulations at the macroscale. Many
classifications exist of multiscale methods, some of which have a much tighter
coupling between the scales than in the present approach, see \eg
\cite{weinan2003,brandt2002} for reviews of multiscale methods.

The outline of this paper is as follows. In \cref{sec:theory-methods} we review the theory and present the
methods used on the
nanoscale and the macroscale, in particular the SAFT-$\gamma$ Mie approach to
coarse-grained molecular dynamics simulation (\cref{sec:mic-theory}), and the hybrid
level-set/ghost-fluid/immersed-boundary method developed in this work for
simulation of two-phase flows with complex interfaces(\cref{sec:mac-methods}). The latter
method is summarised in
\cref{sec_proposed_method}. Subsequently, in
\cref{sec:validation}, we validate these methods using standard test cases for
interface-capturing methods (\cref{sec:macro-val}), and by comparing the
predictions from molecular simulations to experimental results
(\cref{sec:micro-val}). We then present the results obtained with the method in
\cref{sec:results}, discuss the implications of these results in
\cref{sec:discussion}, and finally give some concluding remarks in
\cref{sec:conclude}.

\section{Theory and methods}
\label{sec:theory-methods}
\subsection{Nanoscale: theory}
\label{sec:mic-theory}
The complicated behaviour of liquid-liquid interfaces contaminated with
asphaltenes is caused by the interactions between the molecules at the
interface. It is thus tempting to try and explain the interfacial
phenomena by modelling the molecular interactions. These are, in turn, also
complicated, and one is forced to make simplifications. On the most basic
level, many-body quantum mechanics is what lies behind the nature and
interactions of molecules. Fortunately, on the scale of interactions between
large molecules, one may develop an effective theory which is much simpler. This
theory has roots going back to the dawn of thermodynamics, when pioneering efforts were
made to understand how the microscopic nature of fluids could explain
their behaviour. The works by Maxwell, Boltzmann, van der Waals, Lennard-Jones,
Mie, Chapman, Engskog and others
paved the way to our understanding of simple fluids from the molecular
perspective; see \eg \citet{chapman1991} for an overview. 
For less simple fluids, it was not until the advent of computer
simulations that the molecular perspective was able to provide some insights.

One of the major challenges in simulating the behaviour of chemicals comes
from the need for good models that can accurately predict physical and chemical properties. For
many simple substances, experimental data may already be available or easy to
obtain. However, many interesting systems contain one or
several chemically complex species, e.g. polymers, surfactants, and our
asphaltenes. The quantitative prediction of the thermodynamic
properties of such systems, especially their phase behaviour and mesoscopic
structure, is very challenging. Simple equations of state are typically unable
to make good predictions of the properties of structured fluids. As this much
is clear, one must resort to computer simulations of these more complex fluids.

Many molecular
simulation methods rely on force fields or other empirical parameters that are
fitted to reproduce the properties of particular classes of compounds, or to a
particular large data set. These methods may have trouble predicting the
properties of complex mixtures. Truly \emph{ab initio} prediction, using electronic
structure methods, is possible only for very small systems and processes
spanning very short time scales, and cannot be used directly to predict the
properties of structured fluids or complex materials. It is possible to construct
models for complex systems by first building models for the different (smaller)
component parts using more predictive methods, and then eliminating the
unimportant degrees of freedom.

This leads to a coarse-grained description that
can be used to predict properties of more complex materials. However, it is 
time-consuming to generate the data for the smaller building blocks, and it is
often not clear what the best way is when coarse-graining out the smaller/faster degrees of
freedom. An alternative approach is to construct a coarse-grained model directly
using experimental data for the building blocks (“top-down” coarse graining) -- in
this case the evaluation of the residual when performing the fit depends on a
relatively costly simulation, making the cost of fitting very high.

\begin{figure}[htpb]
  \centering
  \includegraphics[width=0.55\linewidth]{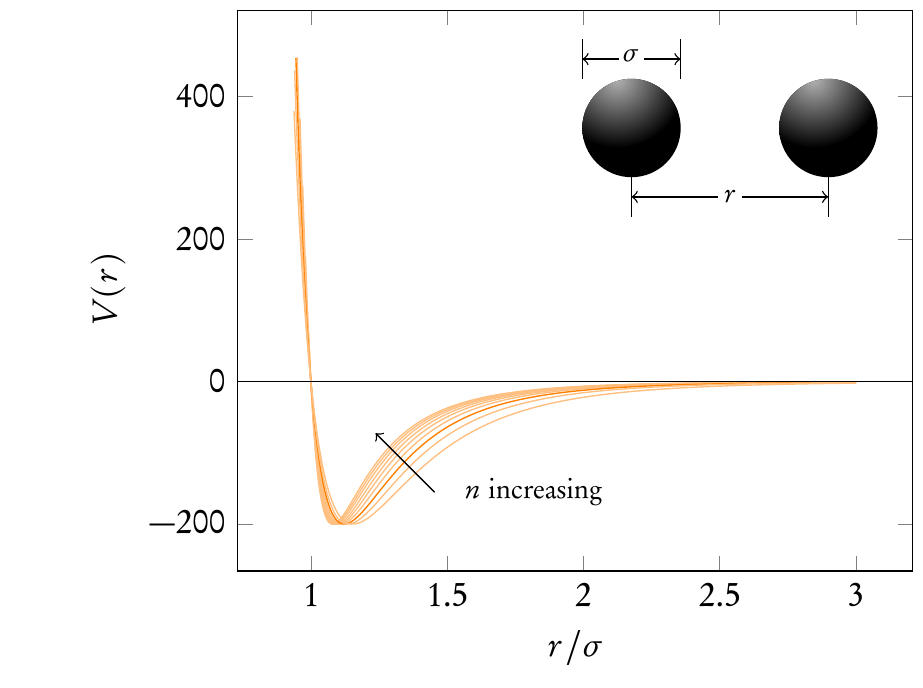}
  \caption{
    The Mie (m,6) potential shown for $\epsilon=200$ (in arbitrary units) and
    $n$ varying from 8 to 24, as a function of the dimensionless distance
    $r/\sigma$.  The Lennard Jones (12,6) potential is shown in a stronger
    colour. In the corner, the two beads are illustrated with a distance $r$ and
    a ``diameter'' $\sigma$.
    }
  \label{fig:mie}
\end{figure}

Recently, however, an
alternative approach has been successfully used to fit coarse-grained models to
thermodynamic data without using simulations, but rather using statistical
mechanical perturbation theory. This novel approach, the SAFT-$\gamma$ method, allows
for the construction of coarse-grained models suitable for molecular simulation
of complex fluids using available experimental data for the constituent blocks.
The SAFT equation of state is a perturbation approach based on a well-defined
Hamiltonian. The reader is referred to several reviews on SAFT that describe the
various stages of its development, up to the current
SAFT-$\gamma$ Mie approach \citep{muller2001,economou2002,tan2008,mccabe2010,lafitte2012,papaioannou2014}. 
In the current approach, a Mie \citep{mie1903} potential gives the forces between the coarse grained
\emph{beads} in the method,
\begin{align}
  V(r) = &C(n,m)\; \epsilon \left[\left(\frac{\sigma}{r}\right)^n - \left(\frac{\sigma}{r}\right)^m \right] \rm{,}\\
       &C(n,m) = \left(\frac{n}{n-m}\right)\left(\frac{n}{m}\right)^{m/(n-m)} \rm{,}
  \label{eq:mie}
\end{align}
where $r$ is the distance between a pair of beads, and $\epsilon$ and $\sigma$ are the adjustable
parameters relating to the energy and distance scales. Each bead typically
corresponds to 2--4 atoms heavier than hydrogen, together with the hydrogen atoms
attached to the heavier atoms. Referring to
\cref{fig:mie}, the parameter $\epsilon$ corresponds to the depth of the potential
well, and $\sigma$ to the distance $r$ where the potential switches from being
repulsive to being attractive. Thus $\sigma$ is taken to correspond to the bead
``diameter'' for visualisation purposes. Note that in the $n \to \infty$ limit,
$V(r)$ becomes a hard-sphere potential with diameter equal to $\sigma$.

It is important to note that in the present approach,
the attractive exponent $m$ is fixed at the value of six, but the short-range repulsion
exponent $n$ takes on different values reflecting the average softness or hardness of the
potential for a given molecule. The fact that $m=6$ is fixed is due to the
observation that these two exponents are not independent parameters
\cite{ramrattan2015}, so fixing one simplifies the parameter space considered.
The effect of varying $n$ is illustrated in \cref{fig:mie}
where the potential is plotted for $n$ ranging from 8 to 24 in steps of 2. The
$n=12$ or Lennard-Jones potential is shown in a stronger orange colour. It is
seen that allowing $n$ to vary allows for using softer or harder potentials for
beads representing different molecules, which is something that cannot be
achieved in many other approaches which rely on only the Lennard-Jones potential.

There are several methodologies proposed in the literature to obtain parameters
for coarse grained models \cite{noid2013} . Most common approaches start from
a fine-detailed model, usually an atomistically-detailed characterization and
integrate out degrees of freedom which are deemed unessential \cite{brini2013}. This procedure, known as a bottom-up approach,
inevitably discards information and produces potentials which can rarely be used
in states points other than those used to develop them.

A pathway to circumvent
these problems is to employ a top-down or thermodynamic approach, where the
force field of the coarse grained sectors is an effective or average potential
capable of reproducing macroscopic thermodynamic properties. These top-down
coarse grained models provide by their own nature parametrizations which are
usually robust, representative or transferrable. Two notable examples are the
MARTINI force field \cite{marrink2007,periole2013}, which employs a group
contribution approach targeted at biomolecular simulations where uniform-sized
coarse grained beads have been fitted to water/octanol solubilities.

In a group contribution approach, molecules which have different parts
with different properties are modelled using combinations of beads which
represent each individual part. As an example, octanol could be modelled with one
bead that represents the head with the alcohol group, based on the
thermophysical properties of ethanol, and the other beads which represent the
aliphatic tail could be based on the properties of hexane.

A more
refined coarse-grained model is the SAFT-$\gamma$ force field \cite{muller2014}, which employs an analytical
equation of state as the bridge between macroscopic thermophysical properties
and the underlying intermolecular potential that can effectively generate them. 
The most direct method of parametrization of coarse grained segments and chemical
moieties is to use the appropriate version of the SAFT equation of state to fit
experimental phase equilibria data, e.g. saturated liquid densities, vapour
pressures, etc. essentially equating the free energy of a coarse grained model
to that obtained from the analysis of experimental data \cite{avendano2011,avendano2013,lafitte2012,lobanova2015}. 

However, if one recognizes the conformal nature of the underlying Mie potential,
one can formulate the equation of state in terms of a corresponding-states model
and can express the properties of any non-associating fluid in terms of a finite
set of defining properties: a critical temperature, the acentric factor and
a well-defined density. This approach \cite{mejia2014} greatly simplifies the parameter
estimation without detriment to the robustness of the methodology, as
exemplified by the predictions of adsorption \cite{herdes2015}, transport and interfacial
properties \cite{muller2014} which are not part of the original fit, and the description of
complex molecules such as surfactants \cite{herdes2015b,theodorakis2015}, resins
and asphaltenes \cite{muller2015}. This approach is used to construct the force
field parameters used in this work. The parameters for different coarse-grained
beads are given in Appendix B.

\subsection{Nanoscale: numerical methods}
\label{sec:mic-methods}
Once the parameters for the intermolecular force field have been established for all
molecules under consideration, one can proceed to study the system. Since only the
two-bead problem has an analytical solution, a numerical approach is required
for any realistic system. There are two fundamental approaches, namely Monte
Carlo methods and Molecular Dynamics (MD) methods. The ergodic hypothesis
asserts that these two are the same, \ie that ensemble averages and time
averages give the same answers. Molecular dynamics is the approach used in this
work, so we will not discuss Monte Carlo methods in any detail.

In molecular dynamics, the equations of motion are solved to evolve the system
in time from some initial state. The equations of motion are Newton's second law for
each of the $N$ beads in the system, with the force given by the Mie potential
in the approach used here. Denote by $\v{x}_i$ the position of bead $i$ with mass $m_i$.
The equations of motion are then
\begin{equation}
  \pdd{\v{x}_i}{t} = \sum_{j\ne i} \frac{F(|\v{x}_j - \v{x}_i|)}{m_i} \frac{\v{x}_j - \v{x}_i}{|\v{x}_j - \v{x}_i|}
  \label{eq:newton2}
\end{equation}
where the forces are assumed to be conservative, and depending only on the
distance between two beads. Thus the forces are given by a potential $V(r)$.
This must be specified; a Mie potential is used here, as discussed previously.
This potential is short-ranged, so a \emph{cutoff} $r_\text{cut}$ is
specified, beyond which $V(r) = 0$. The equations of motion are solved numerically using
a symplectic integration method, since this ensures the simulation is stable over long
times and that the energy drift is very small. An example of such a method is
the velocity Verlet method. Here, the velocities $\v{v}_i$ and positions
$\v{x}_i$ are stored at each time step $n$. The system is integrated forward in
time to the time step $n+1$, with a step length $\Delta t$, according to
\begin{align}
  \v{v}_i^{n+1/2} &= \v{v}_i^{n} + \sum_{j \in \mathbb{N}(i,n)} \frac{\Delta
  t}{2m} F(|\v{x}_j^n - \v{x}_i^n|) \frac{\v{x}_j^n - \v{x}_i^n}{|\v{x}_j^n
- \v{x}_i^n|} \\
  \v{x}_i^{n+1} &= \v{x}_i^{n} + \v{v}_i^{n+1/2}\Delta t\\
  \v{v}_i^{n+1} &= \v{v}_i^{n+1/2} + \sum_{j \in \mathbb{N}(i,n+1)} \frac{\Delta
t}{2m} F(|\v{x}_j^{n+1} - \v{x}_i^{n+1}|) \frac{\v{x}_j^{n+1}
- \v{x}_i^{n+1}}{|\v{x}_j^{n+1} - \v{x}_i^n|} 
\end{align}
Here $\mathbb{N}(i,n)$ is the neighbourlist of bead $i$, \ie the indices of
all beads which are within the cutoff distance of bead $i$ at the time
step $n$. Using a neighbourlist dramatically speeds up the algorithm, but comes
at a storage cost; this data structure accounts for the bulk of the memory used by
a molecular dynamics code. The fact that all interactions are local makes
the method well-suited for parallelisation, both on classical CPUs and on
accelerators such as graphical processing units (GPUs).

The beads are contained in a virtual
simulation box, typically with periodic boundary conditions. The initial state
must be generated somehow. In the present work we start from random initial conditions at a low density. To
go from this initial state to the desired system state, \eg a temperature $T$
and a pressure $p$, the system is evolved in the isothermal-isobaric or \NPT
ensemble, where a thermostat and a barostat are employed to adjust the bead
velocities and the simulation box size, respectively, to obtain the desired
system state. Once this state is reached, the simulation box size is fixed and
the system is subsequently evolved in the canonical or \NVT ensemble. 
The reader is referred to standard textbooks on molecular dynamics, \eg 
by \citet{allen1989,frenkel2001}, for further details.

While the first molecular dynamics simulations \cite{alder1957,alder1959sciam}
were limited to two-dimensional systems and simple potentials, the results
obtained provided fascinating new insights into the molecular world. With the
exponential increase in computing power since the 1950's, molecular dynamics
simulations today have probed systems with hundreds of billions of atoms
\cite{kadau2006}, or entire virus capsules \cite{zhao2013}, using
large high-performance computing systems.

Combining a coarse-grained approach to
molecular dynamics, such as the SAFT-$\gamma$ Mie force field used here, with
the power of general purpose GPUs, we obtain
speedups of more than three orders of magnitude over atomistically detailed
simulations running on usual CPUs \cite{raasaft}. This means the present simulations, even
though they evolve systems with about one million atoms for hundreds of millions
of timesteps in total, are run on computational resources that can fit inside
a desktop computer. The simulations reported here are run in parallel
typically on two or four GPUs; separate simulations are run simultaneously on
different nodes of a GPU cluster, consuming in total 8000 GPU-hours during this
study.

For coarse-grained molecular dynamics simulations we employ our raaSAFT code
\cite{raasaft}, a framework for simulations using the SAFT-$\gamma$ Mie force
field. raaSAFT leverages existing molecular dynamics codes to do the heavy
lifting. Here HOOMD-blue \cite{anderson2008,glaser2015} is used, a modern
GPU-first code that shows excellent performance and scalability on multiple
GPUs. 

HOOMD-blue takes a conservative approach to molecular dynamics, and uses
algorithms that do not sacrifice accuracy for speed. The simulations are 
run in the isothermal-isobaric (\NPT) or the isothermal-isochoric (\NVT)
ensemble, and the system is evolved in these ensembles using the Martyna-Tobias-Klein
approach \cite{martyna1994}. This approach gives dynamics that are provably
time-reversible and energy-preserving. In the light of this, it is interesting
how the simulations show time-irreversible results, such as the clustering of
asphaltenes. This is a variant of
Loschmidt's paradox. Recent work by Hoover \etal \cite{hoover2015a,patra2015,hoover2015b}
might provide some insight into the explanation of this, but so far it appears
to remain an open question.

When systems with immiscible fluids are considered, it is of interest to compute
the interfacial tension. To do this, the simulation box is 
elongated in one direction, and this asymmetry causes the formation of
interfaces along the two shorter dimensions of the box, since this
minimises the free energy of the interfaces. Note that since periodic boundary
conditions are employed, it is a topological impossibility to have an odd number
of interfaces between two liquids, so the desired system state has two slabs of
liquid which are in contact at two interfaces.

For this system, we may compute the interfacial tension from the integral of the
anisotropy in the diagonal elements of the stress tensor, $\sigma_{ii}$, along
the elongated box dimension. To be precise, assuming the box is elongated in the $z$-direction
where the box dimension is $L_z$, the interfacial tension is given by
\begin{equation}
  \gamma = \frac{L_z}{2} \int_0^{L_z} \left(-\sigma_{zz} + \frac{1}{2} (\sigma_{xx} + \sigma_{yy}) \right) dz \rm{.}
    \label{eq:ift-md}
\end{equation}
This is referred to as the mechanical route to the interfacial tension, and goes
back to \citet{kirkwood1949}. For a comparison of this with alternative methods, see \eg
\cite{vega2007}. Note that the integral here can be split into three parts, so molecular dynamics
software such as HOOMD-blue typically compute the values
\begin{equation}
  p_{xx} = \int_0^{L_x} - \sigma_{xx} dx \rm{,}
\end{equation}
and similarly for $p_{yy},p_{zz}$, and refer to these as the diagonal components
of the ``pressure tensor''.
From these three numbers we may compute the interfacial tension using
\cref{eq:ift-md}. Since molecular dynamics simulations are inherently noisy, in
particular for properties related to the pressure,
$\gamma$ computed from this expression for a single point in time will fluctuate
significantly from one time step to the next. To obtain a reliable value 
for $\gamma$, time averages must be employed.

To also compute the elasticity of the interface, we follow refs. \cite{otter2003,boek2005},
where the elasticity $K_a$ is computed from the change in interfacial tension
$\Delta \gamma$ (as computed from \cref{eq:ift-md}) when the interfacial area is changed from $A_0$ to $A$, given by
the expression
\begin{equation}
  K_a = \frac{\Delta \gamma}{A/A_0 - 1} \rm{.}
  \label{eq:elasticity-md}
\end{equation}
These two parameters, $\gamma$ and $K_a$, are subsequently employed as material
parameters in the macroscale simulations.

\subsection{Macroscale: theory}
\label{sec:mac-theory}
\begin{figure}[tbp]
\begin{center}
  \includegraphics[width=0.4\linewidth]{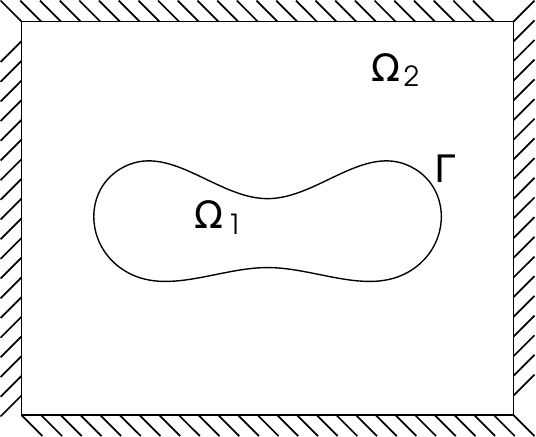}
\end{center}
\caption{Two fluid phases, the inner fluid 1 in $\Omega_1$ and the outer fluid
2 in $\Omega_2$, and the interface $\Gamma$ separating them.}
\label{fig_two_phase_domin}
\end{figure}

The flow inside and around water drops in oil behaves according to the
incompressible Navier-Stokes equations. For a derivation of these and a general
overview, the reader is referred to standard textbooks \eg by
\citet{batchelor2000introduction} or \citet{lamb1945}.
The incompressible Navier-Stokes equations for single-phase flow read
\begin{align}
\rho \left(\pd{\v{u}}{t} + \v{u} \cdot \nabla \v{u}\right) &= - \nabla p +
  \div(\mu\grad\v{u}) + \rho \v{f}, \label{eq:ns}\\
\nabla \cdot \v{u} &= 0, \label{ns2}\\
\v{u}(\v{x},0) &= \v{u}_0(\v{x}), \label{ns3}\\
\v{u}_{\partial \Omega}(t) &= \v{g}(t), \label{ns4}
\end{align}
where $\partial \Omega$ is the domain boundary and $\v{g}(t)$ is the velocity
boundary condition. The initial condition is $\v{u}_0$.
The viscosity $\mu$ and the density $\rho$ are assumed to be constant throughout
the domain. The velocity is denoted by $\v{u}$ and the pressure by $p$, and
$\rho\v{f}$ is a body force such as gravity.
\nomenclature{$\rho_i$}{Density of fluid $i$. \cref{sec_navier_stokes_eqns}}%
\nomenclature{$\gamm$}{Interfacial tension between fluids. \nomunit{\unitfrac{N}{m}}}%
\nomenclature{$\v{u}$}{The Eulerian velocity field.}%
\nomenclature{$t$}{Time. \nomunit{s}}%
\nomenclature{$p$}{The pressure field.}%
\nomenclature{$\mu$}{Dynamic viscosity of fluid. \nomunit{Pa $\cdot$ s}}%
\nomenclature{$\v{f}$}{The Eulerian force field.}%
\nomenclature{$\Omega$}{The spatial domain.}%
\nomenclature{$\partial \Omega$}{The boundary of the spatial domain.}%
\nomenclature{$\Delta$}{The Eulerian grid spacing. \nomunit{m}}%
\nomenclature{$\v{x}$}{A point in the domain $: \v{x} \in \Omega$.}%
\nomenclature{$N$}{Number of grid points.}%
This can be extended to handle two fluid phases, with different
viscosities and densities. Let $\Omega_1$ and $\Omega_2$ denote the domains filled with fluid 1 and fluid 2, respectively
and $\Gamma$ denote the interface separating the two fluids, \ie we have $\Omega
= \Omega_1 \cup \Omega_2$, as illustrated in \cref{fig_two_phase_domin}.

The tension on $\Gamma$ can be modelled as
a contribution to \cref{eq:ns}, localised at the interface. In this work we
consider one-dimensional or axisymmetric two-dimensional interfaces, and in
both cases the interface is parametrised as a one-dimensional curve 
$\v{x}_I(s)$. In the axisymmetric case, this curve is swept around the azimuthal
angle $\phi$ to form the two-dimensional interface. 
The interfacial tension has two contributions in the axisymmetric case, $T_s$
and $T_\phi$, both of which may vary as functions of the position $s$ along the
interface. In the case of a one-dimensional interface, $T_\phi=0$. 

The interfacial force contribution to the body force in \cref{eq:ns} is
then
\begin{equation}
  \label{eq_singular_surface_force}
  \v{f}_s(\v{x},t) = \int_{\Gamma} \left( \pd{T_s}{s} \v{t} + T_s \kappa_s
  \v{n} + T_\phi \kappa_\phi \v{n} \right) \delta(\v{x} - \v{x}_I(s)) \text{d}\v{s},
\end{equation}
where $\pd{T_s}{s}$ is the derivative of the meridional tension along the interface,
$\v{t}$ is the interface tangent, $\kappa_s,\kappa_\phi$ are the curvatures, $\v{n}$ 
is the principal unit normal vector, and $\delta$ is the Dirac delta function. We assume here that the
interfacial curvature is small enough that the interface is approximately flat
on the microscopic level. This is a good approximation for drops with radius $R
\gg 1$nm \cite{lau2015}, which does not pose a significant restriction. The
interfacial tension for pure fluids is then dependent just on the
temperature and the pressure.

\begin{figure}[htpb]
  \centering
  \includegraphics[width=0.7\linewidth]{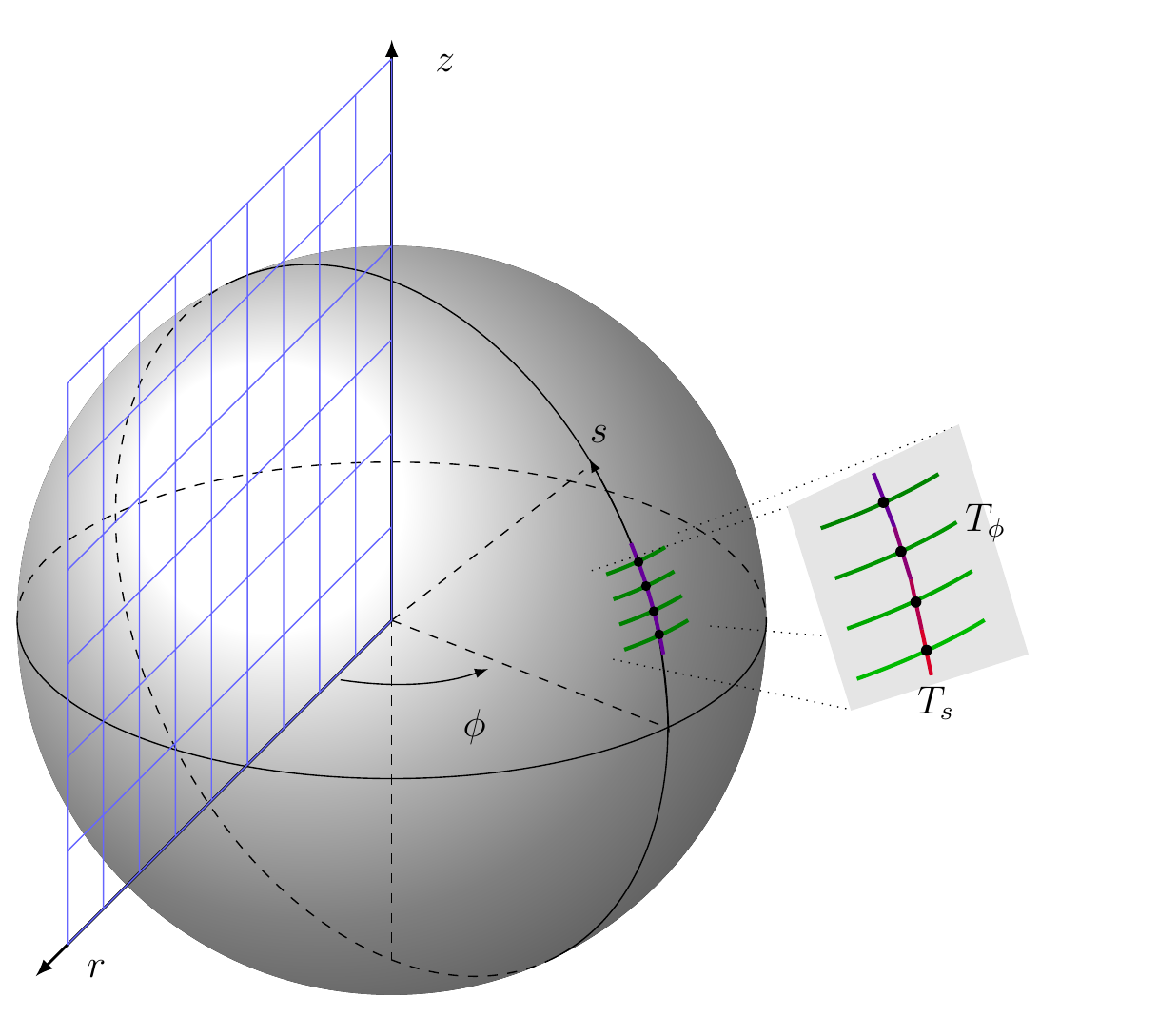}
  \caption{Illustration of the two coordinate systems employed, and the tensions
  on the drop. The blue grid illustrates the $(r,z)$ coordinate
  system, where a structured grid is employed and the Navier-Stokes equations
  are solved. The grey sphere illustrates the shape of a drop. On this drop, the
  coordinates $s,\phi$ are used. Note that $s$ is an arc length, not an angle. 
  On the right, the tensions 
  are illustrated with red ($T_s$) and green ($T_\phi$) line segments. The varying
  colours indicate varying tensions. Note that while both tensions vary from
  point to point, only $T_s$ varies in the direction parallel to the
  corresponding strain $\lambda_s$, while $T_\phi$ is constant in the direction
  parallel to $\lambda_\phi$ (due to axisymmetry).}
  \label{fig:tensions-sphere}
\end{figure}

In \cref{fig:tensions-sphere} we illustrate the tensions for the axisymmetric
case, together with the two
coordinate systems employed, on a spherical drop. Note that the drop is not
restricted to being spherical, so none of the
coordinate systems are spherical, and $s$ is an arc length,
not an angle. The line parametrised by $s$, which represents the drop surface, 
actually lies in the $(r,z)$ plane, but in the figure the line is rotated 
out of the plane in order to avoid clutter.
Since we assume axisymmetry, neither of the coordinate systems will have points
discretising the $\phi$-direction in the numerical methods.

On the right-hand side of this figure, the tensions $T_s$ and
$T_\phi$ are indicated as line segments in the directions of strain, using
varying red and green colours, respectively. Both tensions
may vary along the $s$-direction, but are constant in the $\phi$-direction since
axisymmetry is assumed. As is indicated in \cref{eq_elastic_force_density} in
the next section, only the variation in $T_s$ can give rise to a tangential
force. From the figure we understand this, since any non-normal component of
the force caused by $T_\phi$ would have to be either in the azimuthal direction,
violating the assumption of axisymmetry; or in the meridional direction, but
this force would be binormal to the strain line, which is not possible.

The description of an interface between two fluids
can be as simple as a constant interfacial tension $T_s = T_\phi = \gamma$, or more complicated
due to molecules which are interfacially active.
In any case, the formation of the interface gives an increase in the energy of
the system, and we denote the energy density of the interface by $w$. It should
be noted that while the interfacial energy (in J/m$^2$) and the interfacial
tension (in N/m) are identical for the case of simple fluids, this is
\emph{not} the case for an interface with a more complicated interface which
has elastic properties; see \eg \cite{hui2013} for details. 

Following
\cite{knoche2011}, we write $w$ as a function\footnote{$w$ is technically a functional; see
\eg \cite{ANU:165651} for a more mathematical formalism.} of the interfacial deformations
$\lambda_s$ and $\lambda_\phi$, 
\begin{equation}
  w = w(\lambda_s,\lambda_\phi).
\end{equation}
These deformations refer to the length of an interfacial element $l_i$ at time
$t$ relative to the undeformed length at $t=0$, \ie $\lambda_s = l_i(t)/l_i(0)$.
The interface will tend to deform such that the energy is minimised, while maintaining
a constant volume inside. The minimal energy shape depends on the form of $w$;
for the familiar case of constant interfacial tension the minimal energy shape is a sphere.

We proceed to derive the general tensions for an interface which is described by
a constant interfacial tension $\gamma$ and a Hookean elasticity $K_a$, following again
\cite{knoche2011}. The bending rigidity of the interface is assumed to be zero;
a non-zero bending rigidity may be considered in future work. As noted
previously, we parametrise the interface using the coordinate $s$ along the
interface in the $(r,z)$ plane. All quantities in the system are constant along
the azimuthal direction $\phi$. Under these assumptions the energy $w$ is given by
\begin{equation}
  w = \frac{1}{2} \left[ \frac{K_a}{1-\nu^2}\left((\lambda_s-1)^2 + 2\nu
      (\lambda_s-1)(\lambda_\phi-1) + (\lambda_\phi-1)^2\right) 
+ \gamma \lambda_s \lambda_\phi \right] \textrm{.}
\end{equation}
To obtain the tensions one takes the partial derivatives of the energy with
respect to the deformations; to be precise,
\begin{align}
  \label{eq:tension-s}
  T_s = \frac{1}{\lambda_\phi}\pd{w}{\lambda_s} =  \frac{K_a}{\lambda_\phi(1-\nu^2)} \left( \lambda_s + \nu \lambda_\phi - (1+\nu) \right) + \gamma
  \textrm{,} \\
  \label{eq:tension-phi}
  T_\phi = \frac{1}{\lambda_s}\pd{w}{\lambda_\phi} = \frac{K_a}{\lambda_s(1-\nu^2)} \left( \lambda_\phi + \nu \lambda_s - (1+\nu) \right) + \gamma
  \textrm{,} \\
\end{align}
which are the tensions in the meridional and azimuthal directions, respectively.
These are inserted into \cref{eq_singular_surface_force} to obtain the force on
the interface. In these expressions $\nu$ denotes the Poisson ratio, which is
a material constant that couples the meridional and azimuthal deformations. In
the present work, $\nu = 0.3$ is used, which is a reasonable assumption for
elastic interfaces such as those considered here \cite{knoche2013}.

\subsection{Macroscale: numerical methods}
\label{sec:mac-methods}
Having described the equations governing the system, we now consider how to
solve these numerically. For the numerical methods, the equations must be
discretised in space and time. In
space, a structured, uniform and staggered grid is employed, and the
derivatives are discretised using standard second-order finite differences,
except for the convective term in \cref{eq:ns} which is discretised using the fifth-order WENO
scheme \cite{jiang1996,jiang2000}. Due to the coupling between pressure and
velocity, the Navier-Stokes equations are not a regular set of PDEs, but
technically a differential-algebraic equation with an index-2 constraint given
by the incompressibility equation. To solve this, we employ the pressure
projection method due to \citet{chorin1968}, which leads to a splitting scheme
for the time integration. In this scheme, we solve first for an 
intermediate (non-divergence-free) velocity field using an Euler step, then 
solve a Poisson equation for the pressure based on this intermediate velocity,
and finally use the computed pressure to project the velocity field into the
space of divergence-free velocity fields. The pressure Poisson equation takes up
the bulk of the computation time, and much work has gone into developing fast
numerical methods for this equation. In the present work we employ the
BoomerAMG \cite{henson2000} preconditioned BiCGStab \cite{vandervorst1992}
method, through the Hypre \cite{hypre}
and PETSc \cite{balay1997} libraries, respectively. To obtain a larger stability
domain than with only an explicit Euler step, several Euler steps are combined
to form a Runge-Kutta step (following the approach by \citet{kang2000}),
specifically the SSP-RK(2,2) method is employed here (using the notation
of \citet{gottlieb2009}). The method does, however, remain first-order in time
due to the irreducible splitting error introduced by the projection step; see 
\cite{guermond2006} for a review of error reduction and of higher-order
projection methods for the Navier-Stokes equations.

This summarises how the single-phase Navier-Stokes equations are solved. To
extend this to two-phase flow, several methods are available. In previous work
we have employed the combination of the level-set and ghost-fluid methods, which
gives results that agree well with theory and experiments
\cite{teigen2009,teigen2010,teigen2010a,ervik2014,ervik2014c,ervik2016}, and which can
handle topological changes in the interface, \eg during drop coalescence. This
method can handle a varying interfacial tension, and has been coupled with the
Langmuir equation of state to simulate the effects of insoluble surfactants
\cite{teigen2010a,ervik2016}.

To simulate interfaces with tensions that include a Hookean elasticity, 
this scheme had to be extended, as we will discuss in the following. To this end, a hybrid
level-set/ghost-fluid/immersed-boundary method has been developed. To describe the
hybrid method, we first give a brief overview of each of the methods it is
constructed from. The development of the hybrid method is documented in greater
detail in the MSc thesis of \citet{lysgaard2015} for the
two-dimensional case; the method was extended to axisymmetry later.

\subsubsection{The level-set method}
To solve the Navier-Stokes equations for two-phase flow, knowledge
of the interface is required. For this, a popular choice is the level-set
method, originally proposed
by \citet{founding}. With this method the interface
is encoded as a signed scalar distance field 
\begin{equation}
  \varphi(\v{x},t) =
      \left\{
          \begin{array}{ll}
            \; \min_{\v{x}' \in \Gamma(t)} \| \v{x} - \v{x}' \|& \mbox{if } \v{x} \in \Omega_1 \\
            - \min_{\v{x}' \in \Gamma(t)} \| \v{x} - \v{x}' \| & \mbox{if } \v{x} \in \Omega_2
          \end{array}
      \right.
\end{equation}
This gives an implicit definition of the interface,
\begin{equation}
  \Gamma(t) = \{ \v{x} \in \Omega \mid \varphi(\v{x},t) = 0 \}.
\end{equation}
The interface moves according to the flow of the fluids. This advection is
performed directly with the function $\phi$, and for this reason the level-set
method is referred to as an implicit interface capturing method. The advection
equation is then 
\begin{equation}
  \label{eq_level_set_advection}
  \pd{\varphi}{t} + \v{\hat{u}} \cdot \v{\nabla} \varphi = 0,
\end{equation}
where $\v{\hat{u}}$ is the fluid velocity field at the interface,
extrapolated to the whole domain (as suggested by \citet{Adalsteinsson1995269}), which can be found by solving
\begin{equation}
  \label{eq_velocity_extrapolation}
  \pd{\v{\hat{u}}}{\tau} + S(\varphi)\v{n} \cdot \v{\nabla} \v{\hat{u}} = 0, \quad \v{\hat{u}}|_{\tau=0} = u,
\end{equation}
Here $\tau$ is a pseudo time, and $S$ is a
smeared sign function which is zero at the interface, $S(\varphi)
= \varphi/\sqrt{\varphi^2 + 2\Delta^2}$. We assume here that the Eulerian grid
spacings are equal and denote these by $\Delta_x = \Delta_y = \Delta$. This equation is in principle solved
to steady state, \ie $\tau \to \infty$. In a recent paper,
\citet{sabelnikov2014} presented an alternative approach which appears
promising, since it has a lower computational cost.

As the level-set field is advected by \cref{eq_level_set_advection}
it will become distorted and lose its signed distance-property.
Because of this, the level-set function is reinitialised at regular
intervals by solving
\begin{align}
  \label{eq_level_set_reinit}
  \pd{\varphi}{\tau} + S(\varphi_0)(|\v{\nabla} \varphi| - 1) &= 0,\\
  \varphi(\v{x},0) &= \varphi_0(\v{x}),
\end{align}
to steady state \cite[(7.4)]{osher2012level}.
Even though \cref{eq_velocity_extrapolation} and \cref{eq_level_set_reinit} are defined
for the whole domain, we are only interested in the extrapolated velocity and the
reinitialised field in a neighbourhood around the interface.

Interestingly the characteristics of both \cref{eq_level_set_advection},
\cref{eq_velocity_extrapolation} and \cref{eq_level_set_reinit}
originate at the interface and point outwards. This implies that if one
solves the equations not for $\tau \to \infty$, but rather for $\tau \to
N \Delta$, $N$ \eg equal to 3, it gives a level-set function which is correct in a narrow
band of width $3 \Delta$ around the interface. This significantly reduces
the computational cost of the method; see \eg \citep{Adalsteinsson1995269} for
further details.

The properties required to calculate forces coming from a fluid interface are the
interface normal vectors and curvature.
Both of these are computed directly from $\varphi$,
\begin{equation}
  \label{eq_level_set_curvatures}
  \v{n} = \frac{\v{\nabla} \varphi}{|\v{\nabla} \varphi|}, \quad
  \kappa = \v{\nabla} \cdot \v{n} .
\end{equation}

\subsubsection{The ghost-fluid method} 
\label{sec_ghost_fluid_method}
Given an interface-capturing method such as the level-set method, yet another
method is required to impose the difference in material properties and the
interfacial tension. Different methods are available, and one distinguishes
between sharp-interface and smeared-interface methods. A sharp-interface method (such
as the ghost-fluid \cite{fedkiw1999} or immersed-interface methods
\cite{leveque1997}) is more accurate, but also
more difficult to implement, as compared to a smeared-interface method (such as the
continuum-surface-force \cite{brackbill1992} or immersed-boundary
\cite{Peskin1977220} methods).

With a smeared-out
method, a mollified delta-function is used to spread a singular force out to
several grid cells. With such an approach,
the normal finite-difference approximations to derivatives can be used as there are no
discontinuities in the solution, but rather very steep, smooth transitions.

By contrast, with the ghost-fluid method as used in this work, the discontinuities are incorporated directly
into the numerical stencils. This means that there is an actual jump in the solution,
and jump conditions are used to
relate the values across the interface. For the case of two-phase flow with
a constant interfacial tension, the jumps are given by
\begin{align}
  \label{eq:ujump}
\llbracket \v{u} \rrbracket &=0 \textnormal{,}\\
\label{eq:pjump}
\llbracket p \rrbracket &=2\llbracket \mu \rrbracket \v{n}\cdot
{\grad \v{u}}\cdot\v{n} - \gamma\kappa \textnormal{,}\\
\label{eq:gradujump}
\llbracket \mu {\grad \v{u}}\rrbracket &=\llbracket \mu \rrbracket \Bigl(
(\v{n}\cdot {\grad \v{u}}\cdot\v{n})\v{n}\v{n}+(\v{n}\cdot {\grad
\v{u}}\cdot \v{t})\v{n}\v{t} \nonumber \\
&\qquad\;- (\v{n}\cdot {\grad \v{u}}\cdot \v{t})\v{t}\v{n}+(\v{t}\cdot
  {\grad \v{u}}\cdot \v{t})\v{t}\v{t} \Bigr)
\end{align}
Here $\v{n},\v{t}$ are the normal and tangent vectors at the interface, and we
denote tensors formed by the outer product as \eg $\grad\v{u}$. We take the
normal vector to be pointing outwards on a drop, and then the jump
$\llbracket\cdot\rrbracket$ is the difference between the external and internal
values, \eg $\llbracket\mu\rrbracket = \mu_2 - \mu_1$. 
It should be noted that these expressions have been written in
a form that gives faster code when implemented, see \citet{lervaag2013} for
a derivation and for a more thorough description of the ghost-fluid method.

In the hybrid method developed here, these equations are used without the term
$\gamma\kappa$ in \cref{eq:pjump}, since the tension is then handled by
the immersed-boundary method instead. The term is included when the regular 
level-set/ghost-fluid method is used as a reference for testing the hybrid method.

\subsubsection{Motivation for the hybrid method}
\label{sec_immersed_boundary_motivation}
To compute the tensions using \cref{eq:tension-s,eq:tension-phi} requires knowledge of the
interfacial deformations $\lambda_s,\lambda_\phi$.
We may prove that the level-set function, or any similar scalar marker function, does not contain the
information required to compute this. Equivalently, the marker function
does not contain information about compression or stretching of the interface.
To have compression or stretching of an interface in incompressible flow,
assuming no sources or sinks are present,
the velocity component tangential to the interface has
to be nonzero. By considering the projection operator
\[
    P_{\parallel}(\varphi) = \left( 1 - \nabla \varphi (\nabla \varphi \cdot \; ) \right)
\]
which projects $\v{u}$ into the space of velocity fields that
are tangential to the interface, one may easily prove, using the signed distance property of
$\varphi$, that only the velocity component normal to the interface gives
a non-zero contribution in the advection equation.
In other words, the interface representation $\phi$ is invariant under velocity fields
tangential to the interface. To store information about interfacial compression or
stretching in an interface capturing method which uses an Eulerian marker
function, one must resort to additional data structures to represent interfacial
strain.

Alternatively, we may consider a hybrid interface-tracking method. Such
methods have been successful at combining the best features of several
methods, \eg in the coupled level-set/volume-of-fluid (CLSVOF) method
\cite{sussman2000}. In the present work we have developed a hybrid
level-set/ghost-fluid/immersed-boundary method. The immersed boundary method
provides not only the required information about compression and stretching,
but is also widely used and thoroughly tested with a general tension (\ie
elasticity and interfacial tension). Originally developed
for simulating biological systems, \eg blood flow through a heart
\cite{Peskin1977220}, the immersed boundary method has been successfully applied to the simulation of
red blood cells \cite{fai2013}, which
have similar properties to drops covered
with elastic membranes. 

Another important motivation for using the immersed boundary method is that
it allows for refining the discretisation of the interface independently of the
Eulerian grid. This increases the accuracy of the interfacial representation
while the Eulerian grid remains the same, as indicated by the results of
standard interface-capturing method test cases in
\cref{sec:macro-val}. In particular for the crumpling drop case of interest
here, this represents a large saving of computational cost, since crumpled
interfaces like those discussed in Section 5, represented using the level-set
function, would require at least an order of magnitude more Eulerian grid points
than what is required to represent the flow field with sufficient accuracy. This 
would cause simulation times to be at least two orders of magnitude larger (one order of
magnitude from increased cost in the Poisson solver, and one from the increased
number of time steps required due to the stability condition). The simulations
considered in Section 5 have runtimes of a few days running in serial;
parallelisation would give some improvement in the time-to-solution, but even
state-of-the-art solvers for the pressure Poisson equation show limited
performance gains when the number of unknowns per CPU core is below $O(10 000)$.
Thus if one was to use an extended
level-set/ghost-fluid method, having data structures to represent interfacial
strain, and a parallelised code, the runtime for one of these cases would be of
the order of months and the effort to implement the method would be much larger.

The reason for not completely switching to the
immersed-boundary method is that the handling of density and viscosity
differences across the interface is less accurate than with
the level-set/ghost-fluid method; with the immersed-boundary approach the
contributions from the viscosity difference in the jump conditions
\cref{eq:pjump,eq:gradujump} are typically not taken into account
(see \eg \cite{francois2003}), similar to when the level-set method is used
together with the continuum-surface force
method \cite{brackbill1992}. The reason for retaining the level-set method is
that this eases the implementation, and that it may allow for simulations of
drop coalescence in an extended version of the hybrid method, since the
level-set method handles changes in the interface topology very well.

\subsubsection{The immersed boundary method}
The key point of the immersed boundary method \cite{ANU:165651} is to allow
solving the Navier-Stokes equations, or other continuum equations, on an Eulerian regular grid, while
handling a large class of arbitrary deformable or rigid bodies embedded
in the same domain. These bodies are described by Lagrangian coordinates.
Thus a key element of the method is the transformations between the Lagrangian
and Eulerian coordinates, and vice versa.

In addition to flexible interfaces as considered here,
the immersed-boundary method can be used to simulate rigid
bodies. This has been widely employed for simulations in complex domains. With
this approach, the equations become very stiff, and thus the implicit forcing
method has been constructed \cite{uhlmann2005}. For the case of interest here, namely flexible
interfaces, the explicit forcing method as used in the original immersed-boundary
method is sufficient. Even with an explicit time integration scheme, the method
is stable given that the time step is sufficiently small \cite{tu1992}.

When using the immersed-boundary method to implement a generalised interfacial
tension, we follow the procedure given in \cite{Peskin1995265}.
The interface is imagined as a continuum of elastic fibres immersed in the fluid.
These fibres serve as a device for deriving the model. They do not have a mass
nor do they occupy a volume,
but together with the fluid they are immersed in, they act as a viscoelastic material.
The fibres are arranged in a mesh parametrised by three space
coordinates, which we take to be  $(\phi,r,s)$ with reference to
\cref{fig:tensions-sphere}.
With this framework, fixing two of the space coordinates, \eg $(\phi,r)$, uniquely determines
a fibre. The last coordinate, $s$, is then a parametrisation along the elastic
fibre given by the fixed values of $(\phi,r)$.

For the case of interest here, namely a drop interface that has no thickness, one coordinate
is given by the other two, \ie $r = r(\phi,s)$; for a spherical drop $r$ would
be constant. Moreover, we consider the axisymmetric case, meaning that $r
= r(s)$ and that nothing depends on $\phi$. This means we consider the situation
illustrated in \cref{fig:tensions-sphere}, \ie a single fibre going in the meridional (or $s$)
direction of the drop, which is discretised by many points. For each such point
there is also a fibre going in the azimuthal direction, which is not
discretised.

Using Peskin's notation, we write
the strains as \eg $\lambda_s = |\pd{\v{X}}{s}|$. We then have \cite{Peskin1995265} the forces from the
interface acting on the fluid given as
\begin{equation}
  \label{eq_elastic_force_density}
  \v{f} = \pd{T_s}{s}\v{t} + T_s \left| \pd{\v{X}}{s} \right| \kappa_s \v{n}
  + T_\phi\left| \pd{\v{X}}{\phi} \right| \kappa_\phi \v{n} \textrm{.}
\end{equation}
From this we see that the force consists of
a component along the fibre in the direction $\v{t}$,
as well as a component in the principal normal direction, pointing towards the
centre of the osculating circle of the curve, $\v{n}$. As previously noted, 
there is no force in the binormal direction, $\v{t} \times \v{n}$.

If we assume no elasticity and a constant interfacial tension, $T_s = T_\phi
= \gamma$, \cref{eq_elastic_force_density} becomes
\begin{equation}
  \v{f} = \gamma \left(\left| \pd{\v{X}}{s} \right| \kappa_s + \left|
  \pd{\v{X}}{\phi} \right| \kappa_\phi \right) \v{n}
\label{eq_general_tension_surface_tension}
\end{equation}
and the tangential force disappears. This corresponds to the normal
two-phase flow situation with a simple interface described only by interfacial
tension, and will serve as a test case
for the hybrid method. The deformations entering into this expression serve as
normalisation factors, since the original expressions are derived with reference
to the undeformed coordinate system. This point may be confusing at first; the 
reader is referred to the thorough derivation in \cite{Peskin1995265}. 
The parenthesis in this expression corresponds to the mean curvature of the
drop, so the entire expression corresponds to the familiar Laplace-Young formula.

To implement the tensions numerically requires computing the interfacial
deformations. Let
\begin{equation}
  \| \v{X} \|_i^k = \| \v{X}_k - \v{X}_i \|
\end{equation}
be the Euclidean distance between Lagrangian points $i$ and $k$.
 For simplicity we restrict the disposition here to the fully two-dimensional
 case, where $T_\phi = 0$ and the tension $T$ is given as
 \begin{equation}
   T_s = K_a \left( \left|\pd{\v{X}}{s}\right| - 1 \right) + \gamma
  \label{eq:tension-2d}
 \end{equation}
Discretising this equation gives the following
expression for the tension at point $i$, $T_{s,i}$:
\begin{equation}
  \label{eq_disc_general_tension}
  T_{s,i} = K_{a} \left( \frac{\| \v{X} \|_i^{i+1} + \| \v{X} \|_{i-1}^i}{l_i
  + l_{i-1}} - 1 \right) + \gamma,
\end{equation}
where $l_i$ is the equilibrium length between point $\v{X}_i$ and $\v{X}_{i+1}$.
This gives the tension for each Lagrangian point along the boundary, which is 
then used in a discretised version of \cref{eq_elastic_force_density},
\begin{equation}
  \v{f}_i = \frac{T_{s,i+1}-T_{s,i-1}}{2} \v{t}_i + T_{s,i} \frac{\|\v{X}\|^{i+1}_i
  + \|\v{X}\|^i_{i-1}}{2} \kappa \v{n}
\end{equation}
to compute the discretised interfacial force. This force enters the right-hand
side of the discretised Navier-Stokes equations on the
Eulerian grid points close to the interface. The force is smeared out to these
points using the mollified delta function, which is made up of combinations of
one-dimensional mollified delta functions $\delta(r)$.

In contrast to other
smeared-interface methods, the delta function in the immersed-boundary method is
uniquely determined by six requirements on the properties of this function.
The reader is referred to \cite[sec. 6]{ANU:165651}, \cite{Peskin1995265}, as well as the previously
mentioned MSc thesis by \citet{lysgaard2015}, for details of the delta function
construction, as well as description of the spreading of Lagrangian quantities to the Eulerian
grid, and the interpolation in the opposite direction. The resulting
one-dimensional delta function, which is the basis for both the interpolation
and spreading operations, is
\begin{equation}
  \label{eq_deltafunc}
  \delta(r) =
  \begin{cases}
           \frac{1}{8} \left( 3 - 2|r| +\sqrt{  1 + 4|r| - 4r^2} \right), & |r| \leq 1\\
           \frac{1}{8} \left( 5 - 2|r| -\sqrt{- 7 + 12|r| - 4r^2} \right), & 1 \leq |r| \leq 2\\
           0,& \quad 2 \leq |r|
  \end{cases}
\end{equation}
where $r$ is the distance \eg from a Lagrangian point to the Eulerian grid cell
centre.

We remark that if the distance between two Lagrangian points is too big, the 
spreading operations using the delta function will not approximate the continuous versions correctly.
For this reason it is required that two Lagrangian points never be further
apart than half the width of an Eulerian grid cell.

In the proposed method, cubic splines are used to
generate a smooth analytic parametrisation of the interface. The main advantage of this is that properties
such as the curvature, tangent- and normal vectors are all naturally defined for a cubic spline.
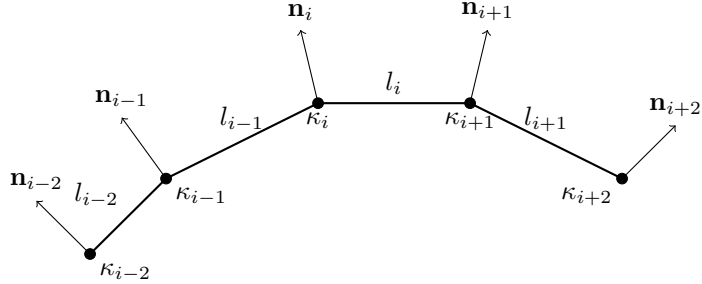
\begin{figure}[tbp]
\begin{center}
\begin{tikzpicture}[
    interface/.style={
        % The border decoration is a path replacing decorator. 
        % For the interface style we want to draw the original path.
        % The postaction option is therefore used to ensure that the
        % border decoration is drawn *after* the original path.
        postaction={draw,decorate,decoration={border,angle=-45,
                    amplitude=0.3cm,segment length=2mm}}},
    arrow/.style={
        ->,
        %shorten <=2pt,
        shorten >=0.1,}
    ]
  \draw[fill] (-1,-1) circle (2pt) coordinate (a) node[below right] {$\kappa_{i-2}$};
  \draw[fill] (0,0) circle (2pt) coordinate   (b) node[below right] {$\kappa_{i-1}$};
  \draw[fill] (2,1) circle (2pt) coordinate   (c) node[below] {$\kappa_{i}$};
  \draw[fill] (4,1) circle (2pt) coordinate   (d) node[below] {$\kappa_{i+1}$};
  \draw[fill] (6,0) circle (2pt) coordinate   (e) node[below left] {$\kappa_{i+2}$};

  \draw[thick] (a) -- (b) -- (c) -- (d) -- (e);

  \coordinate (ab) at ($(a)!0.5!(b)$);
  \coordinate (bc) at ($(b)!0.5!(c)$);
  \coordinate (cd) at ($(c)!0.5!(d)$);
  \coordinate (de) at ($(d)!0.5!(e)$);

  \node[above left] at (ab) {$l_{i-2}$};
  \node[above] at (bc) {$l_{i-1}$};
  \node[above] at (cd) {$l_{i}$};
  \node[above] at (de) {$l_{i+1}$};

  \coordinate (abn) at ($(ab)!1!90:(b)$);
  \coordinate (bcn) at ($(bc)!1!90:(c)$);
  \coordinate (cdn) at ($(cd)!1!90:(d)$);
  \coordinate (den) at ($(de)!1!90:(e)$);

  \draw[arrow]   (a)   --   ++(180-45:1)   node[above]   {$\v{n}_{i-2}$};
  \draw[arrow]   (b)   --   ++(125.8:1)    node[above]   {$\v{n}_{i-1}$};
  \draw[arrow]   (c)   --   ++(103.3:1)    node[above]   {$\v{n}_{i}$};
  \draw[arrow]   (d)   --   ++(76.7:1)     node[above]   {$\v{n}_{i+1}$};
  \draw[arrow]   (e)   --   ++(45:1)       node[above]   {$\v{n}_{i+2}$};
\end{tikzpicture}
\end{center}
\caption{Part of immersed boundary grid showing where different values are located.}
\label{fig_ib_grid}
\end{figure}
\Cref{fig_ib_grid} shows 
the immersed boundary elements around the point with index $i$ together with
their different properties, and where they are defined. The cubic spline 
fitted to the points is only evaluated at the nodes. Because of this
the curvature is only available at these points.
The same applies to the normal vectors, which are directly calculated from the
first derivative of the spline at the nodes. On the other hand, line
segments are computed as the difference in position between two adjacent points.
This means that lengths are defined on the segments, and not on the nodes.
Note that the cubic spline going through the points is not
shown in this figure, but it is used to compute the curvature $\kappa$ and the normal
vector $\v{n}$. 

It should also be noted that the differences in
segment length are exaggerated in \cref{fig_ib_grid}, since the tangential term in the
interfacial force will very quickly eliminate such differences. This is in line with
what one expects from such longitudinal waves, which are known from
theory \cite{lucassen1968} and experiments \cite{lucassen1968b} to be extremely rapid.
This can also be understood intuitively since, in contrast to a regular 
capillary wave, a longitudinal surface wave
displaces essentially no mass, and thus inertial effects are very small.

When computing the interfacial force on each node, all variables are required at
the node. As mentioned, the segment-lengths are not stored at the nodes. An option would
be to use the cubic spline to calculate the length, but this requires the
numerical evaluation as well as inversion of an elliptic integral. To keep
the method simple, we approximate the
segment length as the average of the linear distance from the
node in question to its two neighbours. The curvature and unit normal vectors
remain analytically evaluated from the cubic spline.

In total, this approach encapsulates all the surface effects we need to simulate
in one coherent framework.
If, say, the elasticity of the material is a function of temperature, or the
elasticity is found to be non-Hookean, or some relaxation behaviour is observed that makes
the elastic modulus a function of the applied strain, these effects can easily
be accounted for by modifying \cref{eq:tension-2d}. 

\subsubsection{Computing the level-set function from the immersed boundary}
\label{sec_level_set_reinitialization_from_ib}
When using both the immersed boundary method and the ghost-fluid method to calculate
interface forces, special care has to be taken to make the methods consistent.
The following technique is proposed for that. The geometry is completely determined by the
Lagrangian points along the interface. In each stage of the time integration
method, the shortest distance from the Eulerian points to the Lagrangian boundary is computed.
In other words, we compute the level-set function purely from the immersed boundary.

This has several advantages. First, advection is moved from the level-set function
to the immersed boundary. When no advection of the level-set function is required,
it is no longer needed to reinitialise it, \cref{eq_level_set_reinit}, or extrapolate the velocity, \cref{eq_velocity_extrapolation}. These routines
are somewhat costly, and their saving leads to a $\sim$ 25\% reduction
in wall clock run time for some typical two-phase simulations.
Second, using this approach, the level-set function is always the best possible approximation to the
exact signed distance function for the given Eulerian grid. 
Third, given equal initial conditions for the immersed boundary and the level-set field,
the two descriptions of the interface may not be consistent with respect to each other, meaning
that after some time, $t$, the advection of the level-set function and the Lagrangian points
could cause the two methods to have slightly different locations for the
interface
\footnote{The reason for this is that the immersed boundary points can have
  sub grid details. This means that inside a grid cell there will be
  differences between the level set and the immersed boundary. Over time
  these will grow bigger than one grid cell because of advection. At that
  point, the two interface descriptions are not consistent with each
  other.}.
This is problematic because the interfacial forces
would appear at two different interfaces rather than one. This inconsistency disappears
when reinitialising the level-set function from the immersed boundary at every
timestep.

The algorithm for computing the level-set function from the Lagrangian points is as follows.
\begin{itemize}
  \renewcommand{\labelitemi}{$\hookrightarrow$}
  \renewcommand{\labelitemii}{$\hookrightarrow$}
  \renewcommand{\labelitemiii}{$\hookrightarrow$}

  \item Loop over the Lagrangian points representing the interface
    \begin{itemize}
    \item For each line segment connecting two points, compute its bounding box.
    \item Grow the bounding box such that it contains the widest Eulerian
      stencil used in the discretised reinitialisation equation.
    \item Loop over the Eulerian grid points inside the bounding box
      \begin{itemize}
        \item Compute the shortest distance from this grid point to the line segment using standard formulae.
        \item Compute whether the grid point is inside or outside of the closed
          interface using a standard point-inside-polyhedron algorithm.
        \item From these two results, compute the signed distance.
        \item If this is the smallest distance computed for this grid point so
          far in the outer loop, store it as the signed distance for this grid point.
        \end{itemize}
    \end{itemize}
\end{itemize}

\subsubsection{Penalisation method}
\label{sec_penalization_method}
As will be seen in \cref{sec_pipette_draining_drop}, we want to be able to simulate
solid objects in our domain, in addition to the two-phase flow with complex
interfaces. To achieve this we utilise a standard $L_2$ penalisation
method \cite{angot1999penalization} since it is very easy to implement (literally
just twenty lines of code) and since
it can also be used to enforce a flow field such as the desired suction inside a pipette.
With the penalisation method, the flow field exists inside the solid objects, but is
forced to be approximately equal to zero, or in general equal to a specified
field $\v{u}_{\rm{spec}}$, through a forcing parameter $1/\eta$ which enters in an additional term $(1/\eta) \chi (\v{u}_{\rm{spec}}
- \v{u})$
added to the right-hand side of the momentum equation. Here the scalar field
$\chi$ is a marker function for the solid body, so it is 1 inside the body and 0 outside
it. One can think of the penalisation term as an additional body force which outside the
body is zero and inside the body is proportional to the
difference between the actual and the prescribed flow field.
In \cite{angot1999penalization} proofs of the existence and uniqueness of solutions
with this method, as well as an error estimate, are given. 
The error is of the order of $\eta$.
To get good results, one would naively set $\eta=0$ and get zero error, but as
the time step needed for stability is proportional to $\eta$ there is the
usual trade-off between speed and accuracy.

\subsubsection{The time step restrictions}
In the simulations, the appropriate time step is adjusted
dynamically using the conditions given here, in order to have time steps as
large as possible without causing the method to become unstable.
Following \cite[sec 3.8]{kang}, we take the contribution from the 
advection term into account with
\begin{equation}
  \label{eq_cfl_conv}
  \text{M}_c = \frac{\max u_x + \max u_y}{\Delta} ,
\end{equation}
where $\Delta$ is the width of an Eulerian cell and $\max u_x, \max u_y$ are the
largest magnitudes taken by the velocity components in the simulation domain.
The contribution of the viscous stress to the time step restriction
is taken into account with
\begin{equation}
  \label{eq_cfl_visc}
  \text{M}_v = \max \left( \frac{\mu_1}{\rho_1},
  \frac{\mu_2}{\rho_2} \right) \frac{4}{\Delta^2}.
\end{equation}
These are combined with the contribution from the interfacial force $\v{f} = [f_x\; f_y]^T$ to form
the time step restriction 
\begin{equation}
  \label{eq_cfl_force}
  \frac{\Delta t}{2} \left( ( \text{M}_c + \text{M}_v ) + \sqrt{(
  \text{M}_c + \text{M}_v )^2 + \frac{4 f_x + 4 f_y}{\Delta}}
\right) \leq C
\end{equation}
From smearing out the interfacial force density $\v{F}$ using the mollified delta
function $\delta_\Delta$, we have that  $\v{f} = \v{F} \delta_{\Delta}$
and since $\delta_{\Delta} \le
\frac{1}{\Delta}$, cf.\ \cref{eq_deltafunc}, the time step restriction can be written as
\begin{equation}
  \label{eq_cfl_forcedens}
  \frac{\Delta t}{2} \left( ( \text{M}_c + \text{M}_v ) + \sqrt{(
    \text{M}_c + \text{M}_v )^2 + \frac{4 F_x
  + 4 F_y}{\Delta^2}} \right) \leq C.
\end{equation}
In this final condition, $C$ is the time step safety factor, typically $C=0.5$. 
Finally, as mentioned in the previous section, when the penalisation method is
used, the timestep must also fulfill $\Delta t \le \eta$ where $\eta$ is the
penalisation parameter.

\section{Summary of the proposed method}
\label{sec_proposed_method}
At this point we may assemble the proposed multiscale method in its entirety.
\begin{itemize}
  \item \textbf{At the nanoscale:} consider a tiny patch of the interface, $\sim 300$ nm$^2$:
    \begin{itemize}
      \item A volume around this patch is simulated using coarse-grained
        molecular dynamics.
      \item Accurate models for water, heptane, toluene and asphaltenes are used.
      \item The domain is elongated normal to the interface. Large systems of
        $\sim 10^6$ atoms are simulated.
      \item The interfacial tension $\gamma$ is computed from \cref{eq:ift-md}.
      \item Using volume-preserving deformations, the elasticity $K_a$ is computed from
        \cref{eq:elasticity-md}.
    \end{itemize}
  \item \textbf{At the macroscale:} two-phase flow simulation of drop with complex interfaces:
    \begin{itemize}
      \item Flow is governed by \cref{ns2,eq:ns}, solved using numerical methods
        described in \cref{sec:mac-methods}.
      \item The interface is handled with the new hybrid
        level-set/ghost-fluid/immersed-boundary method.
      \item Level-set/ghost-fluid method gives a sharp handling of density and
        viscosity jumps using \cref{eq:ujump,eq:pjump,eq:gradujump}.
      \item Immersed-boundary method gives accurate interface representation, and
        computes the tension $T$ with \cref{eq:tension-s,eq:tension-phi} 
        using $\gamma$ and $K_a$ from the nanoscale.
      \item The forces caused by $T$ are computed from
        \cref{eq_general_tension_surface_tension} and distributed from the
        Lagrangian points to the Eulerian grid using \cref{eq_deltafunc}.
      \item The level-set function is computed from the Lagrangian points using
        the algorithm in \cref{sec_level_set_reinitialization_from_ib}.
    \end{itemize}
\end{itemize}

\begin{figure}[htpb]
  \centering
  \includegraphics[width=0.8\linewidth]{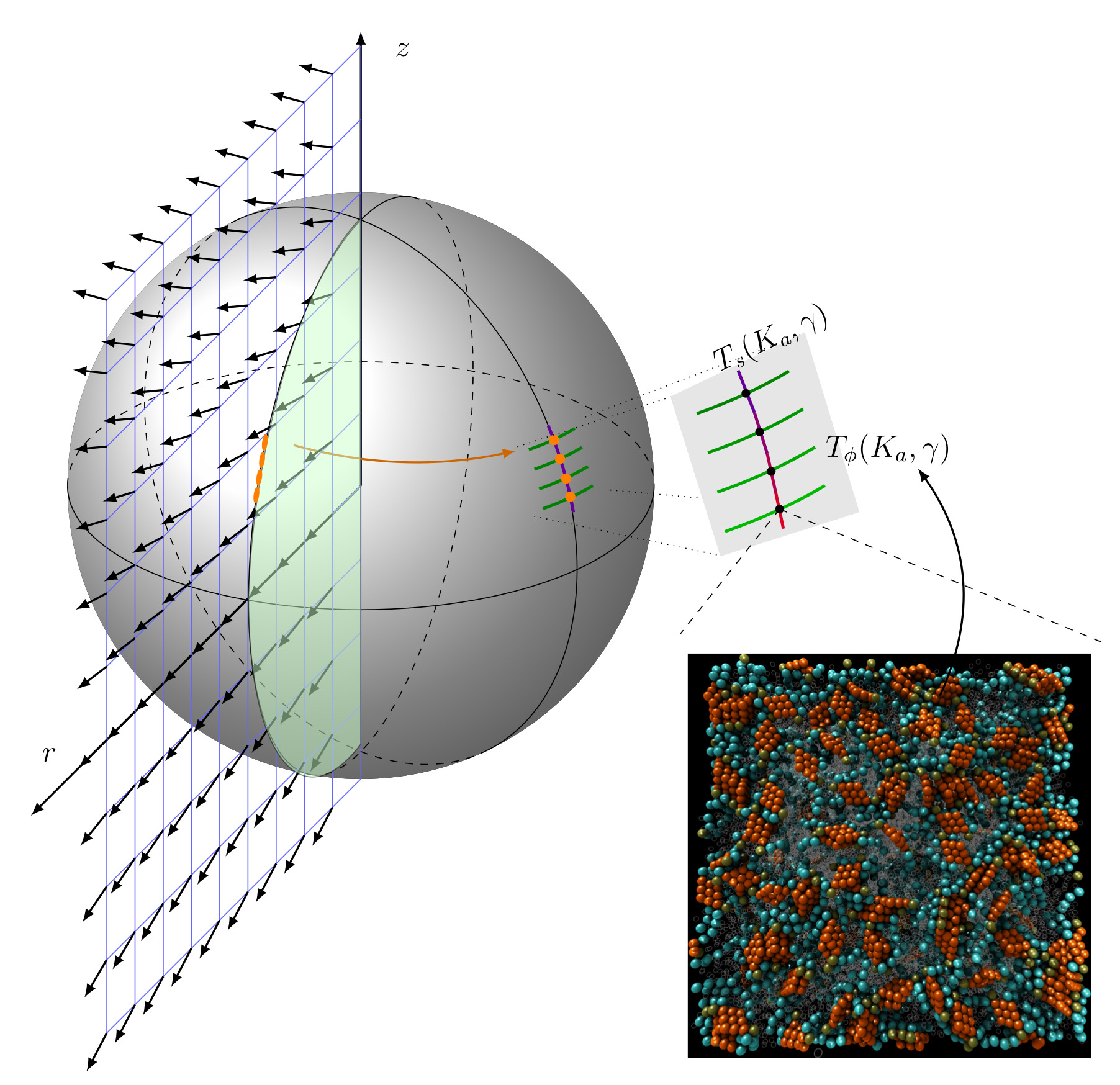}
  \caption{Illustration of the proposed method. On the Eulerian $(r,z)$ grid, the flow field $\v{u}$ (vectors) and the level-set function $\phi$
  representing the green-shaded portion of the droplet are shown. Some of the
  Lagrangian immersed-boundary points are shown in orange, with lines as before indicating
  the tensions $T_s$ and $T_\phi$. For a tiny patch on the interface, we compute
  the interfacial properties $K_a$ and $\gamma$ using molecular dynamics
  simulations (lower right corner). These properties are used in the
  calculations of the tensions. Note that the colours used in the molecular
  dynamics simulation snapshot do not have any relation to the other colours
  in the figure.}
  \label{fig:method-summary}
\end{figure}

In \cref{fig:method-summary} the method is summarised, showing the vectors and
interface representation living on the Eulerian grid, the immersed boundary
Lagrangian points and the tensions acting on them, and in the corner the
molecular dynamics simulation which represents a tiny patch of the interface and
is used to estimate the properties $K_a$ and $\gamma$.

We will now proceed to validate the different components that make up this
method. We begin in \cref{sec:macro-val} with the macroscale method, and
continue in \cref{sec:micro-val} with the nanoscale, where the models for different fluids and
for the asphaltene molecules are considered.

\section{Validation}
\label{sec:validation}

We will now present validation results for both the nanoscale and the
macroscale models. We first consider the hybrid macroscale method and
demonstrate that this method gives the correct results on several test
cases.  Then we consider the nanoscale models for both the simple fluids
and the more complex asphaltene molecules.

\subsection{Macroscale: validation}
\label{sec:macro-val}
In this section we will demonstrate the validity of the developed hybrid method.
We start by demonstrating the superior resolution of the immersed-boundary
method over the standard level-set method, as previously mentioned. To this end
we employ two standard test cases for interfacial advection, namely the drop in
vortex test and Zalesak's disk test.

Following this, we consider the case of an initially elongated spheroidal drop
relaxing under interfacial tension. The case is considered both for
a two-dimensional and for an axisymmetric drop, and it is also considered both
with and without density and viscosity differences across the interface. In
summary, we demonstrate that the methods converge to the same solution under
grid refinement. For further verification, see \cite{lysgaard2015}. For the
continuum simulations presented in this work, we provide tables with details of
the configuration and parameters used in Appendix A.

\subsubsection{Drop in vortex}
\label{sec_drop_in_vortex}
A standard test of advection for interface-tracking methods is the
drop in a potential vortex~\cite{Leveque1996}. Here a drop is placed in the unit box, and a static
potential vortex advects it. The velocity field is given by
\begin{align}
  u_x &= -2 \big[\sin(\pi x-\sfrac{\pi}{2})\big]^2 \cos(\pi y-\sfrac{\pi}{2}) \sin(\pi y-\sfrac{\pi}{2}) \\
  u_y &= -2 \big[\cos(\pi y-\sfrac{\pi}{2})\big]^2 \sin(\pi x-\sfrac{\pi}{2})
  \cos(\pi x-\sfrac{\pi}{2})
\end{align}
The remaining parameters for this test are given in \cref{tab_param_vortex}.

\begin{figure}[tbp]
  \centering
  \begin{subfigure}{0.3\textwidth}
    \includegraphics[width=\textwidth]{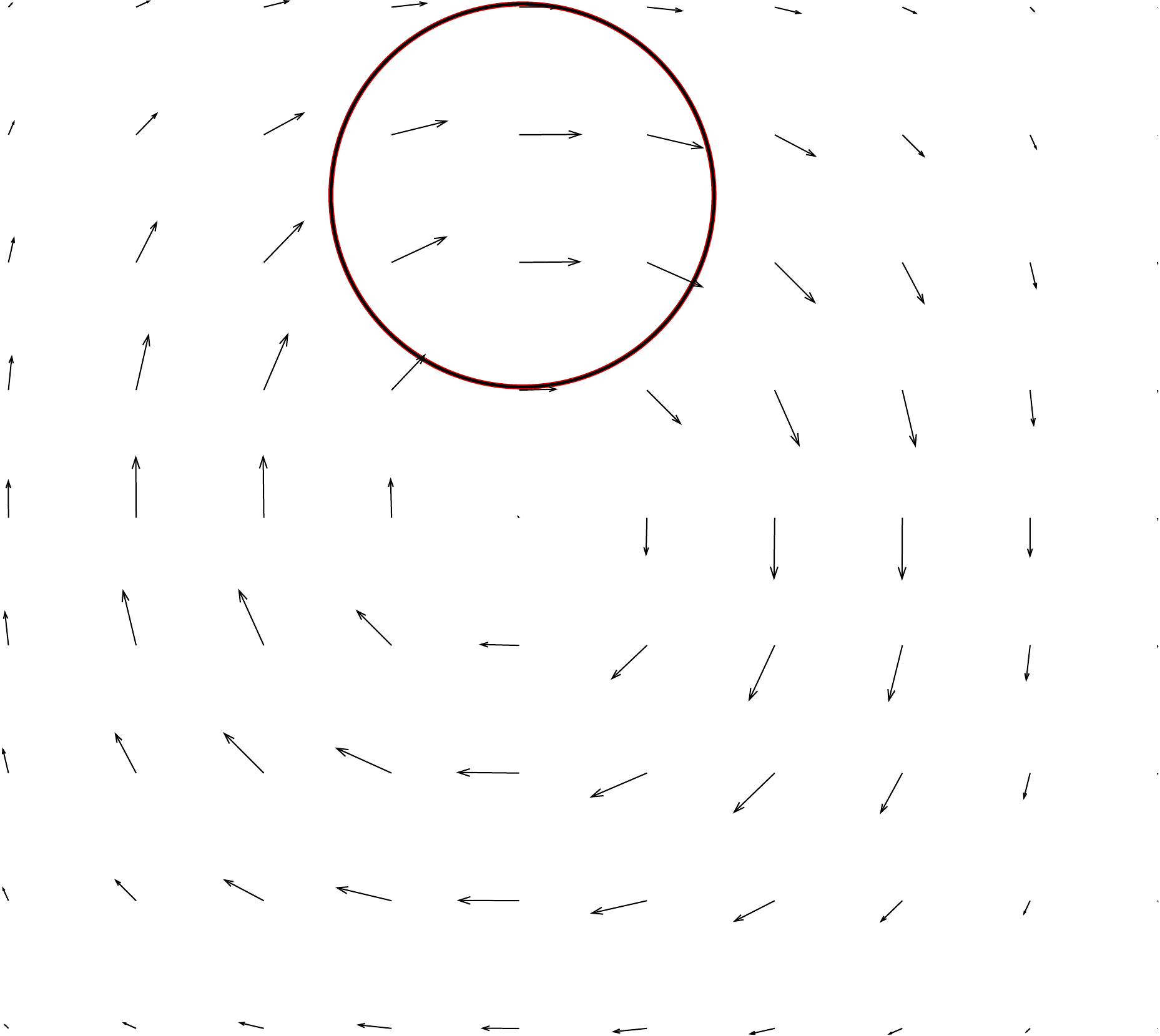}
    \caption{$t=0$}
    \label{fig_drop_in_vortex0}
  \end{subfigure}%
  ~ %add desired spacing between images, e. g. ~, \quad, \qquad, \hfill etc.
  \begin{subfigure}{0.3\textwidth}
    \includegraphics[width=\textwidth]{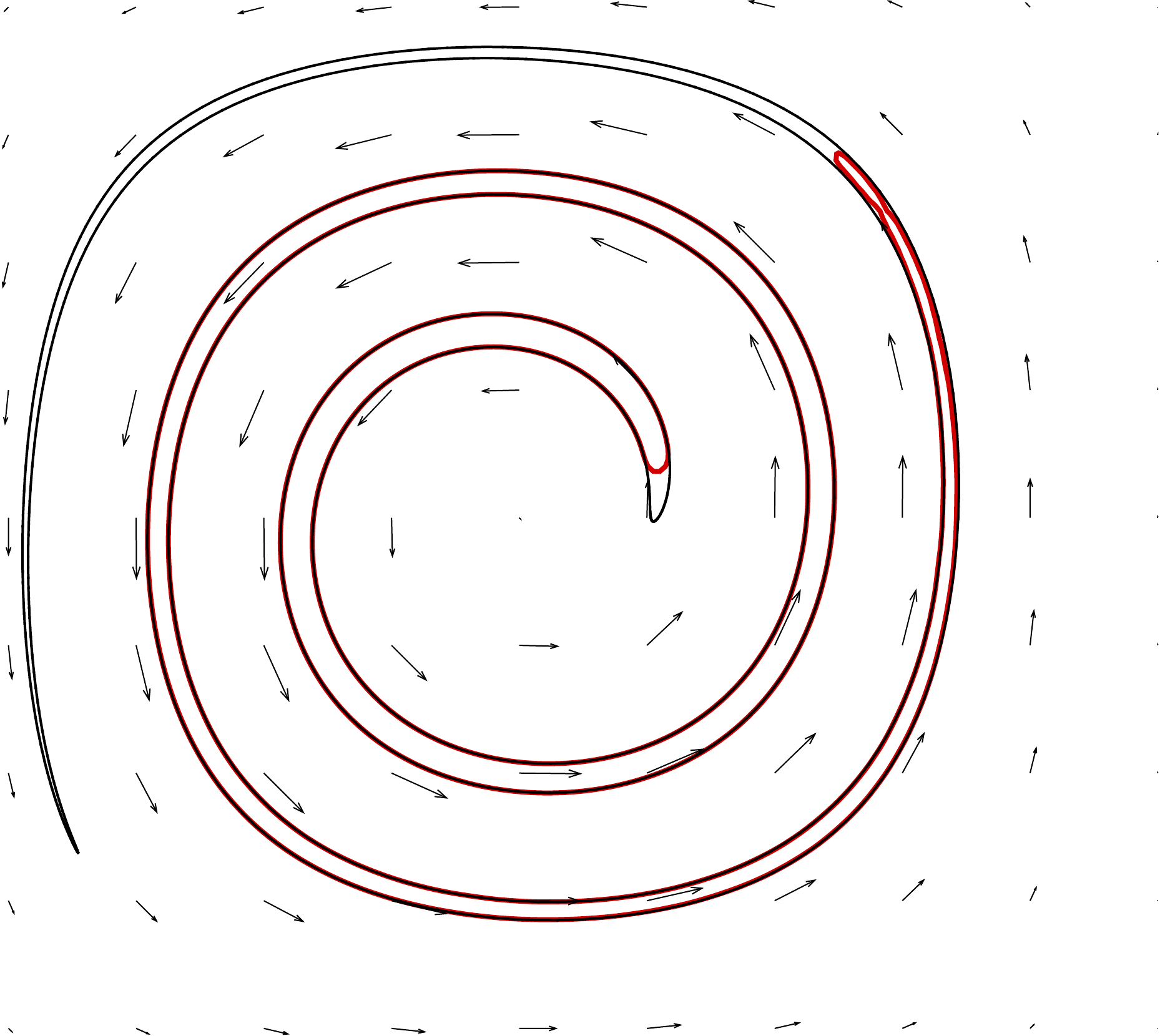}
    \caption{$t=3.5$}
    \label{fig_drop_in_vortex1}
  \end{subfigure}
  ~ %add desired spacing between images, e. g. ~, \quad, \qquad, \hfill etc.
  \begin{subfigure}{0.3\textwidth}
    \includegraphics[width=\textwidth]{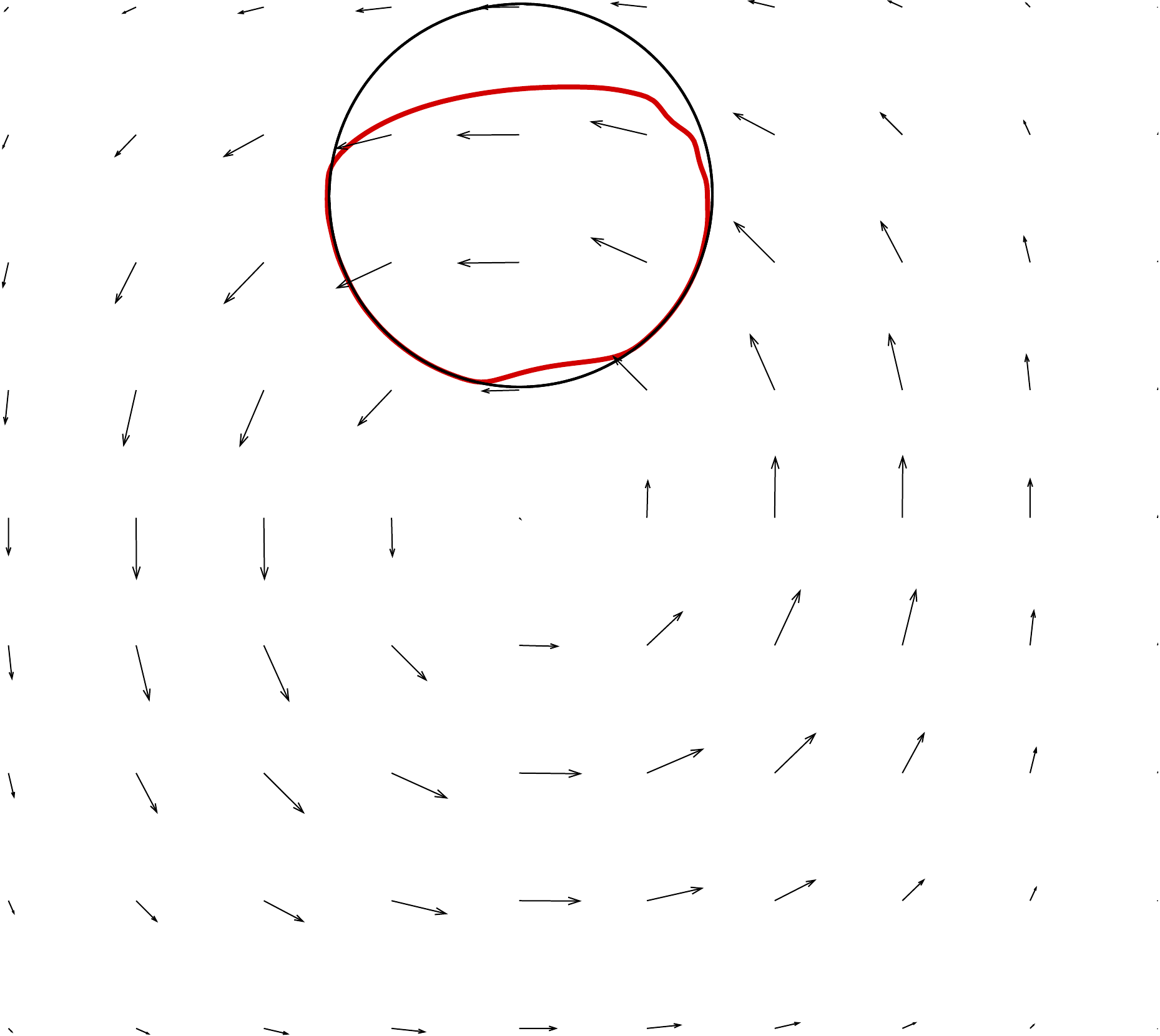}
    \caption{$t=7$}
    \label{fig_drop_in_vortex2}
  \end{subfigure}
  \caption{Drop in potential vortex. Red is level-set solution while black is
  the immersed boundary. Velocity vectors are shown for every 20 grid points in
  each direction.}
  \label{fig_drop_in_vortex_full}
\end{figure}

At some time $t=T/2$, the flow field is 
reversed, and the simulation is run until $t=T$. Then the initial interface is compared with
the final one. \Cref{fig_drop_in_vortex_full} shows the initial condition (a),
the interface at half time (b) where $t=T/2=3.5$,
and the final interface for both the level-set and the immersed boundary method,
in red and black colours, respectively.

We see that the immersed boundary method has no visible mass loss,
while the level-set representation loses mass when the drop gets
stretched thinner than a grid cell. 
The reason for the large level-set mass loss is that when two interfaces come
this close together,
the discrete level-set function does not have the required resolution
to switch sign. This follows from the Nyquist-Shannon theorem.
The immersed boundary method does not have this restriction.
If one wanted to represent the smaller features with level-set representation,
one choice would be to double the grid resolution. For two dimensions this would
make the computational cost increase quadratically. To get the same increase
in resolution with the immersed boundary method, one would need to double the
number of points, this would only double the amount of work needed. Thus
immersed boundary scales considerably better than level-set with respect to the 
interface resolution.

The previous argument makes immersed boundary seem superior to level-set
when it comes to resolution. However, this is not the whole story.
For the immersed boundary to represent a non-smooth sub-grid feature, the
Lagrangian points have to be advected in a sub-grid way. With the immersed
boundary method, the Lagrangian points are advected using an interpolated
of the velocity field from the Eulerian grid. This means that the
highest wave number that can be created in the
immersed boundary representation \emph{by the flow} is proportional to $1/ \Delta$
(where $\Delta$ is the width of
the Eulerian grid cells).
For smooth velocity fields, like this potential vortex, the immersed boundary method
has some sub grid resolution. This is because it can accurately represent
stretching, squishing and other smooth transformations
that lead to sub grid details. Also, in the case of the interface crumpling as
will be discussed in later sections, the immersed boundary representation will
produce wrinkles inside each grid cell.

\subsubsection{Zalesak's disk}
Another interesting difference between an Eulerian and Lagrangian
description of geometry is the effect of grid alignment. This effect can be
seen in the next test, Zalesak's disk~\cite{Zalesak1979335}.  Here, a
slotted disk is put in a velocity field that has constant angular velocity,
corresponding to rigid body rotation. The boundary is advected one or
several revolutions and the result is inspected. Further details of the
case are given in \cref{tab_param_zalesaks}.

\begin{figure}[tbp]
  \centering
  \begin{subfigure}[b]{0.3\textwidth}
    \includegraphics[width=\textwidth]{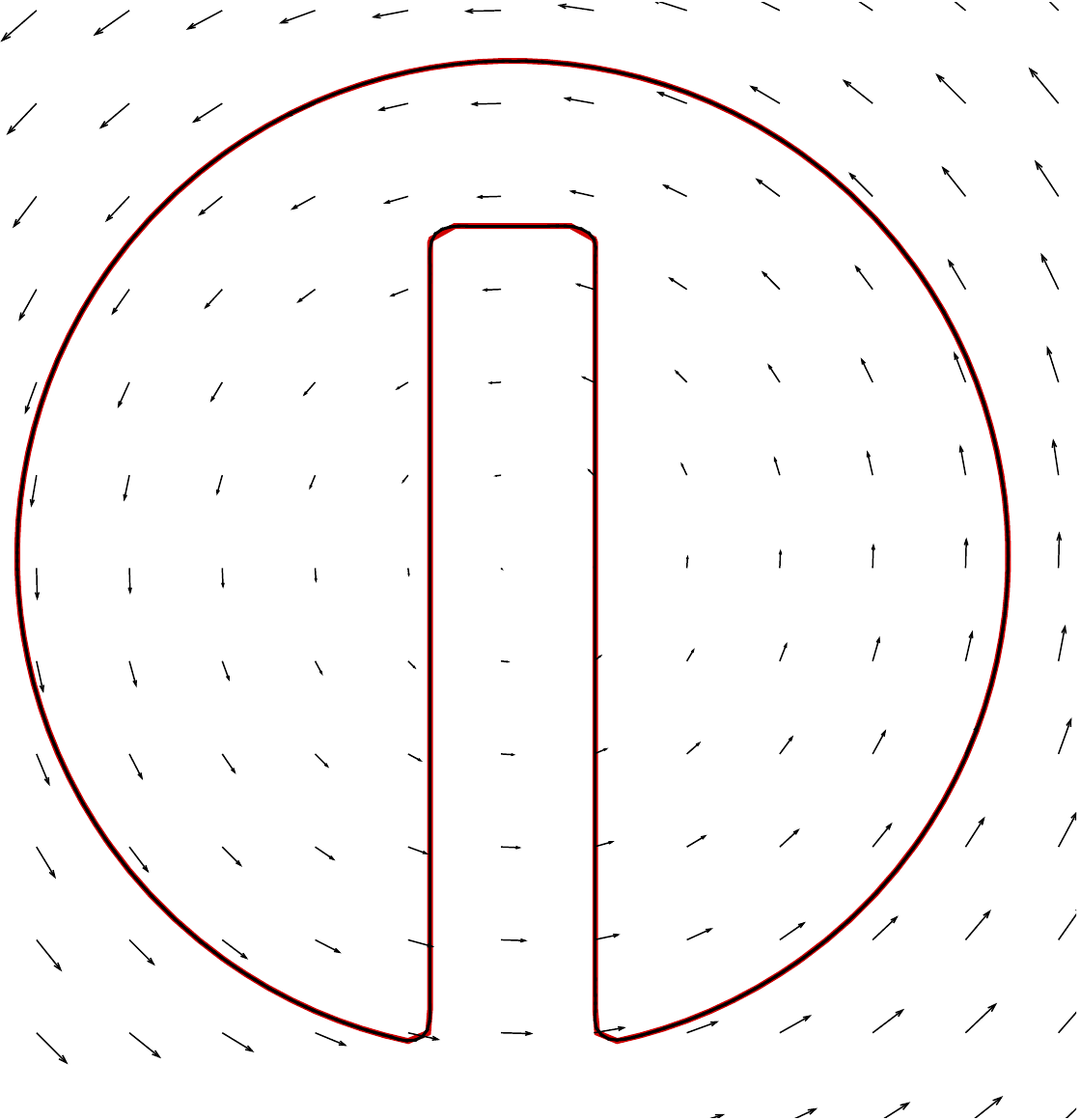}
    \caption{$t=0$}
    \label{fig_zalesaks0}
  \end{subfigure}%
  ~ %add desired spacing between images, e. g. ~, \quad, \qquad, \hfill etc.
  \begin{subfigure}[b]{0.3\textwidth}
    \includegraphics[width=\textwidth]{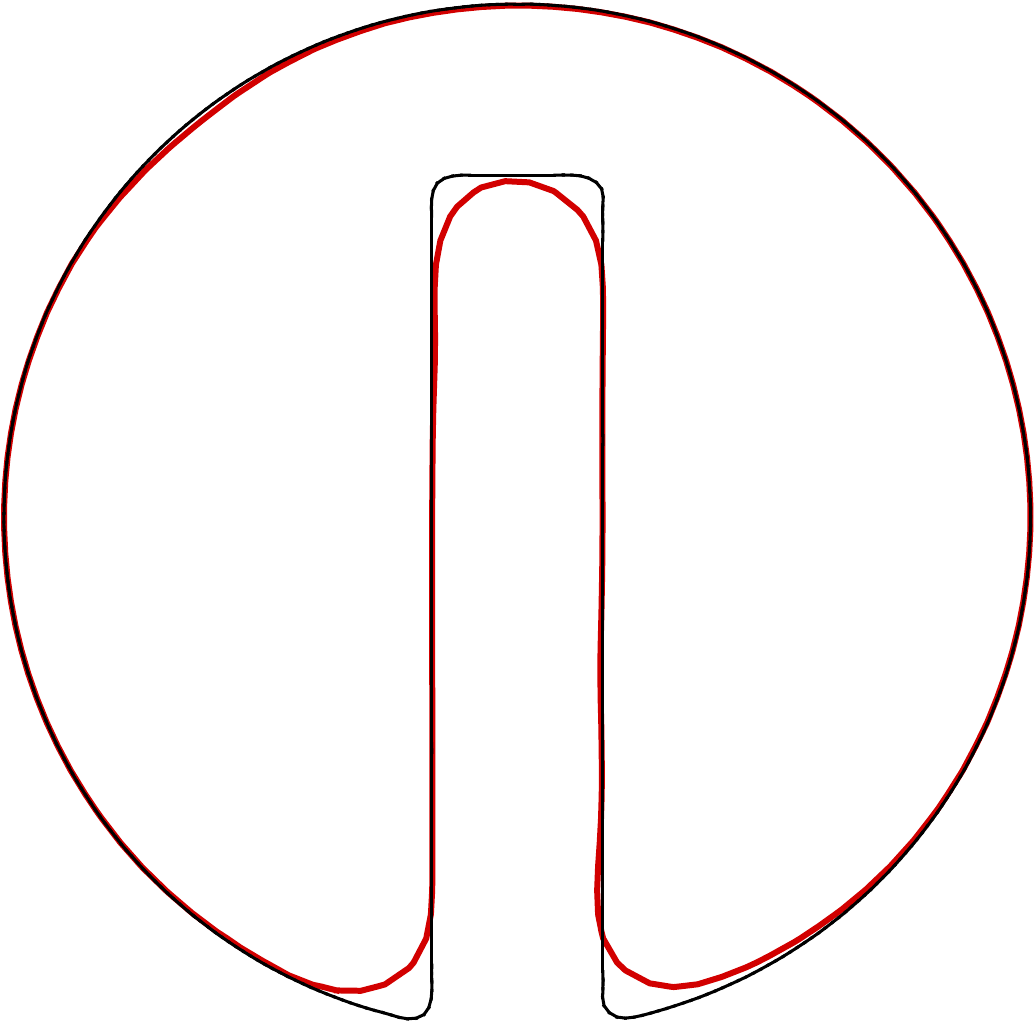}
    \caption{$t=40$}
    \label{fig_zalesaks1}
  \end{subfigure}
  ~ %add desired spacing between images, e. g. ~, \quad, \qquad, \hfill etc.
  \begin{subfigure}[b]{0.3\textwidth}
    \includegraphics[width=\textwidth]{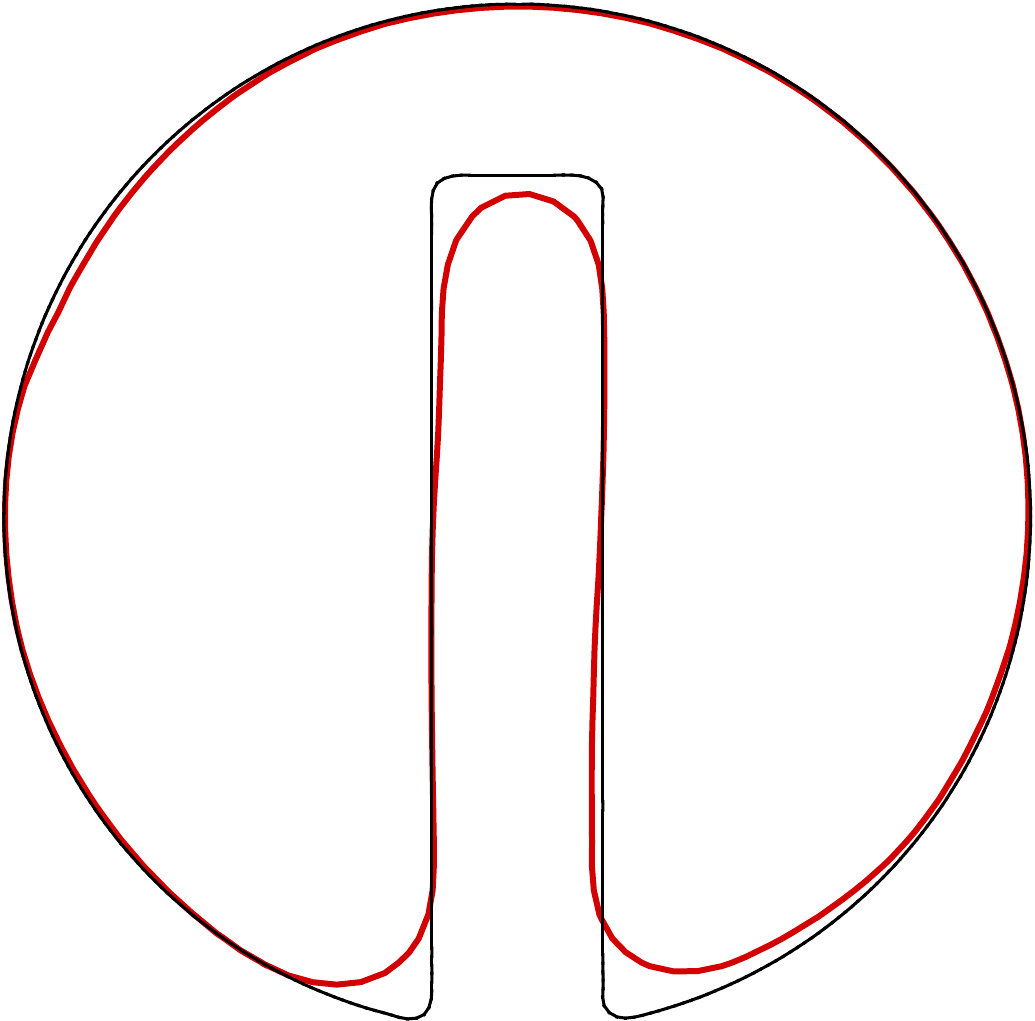}
    \caption{$t=80$}
    \label{fig_zalesaks2}
  \end{subfigure}
  \caption{Zalesak's disk for 0, 1 and 2 revolutions. Red shows level-set
  interface while black shows immersed boundary. The velocity field is constant
  in time and corresponds to rigid body rotation.}
\label{fig_zalesaks_full}
\end{figure}
From \cref{fig_zalesaks_full} it is clear that the immersed boundary 
resolves the rotated disk better than the level-set function.
During the rotation, information is lost in the level set, while the immersed boundary is
virtually not affected. The reason for this is that the level set, based on an Eulerian grid,
cannot represent non-smooth features that are not aligned with the grid perfectly. This means
that over the duration of the rotation, small errors in the interface position creeps in as
a consequence of the interface not being straight and aligned with the grid.
In the immersed boundary method, the grid has no preference about the orientation of the interface.
When it comes to the drop-in-vortex test, \cref{sec_drop_in_vortex}, one may
argue that the difference between the two
methods is exaggerated by the specifics of the test, to the detriment of the
level-set method.
There does not seem to be any such argument for Zalesak's disk. The immersed boundary method is fundamentally
better at preserving non-smooth features like corners without smearing. In real
life, non smooth interfaces occur \eg when two drops coalesce.

\subsubsection{Comparison with reference method}
We have now verified that the immersed boundary method captures the interface correctly
under advection. Next we need to verify that the forces from the boundary on the fluid are
implemented correctly. The handling of the viscosity and density jumps must also be verified
to be correctly coupled with the immersed boundary.
The technique chosen for this was to compare the proposed method, \cref{sec_proposed_method},
with a reference method, the level-set method with the ghost fluid method,
which has previously been verified and validated
\cite{teigen2009,teigen2010,teigen2010a,ervik2014,ervik2014c,ervik2016}.
To have some measure of the drop dimensions during oscillations, 
the horizontal and vertical axis lengths are used. These
parts of the drop are the ones most rapidly advected, with the highest pressure differences
and curvatures. Thus any differences between the two methods would be most
pronounced at these points.

\begin{figure}[tbp]
  \centering
  \includegraphics[width=0.8\linewidth]{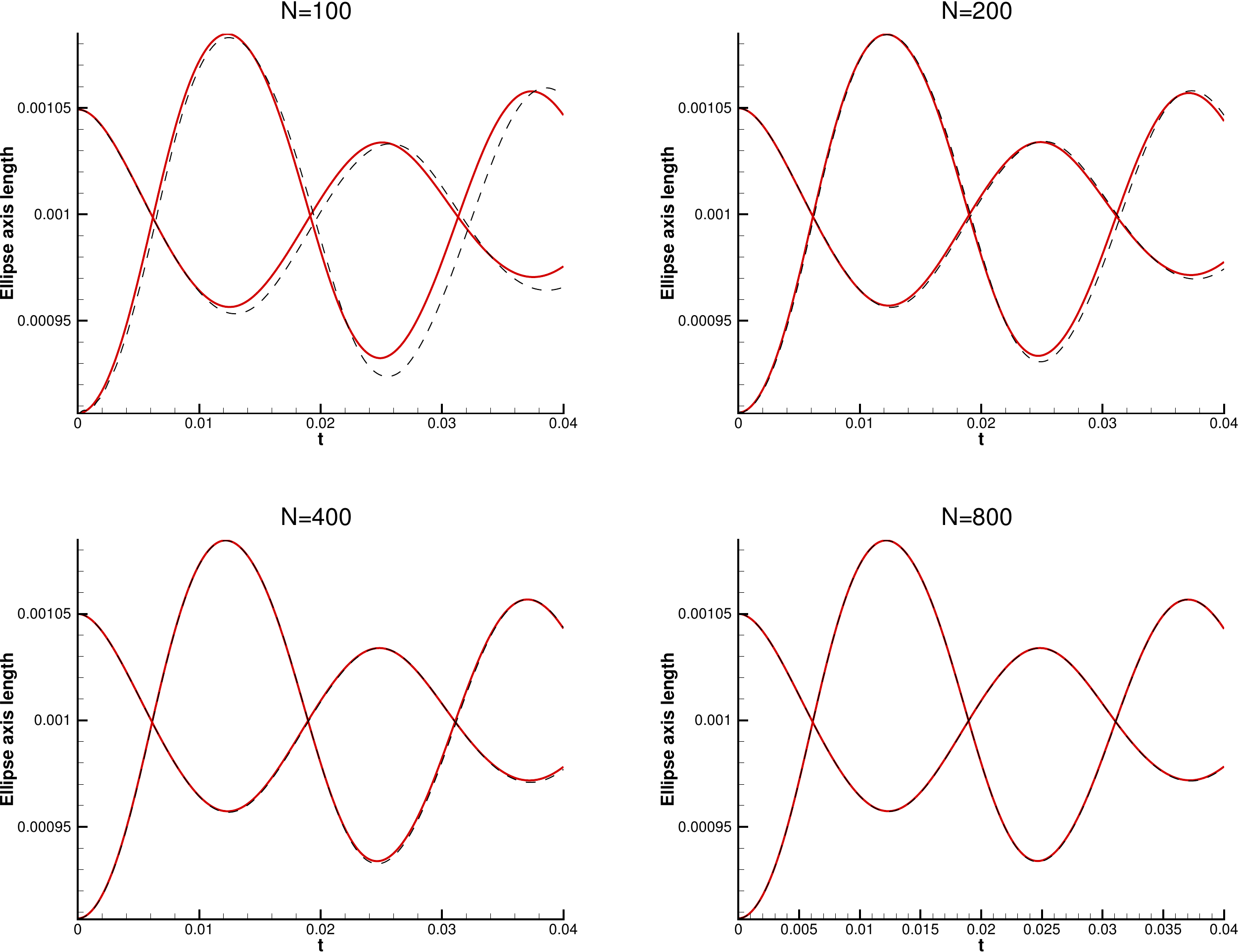}
  \caption{Drop axis lengths for the axisymmetric relaxing drop. Red is the 
  reference solution, dashed black is immersed boundary solution. 
  The two methods converge to the same solution as the grid is refined.}
  \label{fig_relaxing_ellipse_axi_consistency}
\end{figure}

In the first test, an ellipsoidal drop is relaxing to its equilibrium, a sphere,
driven by interfacial tension.
The purpose of this test is to verify that interfacial tension
simulated with the proposed hybrid method gives the same results as when
simulated with the standard level-set/ghost-fluid method. There is no 
gravity and no density or viscosity differences in this test. This way, all
forces are generated by interfacial tension as the ellipse relaxes to equilibrium.
The parameters of the test are listed in \cref{tab_surf_tens_parameters}.
The test was run for increasing grid resolutions to see how the two methods
compare under refinement.
The result for the axisymmetric simulation can be seen in
\cref{fig_relaxing_ellipse_axi_consistency}.
We see that under grid refinement,
the proposed method converges to the same answer as the previously verified and
validated implementation of the level-set/ghost-fluid method.
This demonstrates that the hybrid level-set/ghost-fluid/immersed-boundary method
gives the correct result. For corresponding tests in two dimensions, see
\cite{lysgaard2015}.

This test shows that the proposed method converges to the same solution
as the reference method for a relaxing ellipse driven by interfacial tension.
For the coarsest grid, it appears the hybrid method is the least accurate. This
is likely caused by the disagreement between the Eulerian and Lagrangian
interface representations becoming significant when the curvature of the
interface is no longer much smaller than $1/\Delta$.

This test confirms that the method is consistent, but there is no jump in
density or viscosity in this case.
As discussed previously
the proposed method will treat density and viscosity jumps in a
sharp fashion, while the tension in the interface is handled in a smeared-out
fashion. To verify that this also gives consistent results, a simulation of a similar 
case, \ie a relaxing ellipse driven by interfacial tension was set up.
Instead of equal density and viscosity, the density ratio is now 2 and the viscosity
ratio is 10. These parameters are representative for the case of a water drop in oil.
The simulation was run on a moderately fine grid, $N=400$, where
good agreement was found between the two methods in the previous test.
The full set of parameters for the simulation are listed in \cref{tab_relaxing_jump_params}.
The simulations were run both for 2D and axisymmetric flow for both methods.

\begin{figure}[tbp]
  \centering
  \begin{subfigure}{0.4\textwidth}
    \includegraphics[width=1.0\linewidth]{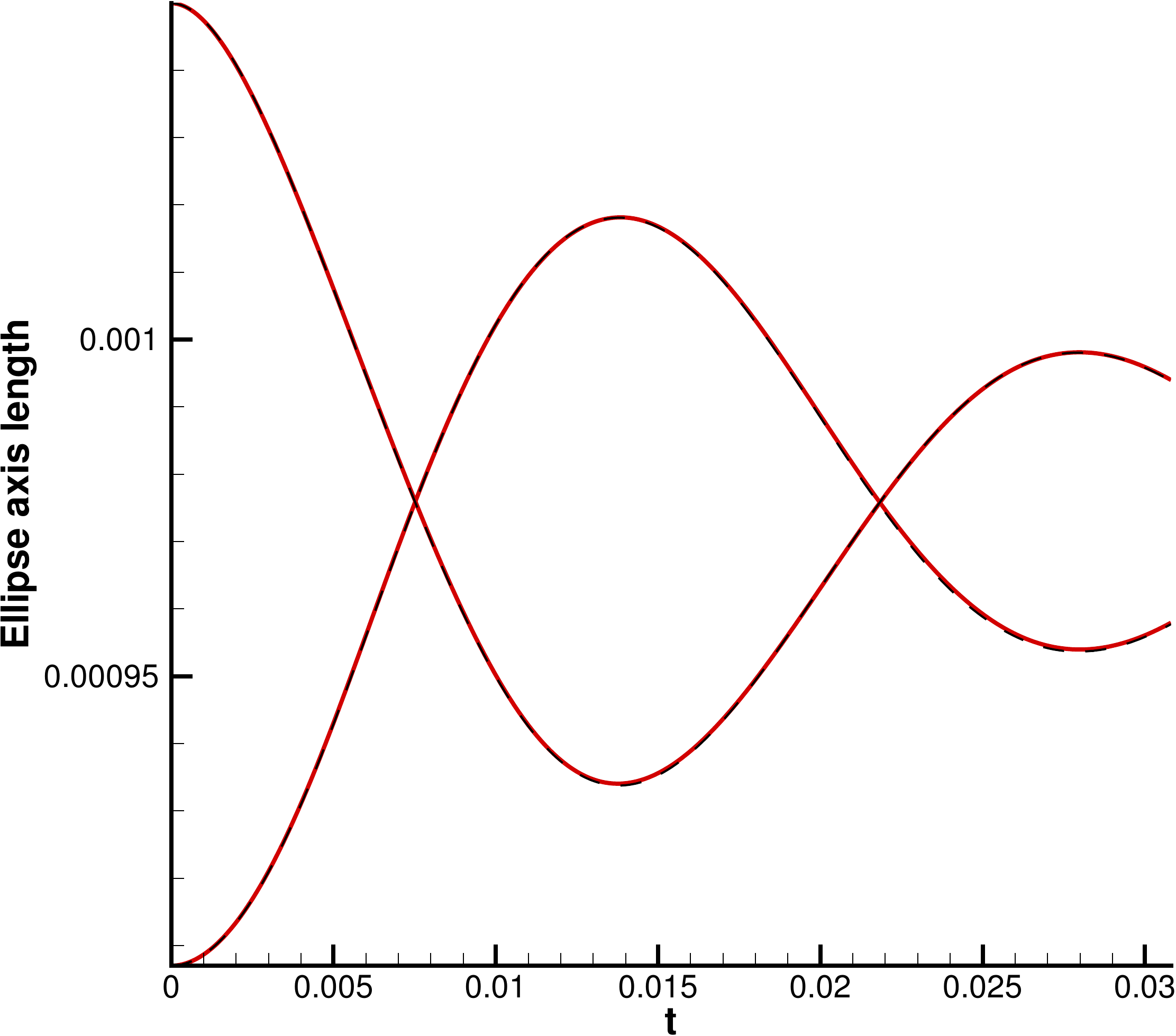}
    \caption{Two-dimensional drop.}
    \label{fig_mu_rho_jump_2d}
  \end{subfigure}%
  \quad %add desired spacing between images, e. g. ~, \quad, \qquad, \hfill etc.
  \begin{subfigure}{0.4\textwidth}
    \includegraphics[width=1.0\linewidth]{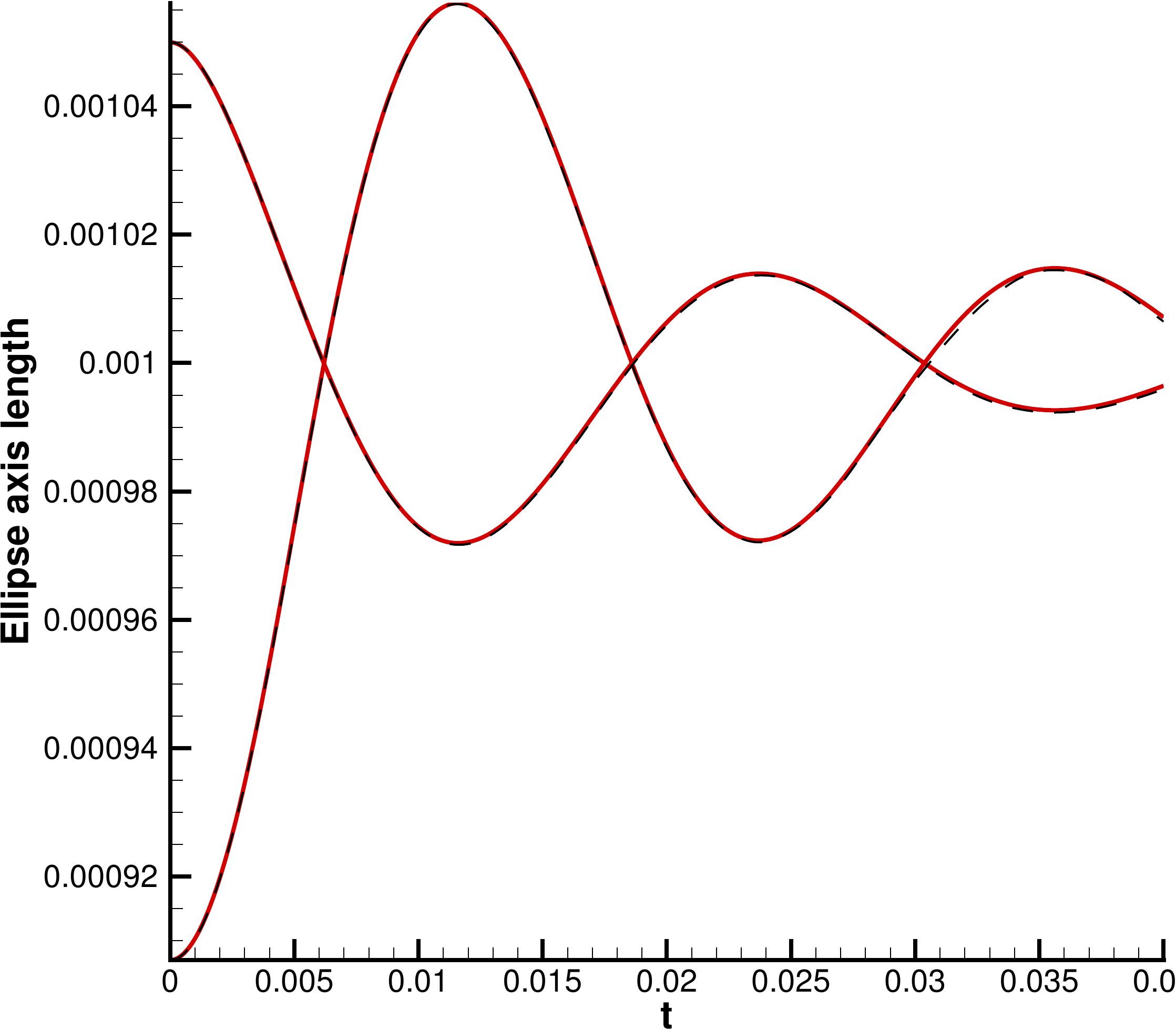}
    \caption{Three-dimensional, axisymmetric drop.}
    \label{fig_mu_rho_jump_axi}
  \end{subfigure}
  \caption{Comparison of drop axis lengths for the reference method and proposed method with a viscosity and density jump. Red is reference method while dashed black is proposed method.}
  \label{mu_rho_jump_full}
\end{figure}
As seen in \cref{mu_rho_jump_full} the two methods are
in agreement both for two-dimensional and axisymmetric flow.
This shows that the proposed method correctly and consistently combines the
interfacial tension
from the diffuse interface, with the sharp handling of viscosity
and density jumps.

\subsubsection{Relaxing drop with elastic membrane}
To test the effect of additional elasticity on the interface, the relaxing
ellipse was again considered.
In this two-dimensional test case, the initial condition is an ellipse with both interfacial elasticity
and interfacial tension, compared to the same case with zero elasticity. 
The parameters for this test are listed in \cref{tab_relax_elastic}.
At the initial state, the deformation $\pd{\v{X}}{s} = 1$ in
\cref{eq:tension-2d}, so the elasticity does not contribute to the interfacial
force. One can say that from the elastic point of view, the interface is neither
stretched nor compressed, but there remains the constant interfacial tension
which produces a force. In this particular case, $K_a$ is set to be ten times larger 
than $\gamma$, which means that $T$ computed from \cref{eq:tension-2d} will be close
to zero when $\pd{\v{X}}{s} \approx 0.9$.
In other words, when the membrane is compressed to 90\% of its original length, elastic forces
and interfacial tension forces will be in balance. 

\begin{figure}[tbp]
  \centering
  \includegraphics[width=\linewidth]{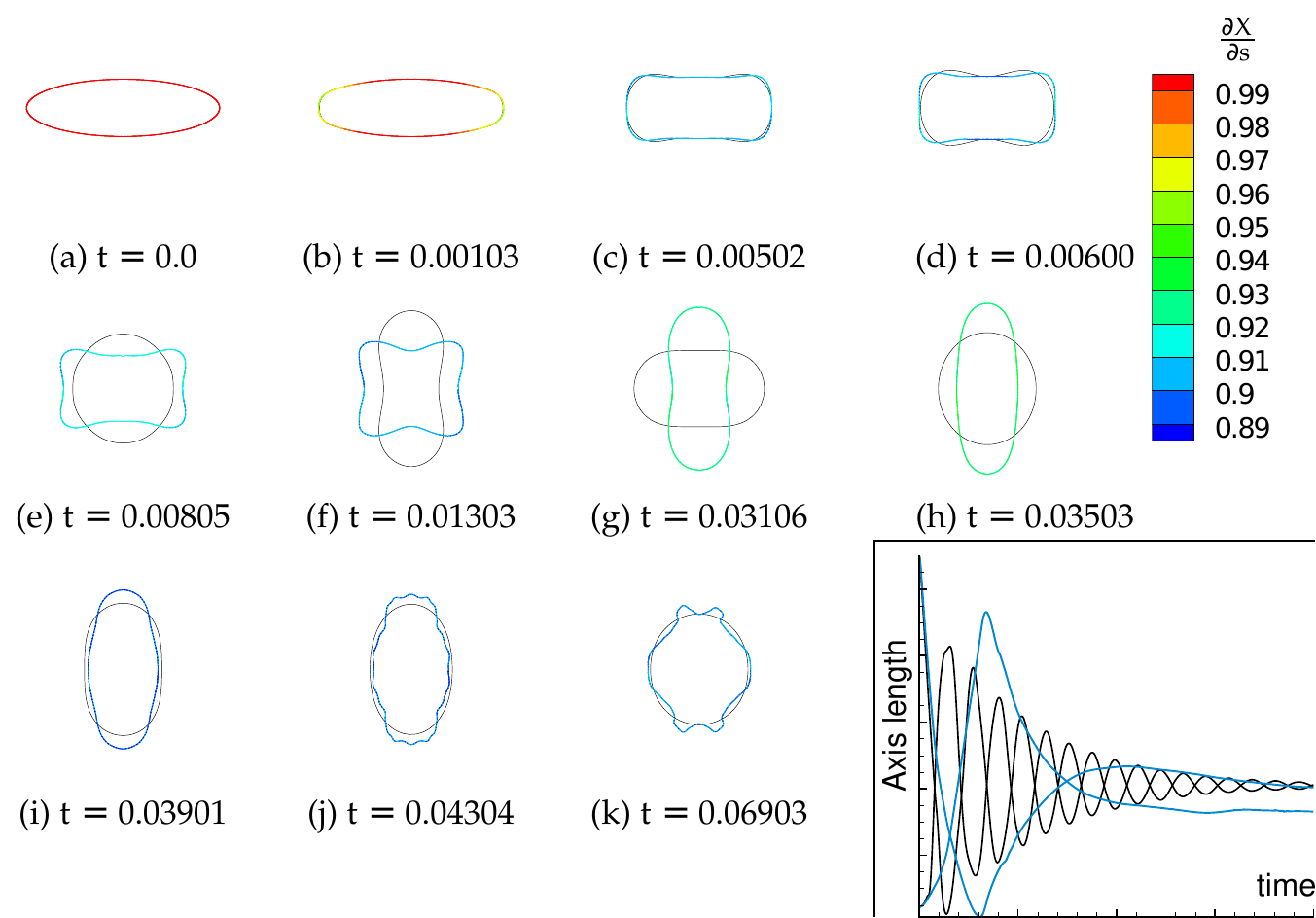}
  \caption{Several frames of the two-dimensional simulation with elastic membrane, (coloured)
  together with the clean interface, (black). The colours indicate the relative
length of the interface compared to its equilibrium length. The inset in the
lower right corner shows the time evolution of the horizontal and vertical axis
of the clean (black) and the contaminated (blue) interface, which indicates that
clean drop undergoes many oscillations during this time, while the contaminated
drop only undergoes one.} 
  \label{fig_elastic_drop_full}
\end{figure}
The interface starts in the initial state shown in red in
\cref{fig_elastic_drop_full} (a).
Because of its eccentricity, interfacial tension is relatively strong on the
left and right sides
of the drop and it is quickly compressed,
\cref{fig_elastic_drop_full} (b). After $5\times10^{-3}$ s
the drop is compressed to approximately 90\% of the initial length
\cref{fig_elastic_drop_full} (c). This means that the interface no longer
introduces any force, and without any viscosity or density differences the simulation
would proceed as if it was momentarily single phase. As there is a significant flow
still present in \cref{fig_elastic_drop_full} (c),
advection of the interface continues. Some of the kinetic energy
which is not dissipated by viscosity goes into deforming and again stretching
the interface. At $t=3.5\times10^{-2}$ s,
\cref{fig_elastic_drop_full} (h), the interface is stretched to
the next maximum again, and the velocities are close to zero.
Now there is not enough potential energy in the membrane to do another
oscillation, it is in some sense overdamped. The interface contracts creating a crumpled drop as seen in
\cref{fig_elastic_drop_full} (i) to
\cref{fig_elastic_drop_full} (k).
As this has happened the drop with a clean interface, in black, has oscillated towards its
equilibrium shape, a circle, by going through about eight oscillation cycles.

It is clear that for these parameters, the elastic membrane has a significant
effect on the time evolution of the drop, fundamentally changing its response.
For clean fluids, the equilibrium interface
is always the one that has the minimal interface area. The interfacial elasticity
changes this situation, and the
equilibrium state is no longer obvious given the initial conditions.
One insight from this simulation is that for a clean drop without
an elastic membrane, there exists a unique spherical equilibrium state,
only given by the initial volume of the drop.
On the other hand, for the drop with an elastic membrane,
the equilibrium is not just a function of the initial volume,
but also of the initial shape. This is because the initial shape affects what parts
of the drop are stretched and compressed, which has a significant impact on the
final steady state. This demonstrates that the evolution of a drop with
an elastic membrane is more complex than one without.

\subsection{Nanoscale: validation}
\label{sec:micro-val}
When running simulations of a complex multi-component system, the first thing to
consider is each of the two-component systems. For the toluene-water and
heptane-water systems, there is one free parameter, the binary interaction
parameter $k_{ij}$, to tune in each
case. There is also an important value to tune this against, namely the interfacial
tension measured in experiments. For the alkane-water system with the models
used here, \citet{lobanova2016} obtained the value of $k_{ij}=0.3205$,
transferable across the alkanes with different lengths. The water
\cite{lobanova2015} and alkane \cite{lobanova2016} models used have been
published previously. The model for toluene used here has not been published
previously; see Appendix B for force field parameters for this model.

For the toluene-water system, such tuning has not been done previously, so it is
done here. The system consisted of 10 000 toluene molecules and 40 000 water
molecules, placed in an elongated box, and simulated at 20\textdegree C and
1 bar. After obtaining the desired temperature and pressure by simulating in the
\NPT ensemble, the system was allowed to phase-separate such that two slabs of
liquid were formed. Subsequently, the system was simulated in the \NVT ensemble
for 50 nanoseconds to obtain the interfacial tension.

\begin{figure}[htpb]
  \centering
  \includegraphics[width=\linewidth]{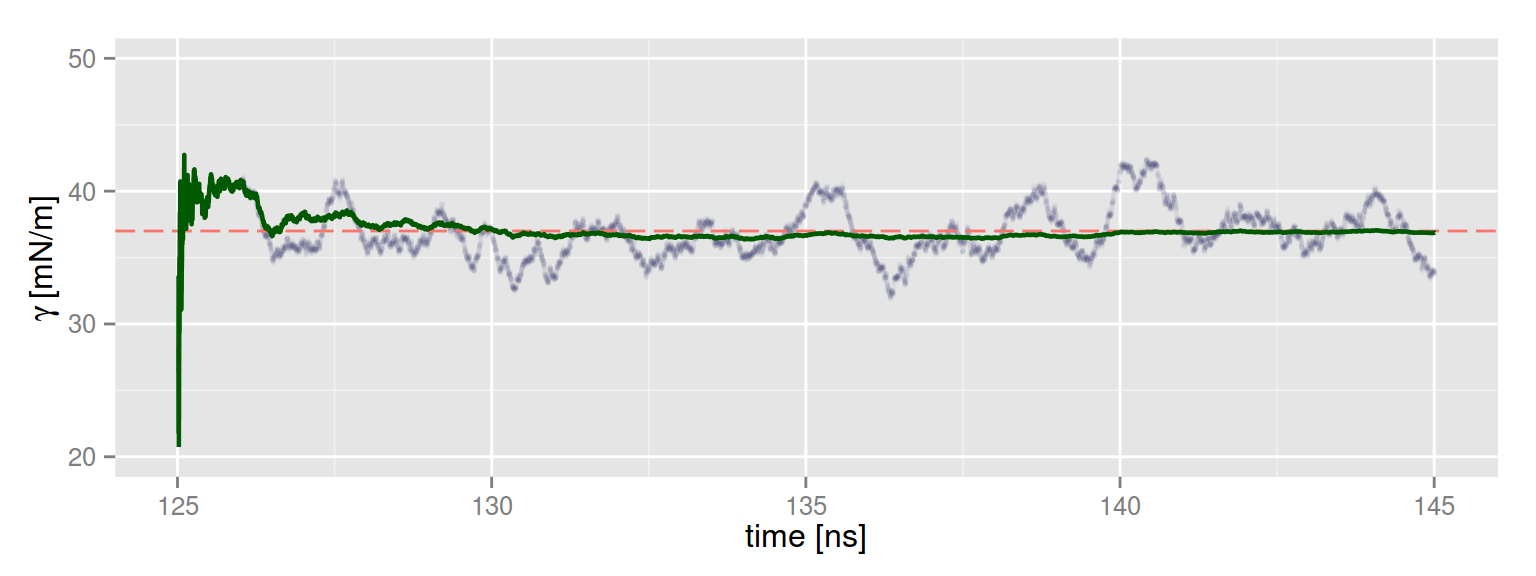}
  \caption{Interfacial tension of toluene and water, using the cross-interaction
    parameter $k_{ij}=0.241$. The green line shows the cumulative average, the
    blue dots show the running average over 1000 points, and the dashed orange
    line shows the experimental value.}
  \label{fig:ift-toluene-water}
\end{figure}

The three values for the cross-interaction parameter, $k_{ij} = (0.1,0.2,0.3)$,
were initially tested, and
subsequent guesses were refined until the value $k_{ij} = 0.241$ was found,
which gave an interfacial tension in very good agreement with the experimental
value of 37 mN/m \cite{saien2006} at this temperature and pressure. See
\cref{fig:ift-toluene-water}, where the cumulative average of the interfacial
tension computed from \cref{eq:ift-md} is plotted as a function of the
simulation time. Also plotted is the experimental result (dashed orange line)
and the running average over 1000 values of the interfacial tension (blue dots).

Having the cross-interactions for the two binary systems established, verifying
that the interfacial tension for the heptane-toluene or
``heptol'' mixture against water agrees with experiments is a good demonstration
of the predictive power of the method. When discussing
a heptol mixture, one must distinguish between molar ratios, convenient in
simulations, and volume ratios, convenient in experiments.
To avoid confusion, we will
refer here to volume ratios using the notation $N:M$, and molar ratios using the
notation $n/m$.
The molar
mass of heptane and toluene are 100.2 and 92.1 g/mol, respectively, and their
densities are 684 kg/m$^3$ and 867 kg/m$^3$, respectively.  This means that a 50/50 molar
ratio gives a volume ratio of 1.38:1, and conversely, a 1:1 volume ratio gives
a 42/58 molar ratio. 

To test the prediction of interfacial tension of the heptol-water system, 
a corresponding simulation was set up with a 1:1 heptol mixture
against water at 20\textdegree C and 1 bar. Experimental data for this
interfacial tension is not available, but an accurate estimate of it can be
obtained following \citet{yarranton1996}. 
This procedure gives the interfacial tension of an organic mixture against
water as  
\begin{equation}
  \gamma_{\rm{mix}} = \gamma_2 - R_u\,T\,\Gamma_m \ln\left(1 +  x_1(\exp\left( \frac{\gamma_2
- \gamma_1}{R\,T\Gamma_m}\right) - 1 ) \right)
\label{eq:ift-mix}
\end{equation}
where $\gamma_1$ and $\gamma_2$ are the two pure-component interfacial tensions
against water, $R_u$ is the universal gas constant, $T$ is the temperature in
Kelvin, $x_1$ is the molar fraction of component 1 in the mixture. We denote
here toluene as component 1 and heptane as component 2. The interfacial tensions
are $\gamma_1 = 37$ mN/m \cite{saien2006} and $\gamma_2 = 51.2$ mN/m
\cite{zeppieri2001}. The parameter 
$\Gamma_m$ is estimated as $\Gamma_m = 0.00415$ mmol/m$^2$
by \citet{yarranton1996}, which is shown to give good results for a wide range of
alkane-aromatic mixtures. With these expressions, the interfacial tension of the mixture is computed as
$\gamma_{\rm{mix}} = 41.9$ mN/m.

\begin{figure}[htpb]
  \centering
  \includegraphics[width=\linewidth]{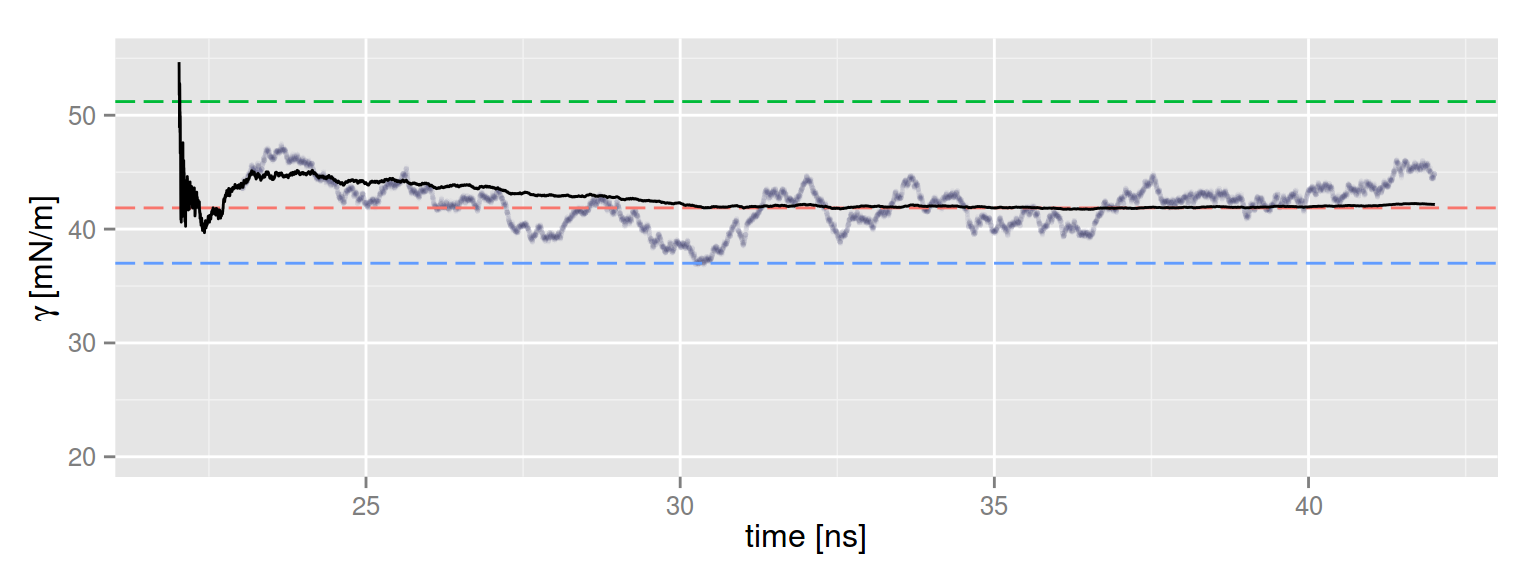}
  \caption{The interfacial tension for a 1:1 heptol mixture against water,
  computed from molecular dynamics simulation. Black line: cumulative average of
  the interfacial tension. Blue dots: running average of the interfacial tension
  over 1000 samples. Orange dashed line: $\gamma_{\rm{mix}}$ predicted from
  \cref{eq:ift-mix}. Blue dashed line: interfacial tension of toluene-water. Green
  dashed line: interfacial tension of heptane-water.}
  \label{fig:heptol-water-ift}
\end{figure}

The simulation result is plotted in \cref{fig:heptol-water-ift}, together with
$\gamma_1$, $\gamma_2$ and $\gamma_{\rm{mix}}$. The simulation result matches
very well the value computed from \cref{eq:ift-mix}. Note that the
cross-interaction between heptane and toluene has not been tuned at all. 
This illustrates the 
predictive power of the SAFT-$\gamma$ Mie approach, \ie that one can compute, with good accuracy,
physical properties that have not been used when constructing the model. 

With the simple fluids taken into account, we may proceed to consider the models
for the asphaltenes. As discussed in the introduction, asphaltenes are immensely
complex mixtures of different molecules that together form a solubility class.
It is likely that an asphaltene molecule in solution never interacts with an
identical molecule, and thus modelling the asphaltenes as a single
molecule is a significant simplification. It is likely that improved asphaltene
models must take the polydispersity explicitly into account, \ie having many
different types of model asphaltenes in the same simulation. This may be
considered in future work.

Molecular dynamics simulations of asphaltenes in the bulk have been considered
in several previous works, see \eg refs.
\cite{rogel2003,pacheco2003,stoyanov2008,kuznicki2009,headen2009,boek2009,headen2010,sedghi2013,ungerer2014}.
In these studies, the association and aggregation of asphaltenes have received
particular focus, as this behaviour is very important for asphaltene deposition
in rocks and in pipelines. Of particular note is the work by \citet{boek2009},
where the technique of quantitative molecular representation (QMR) is used to compile
a set of model asphaltene molecules, based on several experimental sources of
information such as mass spectrometry, NMR, X-ray and neutron scattering studies. These model
asphaltenes are employed by several subsequent authors, and have influenced the
construction of the coarse-grained model asphaltenes used in this work.
\citet{sedghi2013} studied the effect of different side groups and substitutions
on one of these  representative asphaltene molecules, and found that small changes to the
chemical composition caused large variations in the aggregation behaviour.

Only very recently have molecular dynamics simulations been applied to the study
of asphaltenes at the oil/water interface. \citet{mikami2013}, \citet{liu2015}, as well as 
\citet{yang2015} used atomistically-detailed simulations to study this system.
In \cite{mikami2013,liu2015}, the previously mentioned QMR-based model asphaltenes were
employed, whereas in \cite{yang2015}, two different model asphaltenes were
proposed based on experimental measurements and used in simulations. These
studies highlight the significant challenge encountered when using an
atomistically-detailed approach: one is either confined to very few asphaltene
molecules \cite{yang2015} or very short timescales \cite{mikami2013,liu2015} due to the
high computational cost.
A coarse-grained approach, such as that used here, can provide a solution to
this; see \eg the review by \citet{brini2013} for an overview of the advantages
of coarse-grained simulation for soft matter systems in general.
Recent work by \citet{ruiz2015} employed a coarse-grained (dissipative
particle dynamics) approach, but this study is also limited to considering few
molecules and short time scales. Finally, it should be noted that in all
these studies, the asphaltenes are initially placed at the interface.

In contrast to these limitations of either few molecules or short time scales, and
enforced interfacial adsorption, we consider here simulations with up to
two orders of magnitude more asphaltene molecules than considered in
\cite{yang2015} in conjunction with two orders of magnitude longer simulation 
times than considered in \cite{liu2015}. Also the asphaltenes are not initially placed at the
interface, but randomly distributed throughout the oil phase. Thus the
interfacial activity is an inherent property of the model asphaltenes used here.

With this in mind, we briefly discuss the
three different asphaltene models that were used in the
present work. All these models are of the continental asphaltene type, \ie they
have a central core made up of aromatic rings, to which aliphatic tails are
attached. The first model asphaltene, presented by \citet{muller2015}, was used in the
initial studies; we will refer to this as the APCH asphaltene. It is based on using anthracene beads for the
aromatic core, which also contains a pyridine ring, and three tails each made
from dodecane. The model behaves reasonably like an asphaltene. Results obtained with this model are discussed in
\cref{sec:results-apch}.
 
\begin{figure}[htpb]
   \centering
   \includegraphics[width=\linewidth]{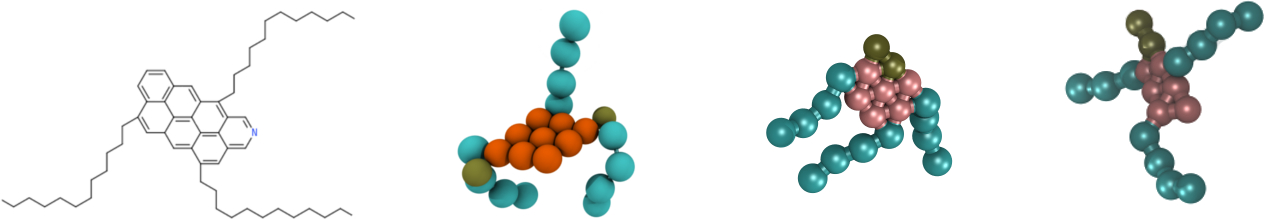}
   \caption{From left to right: corresponding chemical structure, the APCH model, the APCE
     model and the APCL model. Blue/green beads indicate the aliphatic tails,
   orange/red beads indicate the aromatic rings forming most of the core, and the tan/gold
   beads indicate the pyridine.}
   \label{fig:apc-comp}
 \end{figure}

If anything, the APCH model appears slightly too self-associative.  This,
together with the some advances in the SAFT-$\gamma$ Mie theory
for aromatic compounds, led to the second flavor, developed partly in this work.
The main difference with
the first flavor is that the aromatic beads, which make up most of the core, are
obtained using a version of the SAFT-$\gamma$ Mie approach that is tailored for 
ring-shaped molecules, as opposed to the standard version which assumes they are
linearly shaped. The structure is otherwise very similar to the APCH model
asphaltene. The parameters for the aromatic core beads are given in Appendix B.

Two different
architectures were considered for this new flavor of asphaltenes,
built from the same coarse-grained beads, but with different
shapes for the aromatic core. Both have three aliphatic tails made from dodecane,
and cores made from six beads representing six hexagonal aromatic
rings, together with two beads representing a pyridine, \ie a hexagonal aromatic ring
with a single nitrogen substitution. The difference lies in that ``APCE'' has a more
circular core, while ``APCL'' has a more elongated core. See 
\cref{fig:apc-comp} for a comparison of the three different model asphaltenes
considered in this work.

As mentioned in the introduction, the only usable definition of asphaltenes is that they are
insoluble in heptane and soluble in toluene. Accordingly, it should be verified
that the new model asphaltene molecules behave in this way. To test this,
simulations were run with 240 asphaltene molecules in a system with 40 000
solvent molecules (heptane or toluene) at 20\textdegree C and 1 bar.

\begin{figure}[htpb]
  \centering
    \includegraphics[width=0.8\linewidth]{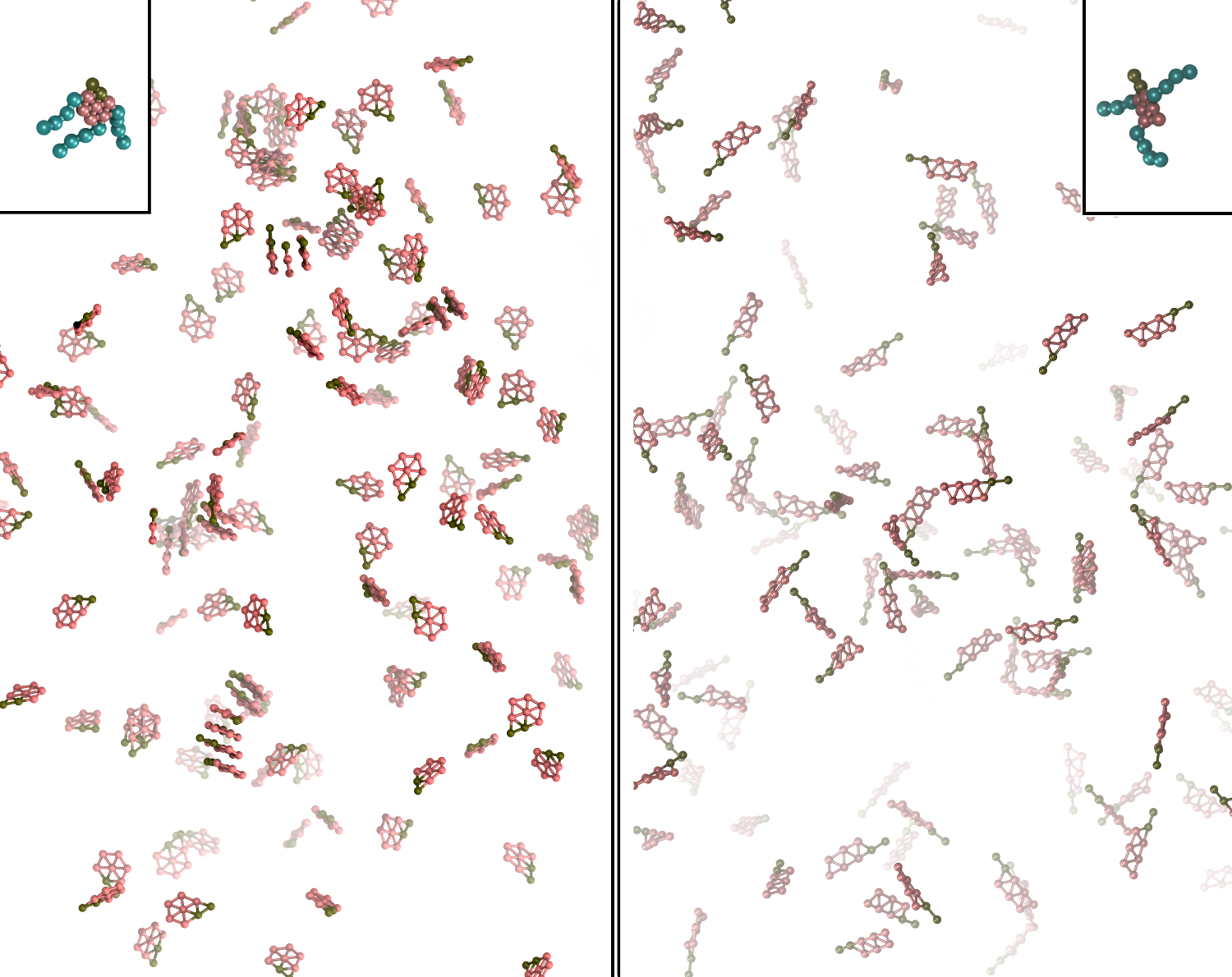}
  \caption{Comparison of the model asphaltenes in heptane. Left: APCE with a circular
    core. Right: APCL with an elongated core. In both cases, the inset shows
  the usual coarse grained beads for the entire molecule, while the main figures
  shows only the aromatic beads and their bonds, with the diameter set to $0.5\sigma$ to make the
  visualisation clearer.} 
  \label{fig:asphaltene-heptane-difference}
\end{figure}

These simulations gave some interesting results which demonstrate how delicate
the energy balance is in systems containing asphaltenes.
The two models APCE and APCL appear very similar; however, 
their solubilities in heptane are completely different, as illustrated in
\cref{fig:asphaltene-heptane-difference} which shows the behaviour in heptane. 
In this figure, the solvent and
the aliphatic tails are omitted for clarity, so one can easily see the
stacking of aromatic rings where the molecules cluster. It is readily apparent
that the circular core asphaltenes cluster in heptane, while elongated core ones
are soluble in heptane.
It is also seen that the circular-core asphaltenes
cluster with a distribution of a few large clusters, many small clusters, and
some monomers; in particular, the asphaltenes do not all gather in one
big cluster. This is in agreement with results from experiments
(\eg \cite{fenistein2001,rahmani2005,lisitza2009}) and
atomistic simulations \cite{headen2009}; and it is in general agreement with the
accepted Yen-Mullins model of asphaltene behaviour \cite{mullins2012}.

\begin{figure}[htpb]
  \centering
  \includegraphics[width=\linewidth]{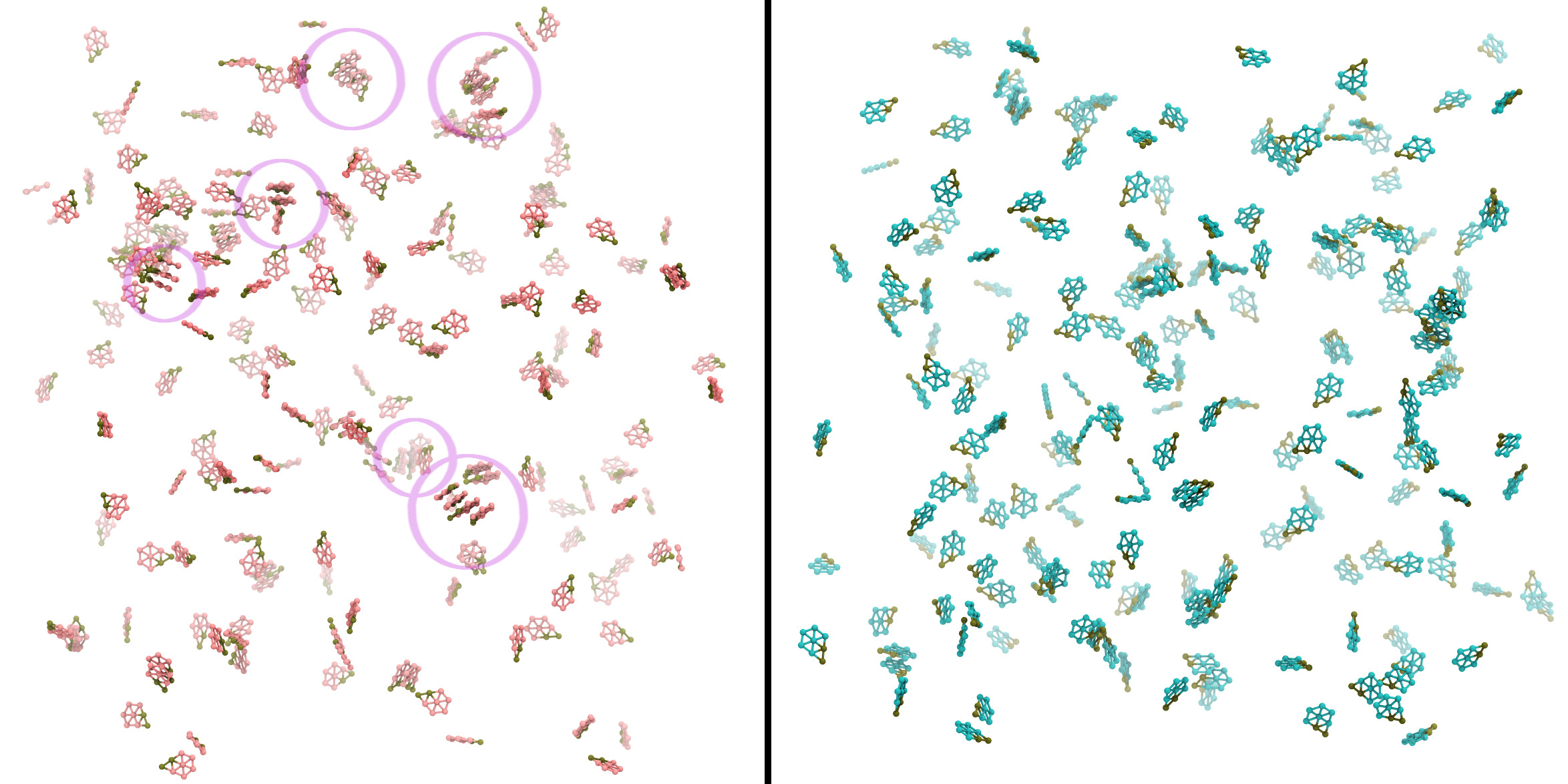}
  \caption{Comparison of the APCE model asphaltenes in heptane (left, red) and in
    toluene (blue, right). As in the previous figure, only the aromatic cores
    are shown. Purple circles highlight clusters with three or more molecules
    (there are none on the right-hand side). It is evident that the molecule is poorly soluble in heptane, where
    two-, three- and four-molecule clusters can be seen. In toluene, there is no
    clustering apart from the occasional contact between two asphaltene cores.
    }
  \label{fig:APCE-hep-tol-50ns}
\end{figure}

The APCE model thus passes the first hurdle, being insoluble in heptane.
Simulations of this model in toluene also showed the correct behaviour, as
illustrated in \cref{fig:APCE-hep-tol-50ns} where the system is compared in
heptane (red)
and in toluene (blue) after 50 ns of equilibration. In toluene, some pairs of asphaltenes
occasionally come in contact, but inspection of the trajectories showed that these pairs only
stay together for about one nanosecond before they break apart again, and thus
no larger clusters have time to form.

\begin{figure}[htpb]
  \centering
  \includegraphics[width=0.4\linewidth]{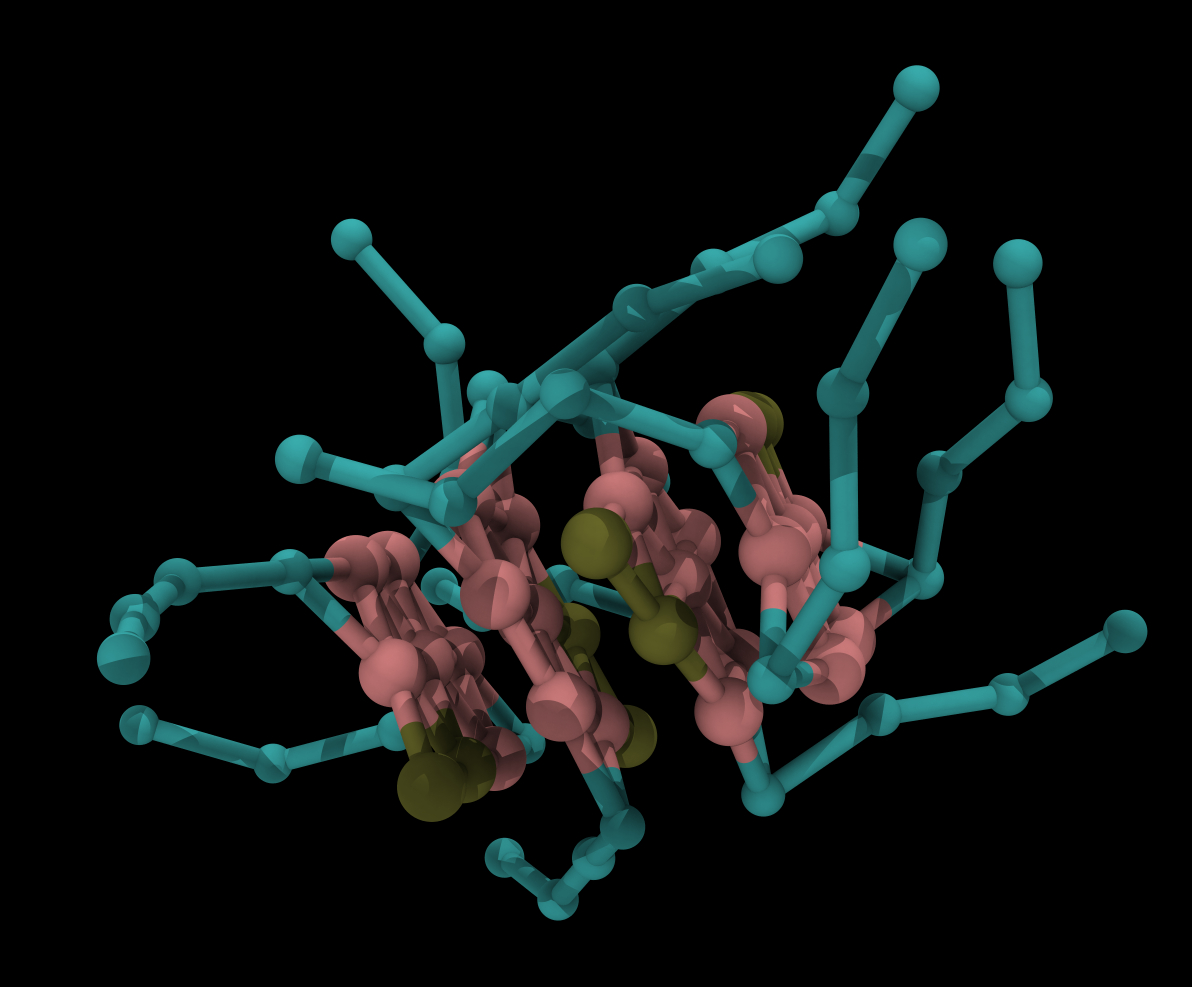}
  \caption{A closeup of a four-asphaltene APCE cluster in heptane. The beads
  are also here shown at reduced size for clarity; aromatic beads with diameter
  set to $0.5\sigma$ and alkane beads with diameter set to $0.25\sigma$. The
colours indicate aromatic beads (red), pyridine beads (gold) and aliphatic beads
(blue)}
  \label{fig:APCE-cluster}
\end{figure}

To summarise, the APCE model asphaltene fits the definition of an asphaltene
molecule, forming clusters in heptane but staying in solution in toluene. In
heptane, clusters of up to four molecules form after 50 ns. It is likely that
five- or six-molecule clusters may form after even longer times or at higher
concentrations. A closeup of a four-molecule cluster is shown in
\cref{fig:APCE-cluster}.

\section{Results}
\label{sec:results}
\subsection{Molecular dynamics simulations of asphaltenes in oil/water systems}
The crude oil/water system has a very large (and not fully understood) parameter space, so we restrict
our attention here to only a few simplified cases. The parameters that can be varied
include temperature, pressure, the mixture of alkanes and of aromatics for the
fluid components, the ratio between alkanes and aromatics in the fluid
components, the amount of resins, the amount of asphaltenes, etc. The liquid
components of a crude oil contain a range of alkanes that may extend all the
way from methane (liquid at high pressure) to alkanes with 20-30 carbon atoms
(liquid at high temperature). Effective models for these mixtures can be
obtained by taking a true boiling point curve and dividing it into classes of
pseudo-components, \eg a mixture of C$_5$H$_{12}$, C$_{10}$H$_{22}$,
C$_{15}$H$_{32}$, C$_{20}$H$_{42}$. The very simplest version of this, which we
adopt here, is to use just one alkane, for which heptane is a common choice.
Similarly, for the aromatic liquids there is a range to choose from, \eg
benzene, pyridine, toluene etc., and again we limit ourselves to just toluene. This
combination, heptol, has been widely used in experimental studies as a model system in
which asphaltenes can be dissolved (\eg refs.
\cite{mclean1997,djuve2001,spiecker2003,spiecker2003b,spiecker2004,yarranton2007,wang2010}).
Heptol is the simplest solvent for asphaltenes where the aromatic to aliphatic ratio of the
solvent can still be varied.

Given heptol as a base, the mixture ratio can be varied, and asphaltenes can be
added to the system in varying concentration. Previous work \cite{mclean1997} has indicated that
a heptol ratio around 70/30 gives the most stabilising (and thus most
interesting) interfacial properties, so
the 70/30 heptol mixture is used as a base fluid here. The temperature and pressure are 
set to 20\textdegree C and 1 bar, respectively. The asphaltene concentration is varied, to
study the effect this has on the system.

It is notable that in all the simulations reported here, the systems are started
from random initial conditions. In particular, the asphaltenes are not placed at
the oil/water interface, but they are themselves interfacially active. The
tuning of cross-interactions is as reported in \cref{sec:micro-val} for the
heptane-water and the toluene-water cross interactions. For the asphaltenes, the
alkane tail-water cross interactions use the same $k_{ij}$ as for heptane-water,
otherwise the cross-interactions have not been adjusted.

These studies using the various model asphaltenes were performed at different
concentrations from 240 to 960 asphaltene molecules in 40 000 molecules of solvent (70/30
heptol), together with 160 000 molecules of water. These systems were simulated
in the \NPT ensemble using an elongated simulation box with a 3:1:1 side ratio,
and the desired state (20\textdegree C, 1 bar, oil/water phase-separated) was
obtained. Subsequently, the system was evolved in the \NVT ensemble in order to
let it equilibrate.

At this point, some remarks are in order with regards to the relaxation times in
these systems, as compared to experiments. Looking at the time it takes to reach
equilibrium for the interfacial tension in asphaltene-heptol-water systems in
experiments (\eg \cite{spiecker2004,sztukowski2005}) the timescales are of the
order of hours, which is completely out of reach for molecular
simulation. However, these long timescales are caused by the slow diffusion of
asphaltene molecules from the far-away regions of the bulk to the asphaltene-depleted oil layer
close to the interface. Moreover, these timescales are not representative of the situation
when water drops travel through oil in a pipeline or a separator, where the 
ratio of the diffusion boundary
layer width to the  viscous boundary layer width, \ie the Schmidt number, can be
of the order of 10$^6$, meaning that the speed of diffusion is increased by this order of
magnitude for a falling drop as compared to a drop at rest in a tensiometer.
Comparing this to the situation in simulations, it is clear that the diffusion
cannot happen over
a longer scale than the simulation box size, which is of the order of 10
nanometres. Experimental measurements indicate the diffusion coefficient of
asphaltenes in solvents like heptol is of the order of $10^{-10}$ m$^2$/s
\cite{norinaga2001}. This means the characteristic time for an asphaltene to
diffuse across the simulation box is of the order of $10^{-6}$ seconds.

There are two other relaxation time scales of importance, in addition to the
time scale for diffusion. The second is the time scale
for the adsorption of an asphaltene at the interface. With a coarse-grained
description such as here, this time scale is very short, of the
order of 1 nanosecond. The third is the time scale for reorientation of the
asphaltenes at the interface, \ie that those molecules already adsorbed find
a tighter packing, which allows further molecules to adsorb at the interface.
This time scale depends on how the asphaltenes associate at the interface, and
is thus difficult to estimate \emph{a priori}.

\subsubsection{Studies using the APCH model asphaltene}
\label{sec:results-apch}
The APCH model asphaltene was used in the first studies of
asphaltene-heptol-water systems. Two concentrations were considered, namely 240
and 720 asphaltene molecules in 40 000 molecules of 70/30 heptol. The former
corresponds to an asphaltene concentration of about 5\%. Note that in these
simulations, a simpler two-bead model for the toluene molecule \cite{mejia2014} was used.

\begin{figure}[htpb]
  \centering
\includegraphics[width=\linewidth]{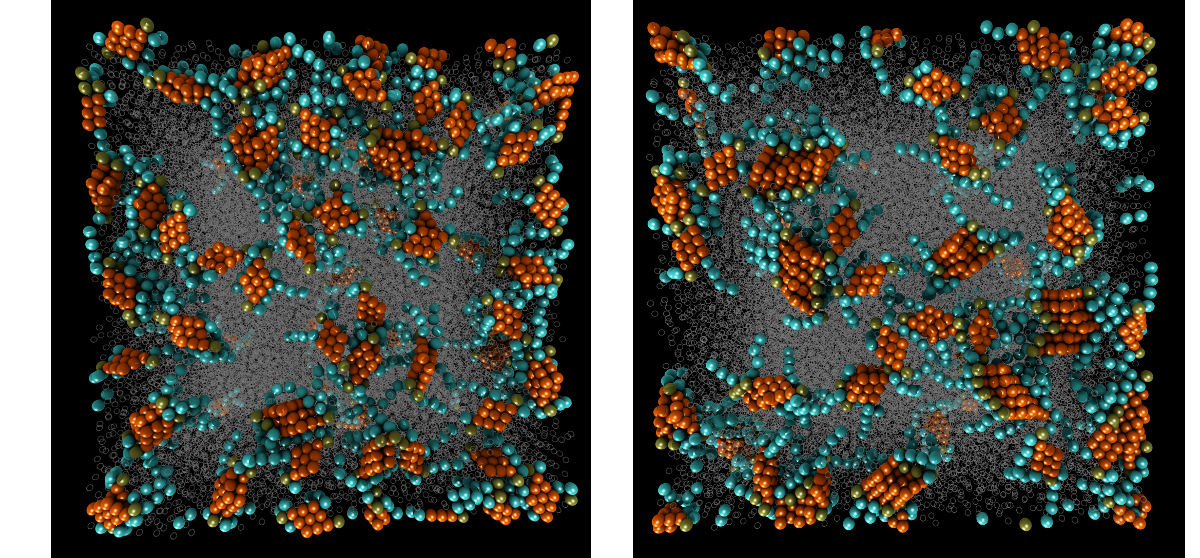}
  \caption{Snapshot of the interface seen from the water side, for the system
  with 240 APCH model asphaltenes after 35 ns (left) and 350 ns (right). Colours
as in previous figures. Water molecules are omitted for clarity, and the heptane
and toluene beads are shown in gray.}
  \label{fig:apch-240}
\end{figure}

Snapshots of the oil-water interface from simulations at this lower
concentration are shown in \cref{fig:apch-240}. This figure shows the interface
from the water side after 35 ns (left) and 350 ns (right). The water beads are
not shown, and the solvent beads are shown only as grey outlines. It is seen
from this figure that the interfacial configuration of asphaltenes changes
during this time, with clusters forming at the interface. The number of
asphaltenes at the interface increases by about 30\% from 35 to 350 ns, and at
350 ns there are 0.25 molecules per square nanometre. Inspection of this system
revealed that the interfacial adsorption left the bulk oil almost depleted of
asphaltenes. Thus this situation with a lower concentration of asphaltenes
is representative of a freshly-formed interface
where diffusion has not yet had time to bring in more asphaltenes from the bulk.

\begin{figure}[htpb]
  \centering
\includegraphics[width=\linewidth]{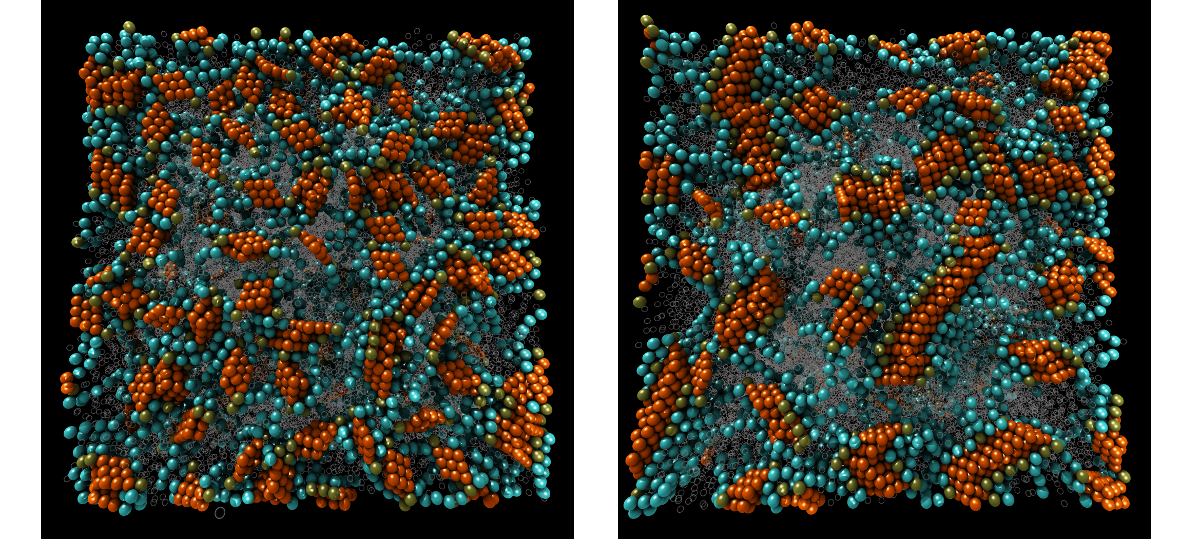}
  \caption{Same as the previous figure, for the system with 720 APCH asphaltene
  molecules. At this higher concentration, the tendency for the asphaltenes to
cluster together at the interface is even more pronounced.}
  \label{fig:apch-720}
\end{figure}

As previously mentioned, the time scales for diffusion can be of the order of
minutes and hours, which is not tractable in these simulations. However, the
effect of diffusion may be taken into account by starting the simulation with an
initially higher asphaltene concentration. The higher concentration case, using
720 asphaltene molecules, gave about 5\% asphaltenes remaining in the bulk after
350 ns, and is thus representative of the same system having reached equilibrium
after diffusion from the bulk. Snapshots of this system are shown in
\cref{fig:apch-720}, again with 35 ns at the left and 350 ns at the right.
Again, there is an increase of about 30\% in the number of adsorbed asphaltenes
from 35 to 350 ns, and at 350 ns there are 0.5 molecules per square nanometre.
It is evident from the snapshot that these asphaltene molecules are highly
self-associative at the interface, \ie that they cluster together.
It is likely that this behaviour is not
representative of real asphaltenes, which are believed to adsorb with the
aromatic core towards the water \cite{andrews2011}.

\begin{figure}[htpb]
  \centering
  \includegraphics[width=0.5\linewidth]{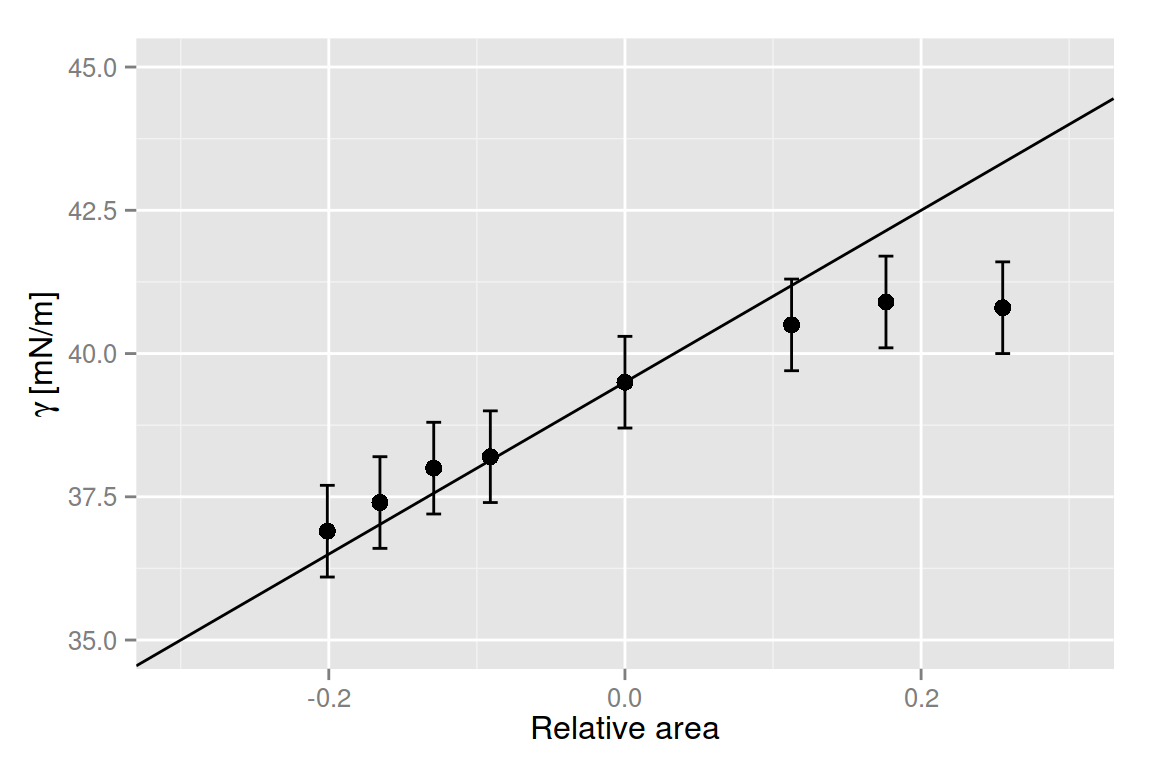}
  \caption{The change in interfacial tension versus the relative area $A/A_0
  - 1$ for compressions (negative relative area) and expansions of the
    interface. It is seen that compressions give a reasonable elastic
  behaviour, but under expansion the change in tension levels off. The line
  corresponds to $K_a$ = 15 mN/m.}
  \label{fig:apch-elasticity}
\end{figure}

For the system with 720 asphaltene molecules at 350 ns, the interfacial
tension and elasticity was calculated using the previously described methods.
The interfacial tension was found to be 39.5 $\pm$ 1 mN/m. The elasticity was estimated to be
15 $\pm$ 10 mN/m; in \cref{fig:apch-elasticity} the change in tension versus the
change in interfacial area is shown together with a line showing the 15 mN/m
slope.

\subsubsection{Studies using the APCE model asphaltene}
\begin{figure}[htpb]
  \centering
  \includegraphics[width=\linewidth]{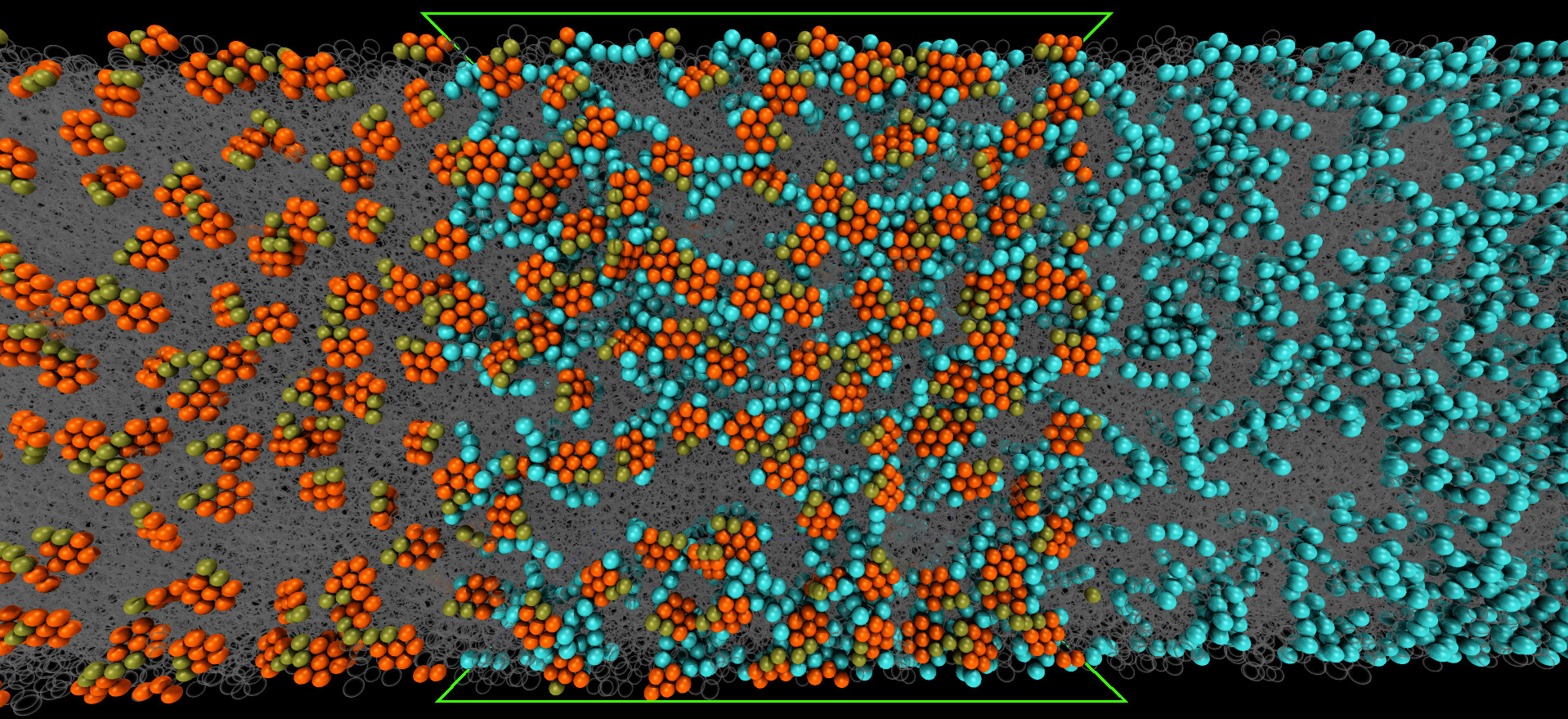}
  \caption{Snapshot of the APCE model asphaltene at the interface seen from the
    water side, for the system with 240 asphaltenes, after 50 ns of
  equilibration. Colors as in previous figures. On the right and left sides,
  periodic images of the interface are shown, where the aromatic cores and the aliphatic tails are
  omitted, respectively. It is seen that the aromatic cores are predominantly 
  oriented in parallel with the interface, while the aliphatic tails are pointing
  away into the oil phase.}
  \label{fig:apce-interface-240}
\end{figure}

\begin{figure}[htpb]
  \centering
  \includegraphics[width=\linewidth]{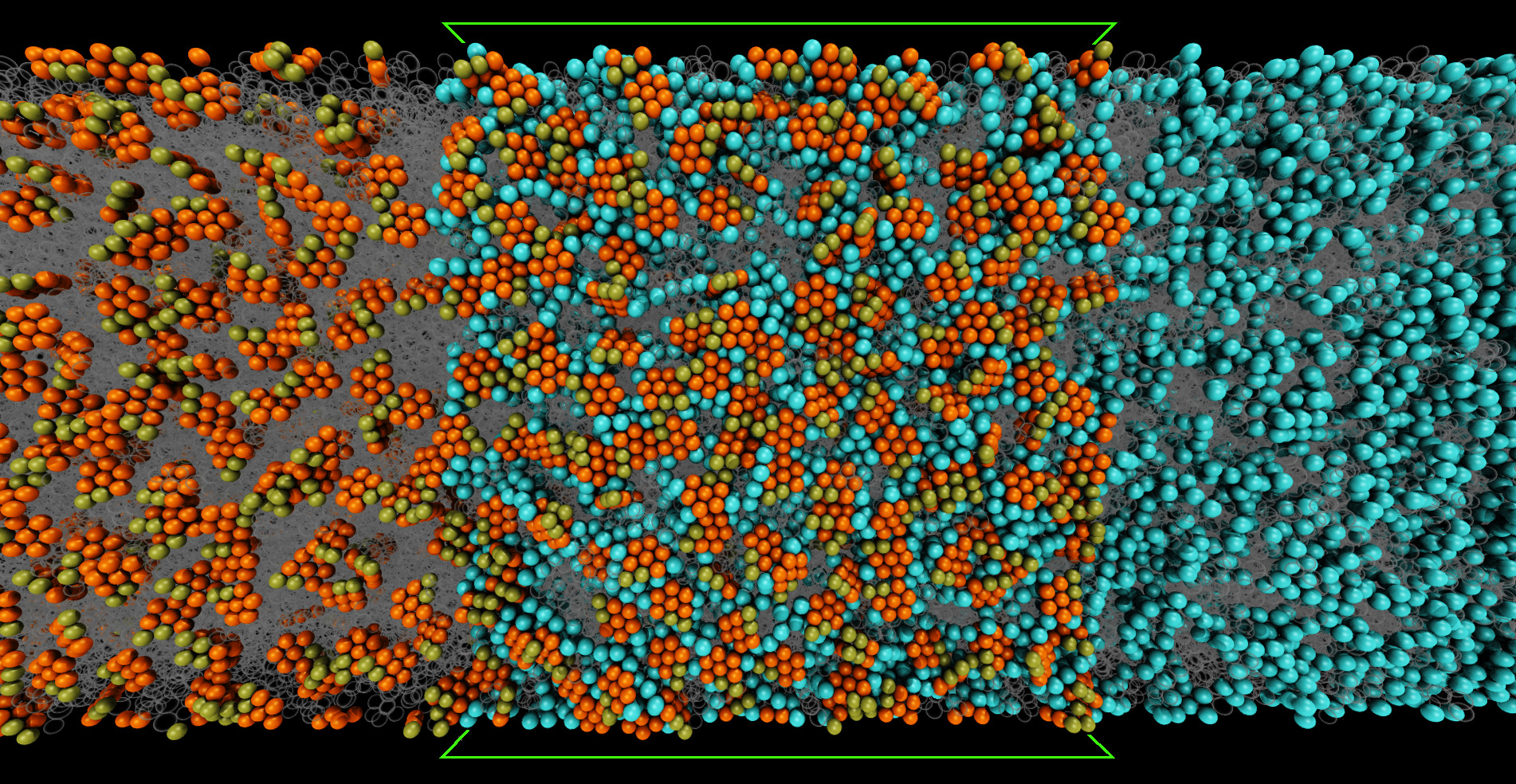}
  \caption{Same figure as \cref{fig:apce-interface-240} for the system with 720
    asphaltene molecules. It is seen that even at this high concentration, the
  asphaltenes show essentially no clustering at the interface. Some asphaltenes
  here have their cores oriented perpendicularly to the interface, but the great
  majority still have cores parallel to the interface.}
  \label{fig:apce-interface-720}
\end{figure}

As previously mentioned, the new model asphaltenes were developed partly in
response to the APCH model asphaltenes being more self-associative than what is
expected based on experimental evidence. The APCE asphaltenes adopt a very
different configuration at the interface, namely with the polycyclic aromatic core aligned
with the water interface, and the aliphatic tails stretched back towards the oil
phase. This is shown in \cref{fig:apce-interface-240}. In this figure, the interface is seen from
the water side, in the system with 240 model asphaltenes after 50 ns of
equilibration. In the centre, highlighted by green planes above and below, 
the interface is shown with the asphaltene molecules as in the previous figures. On the left
side, a periodic image is shown with only the aromatic cores. On the right
side, a periodic image is shown with only the aliphatic tails. In all three
images, grey beads indicate heptane and toluene. It is seen that
the aliphatic tails are protruding into the grey oil phase, indicated by the
shadows these beads are casting on each other. In \cref{fig:apce-interface-720},
the same illustration is shown for the system with 720 asphaltene molecules.

These results are in very good agreement with experiments performed by
\citet{andrews2011} using sum frequency generation (SFG) spectroscopy to study
the orientation of real asphaltenes at
interfaces. They find that ``SFG clearly indicates that asphaltene polycyclic
aromatic hydrocarbons are highly oriented in the plane of the interface
and that the peripheral alkanes are transverse to the interface.'' When
comparing \cref{fig:apch-240} to \cref{fig:apce-interface-240}, it is clear 
that the APCE model asphaltene has an interfacial behavior closer to that
observed in experiments, while results with the APCH model asphaltene show
a qualitatively different interfacial orientation. It is noteworthy that
previous atomistically detailed molecular simulations
\cite{mikami2013,liu2015,yang2015} have shown similar interfacial behaviour
to the APCH model, \ie have found the asphaltenes to be
highly associative at the interface and stacking with cores orthogonal to the
interface, in contrast with the experimental findings using real asphaltenes.
Thus the results presented here with the APCE model asphaltene are the first
simulations to be consistent with the experimental results from SFG spectroscopy
\cite{andrews2011}.

This absence of clustering at the interface means that there is no long time
scale for reorientation at the interface, unlike with the APCH asphaltenes. This
means that extending simulation runtimes to 350 nanoseconds, as for the APCH
asphaltenes, is not necessary. To confirm this, a simulation with the highest
number of asphaltene molecules considered here was run until 150 ns, and no
change in configuration was observed between 50 and 150 ns.

\begin{figure}[htpb]
  \centering
  \includegraphics[width=0.6\linewidth]{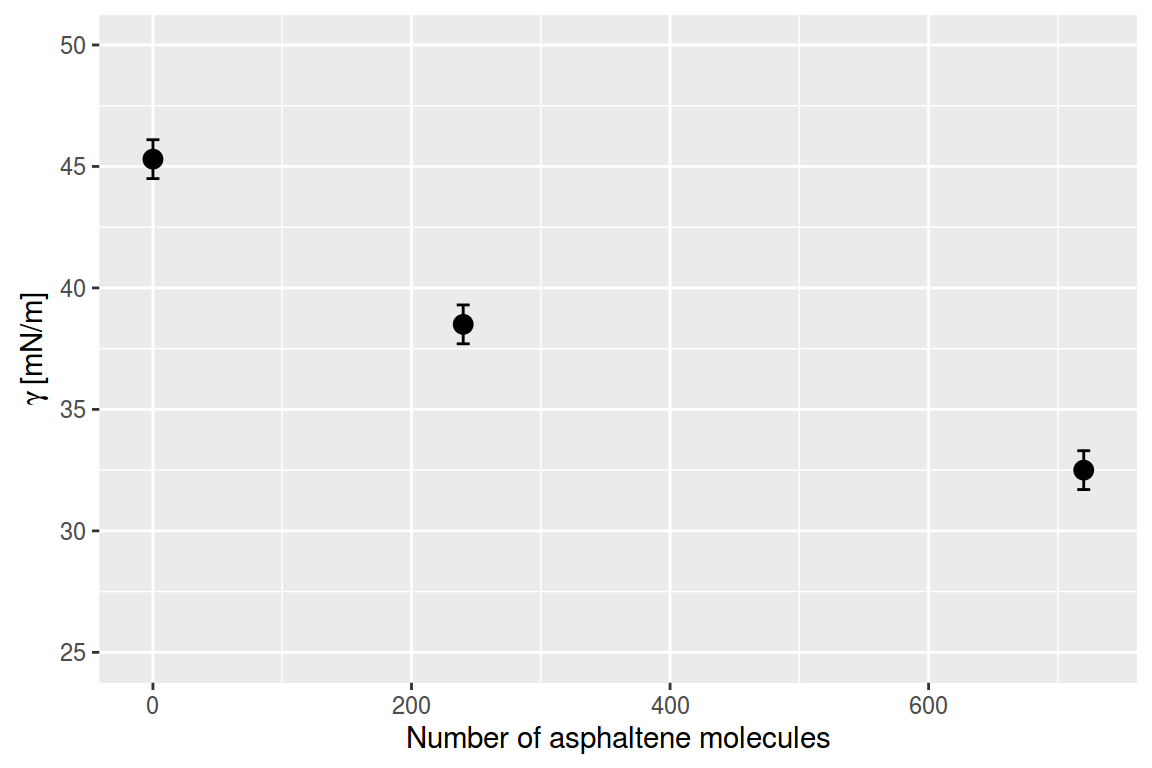}
  \caption{The interfacial tension $\gamma$ computed from molecular simulations, for
  varying concentration of asphaltenes. The decrease in $\gamma$ with increasing
  asphaltene concentration corresponds well with experimental measurements.}
  \label{fig:apce-ift}
\end{figure}

With this model giving the interfacial behaviour as expected from experiments,
it is interesting to see how the interfacial tension varies with the asphaltene
concentration. This is plotted in \cref{fig:apce-ift}. It is seen that adding
asphaltenes to the system decreases the interfacial tension, with what appears
to be an exponentially decaying trend.
Note that the concentration in this figure is given as the number of molecules,
since interpretation of this number into bulk concentration is not immediately
obvious. The variation in the number of molecules in the simulation can be interpreted either as
snapshots in time during the diffusion process for one bulk concentration, or as the
equilibrium interfacial tension for varying concentration of asphaltenes.
In general, these values and variation in interfacial tension corresponds
well with experimental observations, \eg in \cite{fossen2007}. 

\begin{figure}[htpb]
  \centering
  \includegraphics[width=0.6\linewidth]{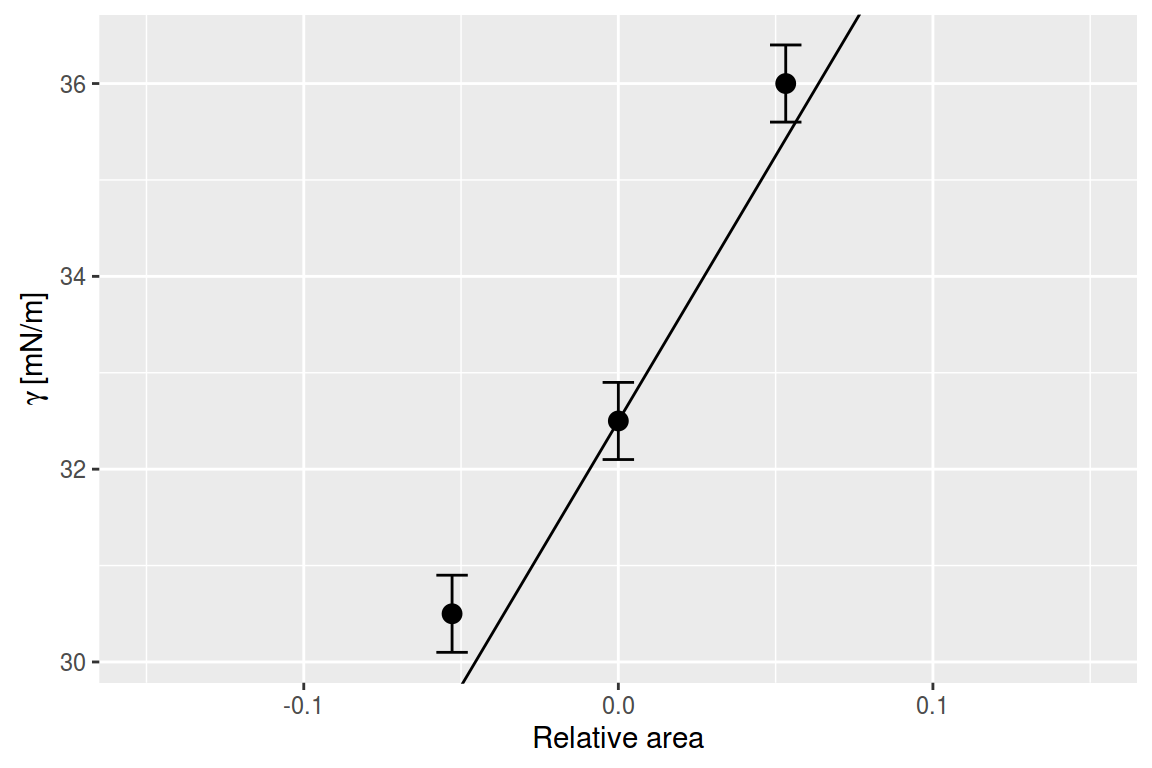}
  \caption{The change in interfacial tension versus the relative area $A/A_0
  - 1$ for compression (negative relative area) and expansion of the
    interface, with the 720 asphaltene molecule system. 
    It is seen that the behaviour is linear, \ie elastic. The line
  corresponds to $K_a$ = 55 mN/m.}
  \label{fig:apce-elasticity}
\end{figure}

The interfacial elasticity was also estimated with this model, 
using the system with 720 molecules after 50 ns, and found to be 55 $\pm$ 20
mN/m. The variation in tension with the interfacial area is
shown in \cref{fig:apce-elasticity}, together with a line having a slope of 55
mN/m.

\subsection{Macroscopic simulations of the draining of water drops in oil}
\label{sec_pipette_draining_drop}
Given the immense potential for variations in even model crude oil/water
systems, and the uncertainties still surrounding the asphaltenes despite the use
of state-of-the-art analytic chemistry techniques, it is not possible in the
experimental literature to provide enough detail for an exact comparison to be
made here between simulations and experiments. It is also noteworthy that the elasticity as
defined here has not been given consideration in the experimental literature.
There, studies of interfacial properties in crude oil/water systems  have either
focused on the shear rheology (\eg \cite{verruto2009,fan2010}), or in the cases
where dilatational elasticity has been considered (\eg
\cite{kumar2001,aske2002,daniel2005}), the authors have presupposed this
elasticity to be of the type considered by Gibbs in his seminal
work \cite{gibbs1878}. As we will demonstrate in the following, a Gibbs-type
elasticity cannot cause the phenomena observed in experiments, for several
reasons: Gibbs elasticity is given as the change in interfacial tension when
a change in area causes a change in the concentration of a surface-active
material. It follows that this elasticity is isotropic by definition, and
furthermore that it cannot reduce the tension to zero. The reader is referred to pp. 467--482 of
the very readable paper by \citet{gibbs1878} for further details.

All in all, these facts imply that our
comparisons with experiments must be of the qualitative kind. A very interesting
case for qualitative comparisons is the crumpled drop phenomenon, which occurs
when a water drop in crude oil is drained by use of a needle. The appearance of
wrinkles or crumples in the drop interface upon contraction is clearly evident
in photographs and is thus not subject to
interpretation, and it is a phenomenon fundamentally different from those
observed in water/oil/surfactant systems.

In the experimental literature there are two different categories of experiments
with draining water drops in crude-oil (or model systems). The first category
concerns
the micropipette experiments as reported \eg by \cite{yeung1999}. In this case, the
drop size is representative of drops found in a real water/crude-oil emulsion,
with a drop diameter of 50 $\mu$m. For these drops, buoyancy effects are
small when compared to tension effects. In the second category of experiments 
the pendant drop tensiometer is employed, with much larger drops of diameter around 5 mm,
\eg as reported by \citet{pauchard2014}\footnote{The drop diameter is not given
in this publication, but the tensiometer needle outer diameter is given as
1.65 mm.}. In this case, buoyancy effects are larger and tension effects smaller than
in the micropipette case. 

To get a quantitative measure of this difference, 
we may consider the Eötvös number, using
the inverse drop curvature $1/\kappa$ as the length scale, \ie
\begin{equation}
  Eo = \frac{\Delta \rho g}{T \kappa^2}
\end{equation}
where $\Delta \rho$ is the water-oil density difference in kg/m$^2$, $g = 9.81$ m/s$^2$
is the gravitational acceleration, and $T$ is the total tension, which is of the
order of 0.01 N/m. This means the Eötvös number is $Eo \approx 5 \cdot 10^{-5} $ for the micropipette
case, while it is $Eo \approx 0.5$ for the pendant drop case, meaning that in the
former case buoyancy is completely negligible, while in the latter case
buoyancy has an effect. This means that in the latter case, the interface is
deformed both by the action of the tensiometer and by the effect of buoyancy.

\subsubsection{The pendant drop case}
With this in mind, we consider first the situation analogous to the pendant drop
tensiometer, where drops are influenced both by buoyancy and by tension effects.
In the experiments reported \eg in \cite{pauchard2014}, the drop remains
axisymmetric up to the point of crumpling. It is seen that the crumpling only
occurs in the azimuthal direction, thus the interface remains in tension in the
meridional direction. This is due to the negative buoyancy of the water drop in oil.
The situation can be compared to wrapping a handkerchief around an apple and
lifting the corners of the handkerchief together: the cloth will wrinkle in the
azimuthal direction, but there will obviously be tension in the meridional direction,
since the cloth supports the weight of the apple.

After crumpling has occurred, as long
as the crumple depth remains small compared to the drop radius, the
axisymmetric simulation remains valid also in the crumpled region \cite{knoche2011}, where the
interface then corresponds to an azimuthal mean interface. This is precisely because the
azimuthal tension is zero in the crumpled region, and so this
tension (including the effect of azimuthal curvature of the crumples) does not exert a force.
Thus axisymmetric simulations can provide insight into this situation.
Unfortunately, neither drop diameters prior to deflation, nor the composition
(and thus density) of the oil phase, are given in \cite{pauchard2014}; moreover,
the interfacial elasticity is not known, and thus an
exact comparison to these experiments cannot be made.

The case considered is a drop of water with an initial diameter of 6.6 mm
immersed in oil ($\rho_2 = 830$ kg/m$^3$) and attached to a needle which sucks
the water out at a rate of 0.059 mL/s. The initial drop volume is 1.2 mL. The
penalisation method is used to enforce no flow inside the needle walls, and
a Pouseille flow inside the needle. There is no explicit handling of the
drop/needle contact angle, but the interface inside the needle wall is forced to
remain in its initial position by the penalisation. With this approach, the drop
interface overlaps with the needle walls, as seen in the following
Figures~\ref{fig:pendant-drop-draining-nice} and~\ref{fig_pipette_draining_drop_full}.
Since the flow field is forced to
be zero inside the walls, the interface points inside the walls (including at
the needle wall boundary) will not move. This gives the effect of pinning the
contact line location, but leaving the contact angle free to vary according to
the drop interfacial dynamics. 

The
tension in this case is given by $\gamma = 30$ mN/m and $K_a = 50$ mN/m. These
specific values are chosen based on the results from the molecular simulations,
which have some uncertainty; further
details are given in \cref{tab_param_pendant}. Thus the Eötvös
number for the initial condition (where $T=\gamma$) is $Eo = 0.61$, and
the simulation corresponds very well to the pendant drop regime.

As the simulation begins, the drop starts to fall, and a balance is 
quickly established between the drop tension and the drop negative buoyancy,
forming a pendant drop. As the needle removes water from the drop, it shrinks,
and eventually the ``neck'' of the drop has shrunk to the point that it becomes
tensionless in the azimuthal direction, \ie $T_\phi = 0$. This is shown in
\cref{fig:pendant-drop-draining-nice}.

\begin{figure}[htpb]
  \centering
  \includegraphics[width=0.8\linewidth]{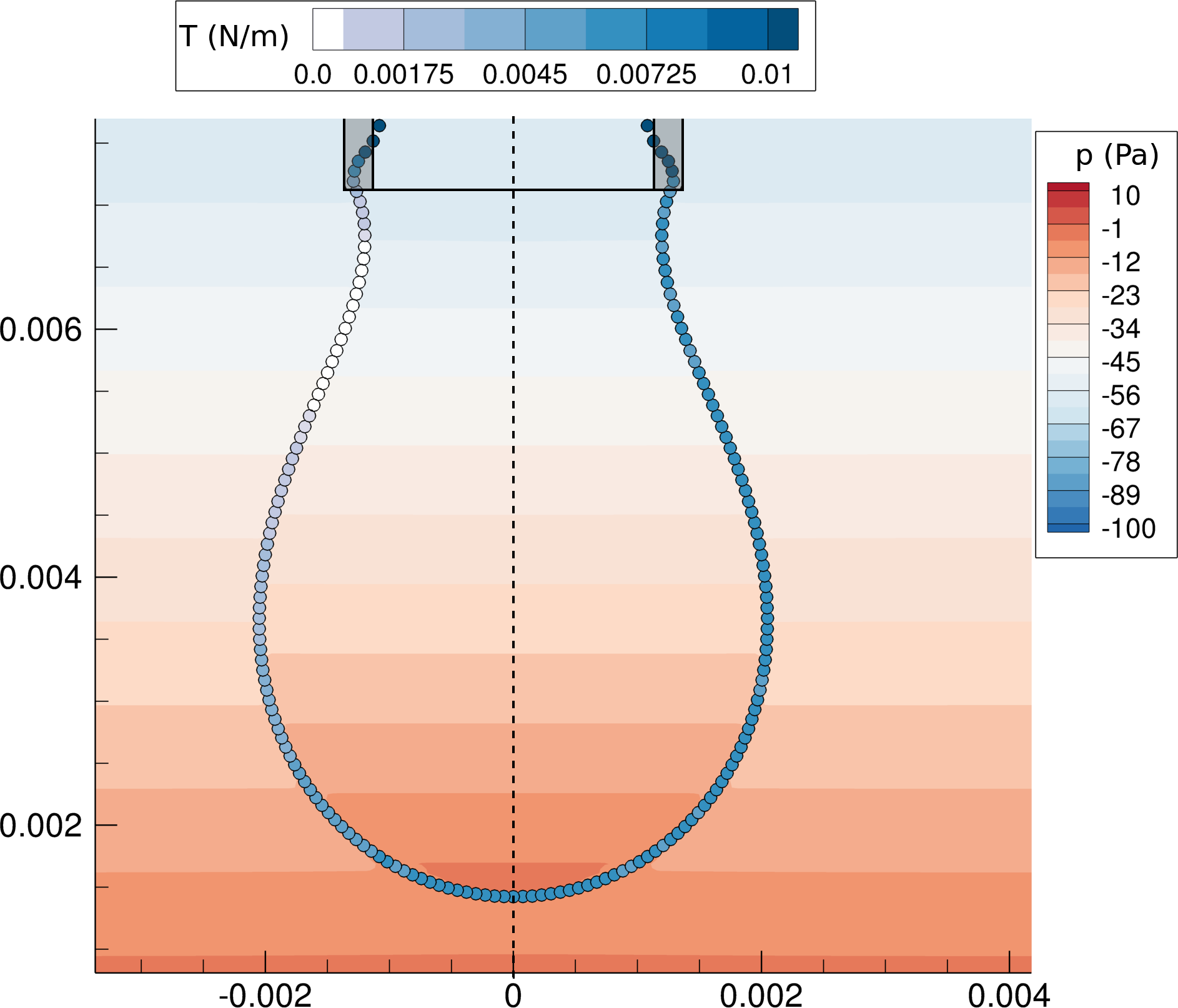}
  \caption{The axisymmetric drop profile (points) mirrored across the symmetry
  axis (dashed line). Only every 20th Lagrangian point is shown. The colour of
the points corresponds to $T_s$ on the right-hand side and $T_\phi$ on the
left-hand side of the plot. The pressure is as coloured contours inside and
outside the drop, showing the varying pressure jump across the interface.}
  \label{fig:pendant-drop-draining-nice}
\end{figure}

In this plot, the
axisymmetric drop profile is mirrored around the symmetry axis, and the tensions
$T_s$ and $T_\phi$ are shown on the right- and left-hand sides, respectively, as
colours on the interface. Inside and outside the drop, contours of the pressure
field are shown. It is evident that even though $T_\phi$ is zero along
a portion of the interface, there is still a pressure jump across the interface.
It is also seen that the pressure difference varies along the interface, since
$T_\phi$ varies in the meridional direction; and as previously pointed out, there
\emph{cannot be} a force which cancels out this variation. 
This is true even in the static situation.

\begin{figure}[htpb]
  \centering
  \begin{subfigure}{0.428\textwidth}
    \includegraphics[width=\linewidth]{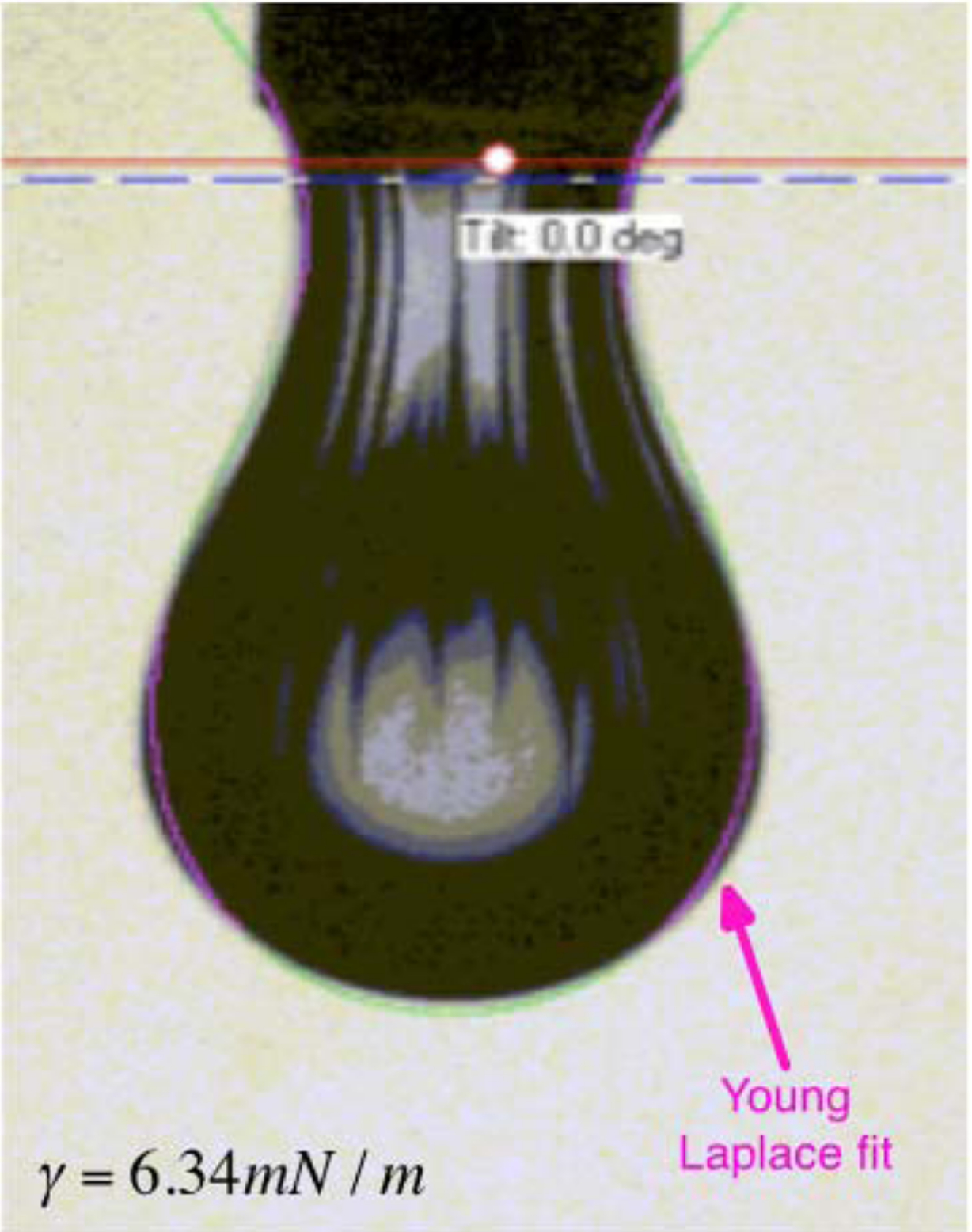}
  \end{subfigure}
  ~
  \begin{subfigure}{0.54\textwidth}
    \includegraphics[width=\linewidth]{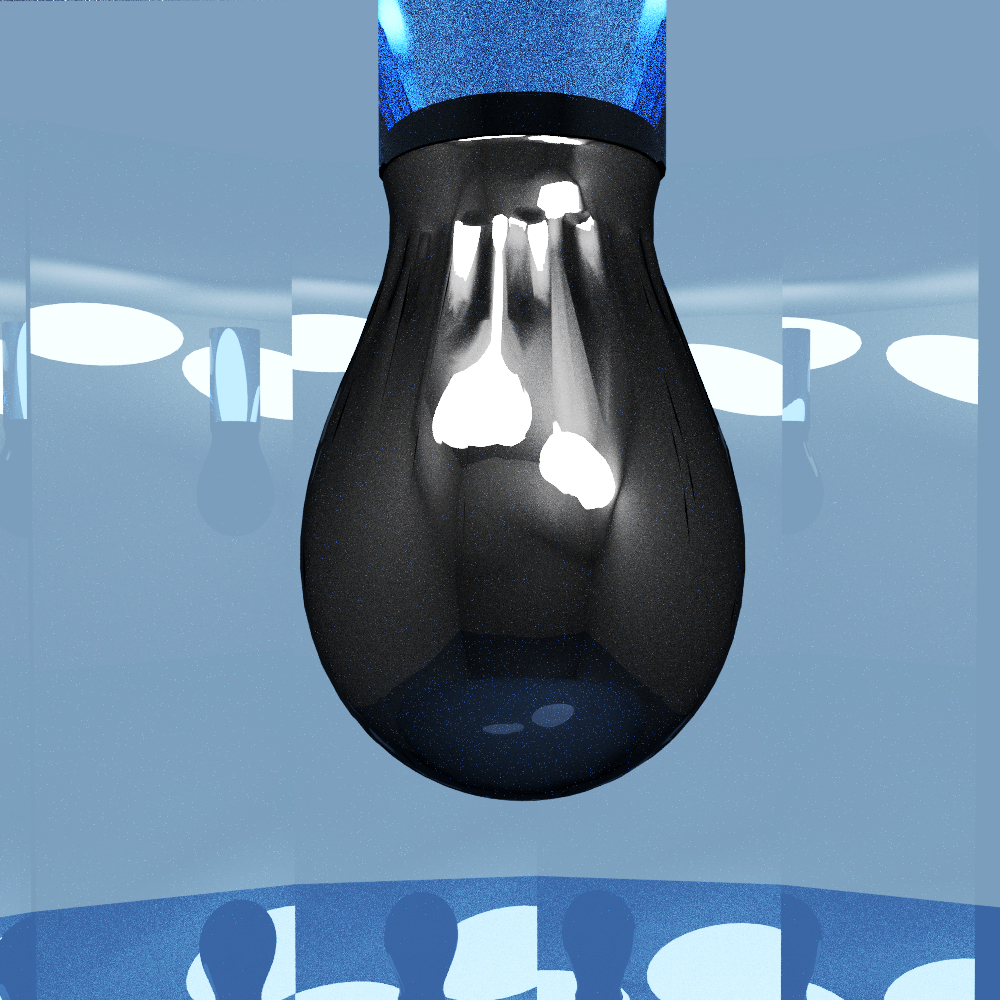}
  \end{subfigure}
  \caption{Right: raytrace of the drop profile from simulations, revolved around
    the symmetry axis and with crumples inserted into the
  region where $T_\phi=0$. Left: experimental result showing a deflated
  asphaltene-stabilised water drop in oil. {\scriptsize (Left figure reprinted with
  permission from: V. Pauchard, J. P. Rane, S. Banerjee, Asphaltene-laden
  interfaces form soft glassy layers in contraction experiments: A mechanism
  for coalescence blocking, Langmuir 30 (2014) 12795–12803. Copyright (2014) American Chemical
Society.)}}
  \label{fig:crumpled-drop-raytrace}
\end{figure}

To compare the simulation result to images from experiments \eg in
\cite{pauchard2014}, it is
illustrative to construct a three-dimensional representation from the
axisymmetric interfacial profile. To this end, the
axisymmetric profile was imported into the Blender 3D graphics software and revolved around the
symmetry axis to create a whole drop. Crumples were then inserted manually
in the region where $T_\phi = 0$. Raytracing was used to create a realistic rendering of the
drop and the needle to which it is attached. The result is shown in 
\cref{fig:crumpled-drop-raytrace} next to the experimental result from
\cite{pauchard2014}. The similarity is striking, indicating that elasticity
in the interface is a very likely explanation of the crumpled drop
phenomenon seen in experiments.

\subsubsection{The micropipette case}
Considering the small Eötvös number regime, analogous to the micropipette drops, the
situation after crumpling occurs is clearly not axisymmetric, and thus full
three-dimensional simulations are required for quantitative studies of this.
While the hybrid method discussed in this paper is readily extendable to
three dimensions, this has not been done here due to time constraints, but may
be considered in future work. One
can, however, study this case from the qualitative perspective by considering
a purely two-dimensional simulation, corresponding to a drop and pipette which
are significantly elongated in one direction (perpendicular to the simulation
domain).

\begin{figure}[htbp]
  \centering
  \includegraphics[width=\linewidth]{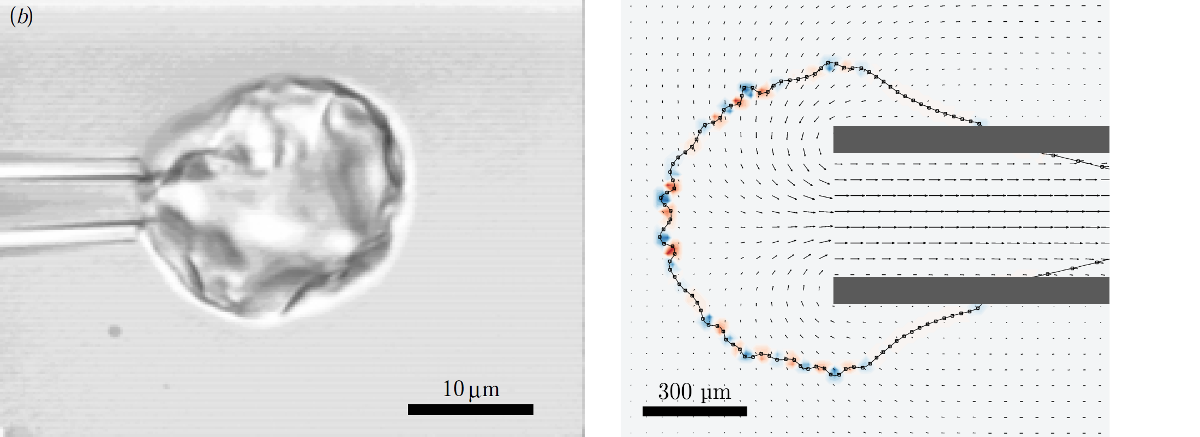}
  \caption{Right: two-dimensional simulation, with red and blue
    colours indicating interfacial curvature. Velocities are
  plotted for every 5th grid point and every 10th Lagrangian point is plotted.
  Left: experimental result showing the crumpled drop attached to a micropipette.
  {\scriptsize (Left figure reprinted with permission from: A. Yeung, T. Dabros, J.
    Czarnecki, J. Masliyah, On the interfacial properties of micrometre–sized water droplets in crude oil,
  Proceedings of the Royal Society of London A (1999) 3709–3723.}}
  \label{fig_pipette_draining_drop_full}
\end{figure}

The case considered for this case was with an initial diameter of 2 mm,
corresponding to $Eo = 0.06$. While this drop (and $Eo$) is substantially larger
than in the experiments, it is small enough that gravity becomes unimportant.
Accordingly, the simulation is with zero gravity. Further details of the case
are given in \cref{tab_param_pipette}; see also \cite{lysgaard2015}.
%The reason
%for using a larger drop than in the experiment is that the viscous stability
%condition
%(\cref{eq_cfl_visc}) means that the time step size in the simulation decreases
%proportionally to the square of the grid spacing, and thus very small
%systems require very long simulation times. Already for the drop size considered
%here, the simulation takes several days to run. The important point is to have
%$Eo \ll 1$, which is obtained with this drop size.
A comparison of
the simulation result with the photograph from experiments by \citet{yeung1999},
both from after crumpling has occurred, is shown in
\cref{fig_pipette_draining_drop_full}. A good qualitative similarity between the
two is seen.

\section{Discussion}
\label{sec:discussion}
This paper presents a multiscale approach to simulations that shed light on 
the behaviour of asphaltene-covered water drops in crude oils. The simulation
results at both the molecular and at the continuum scale showcase several
interesting phenomena which appear in these systems. Ultimately, a better
understanding of these phenomena will lead to a better understanding of crude
oil/water emulsions, which are made up of trillions of tiny water drops, each
covered with asphaltene molecules. This improved understanding can
be utilised to reduce the cost and footprint of oil-water separation equipment,
as well as to reduce the use of chemicals and heat in these separators. This
in turn leads to reductions in both capital and operational expenses, as well as reduced
emissions from oil production.

In the molecular simulations it is found that coarse-grained model asphaltenes
are highly sensitive to the molecular architecture, to the point that molecules
which are isomers both on the atomistic and on the coarse-grained scales have
very different solubilities in the solvents in question (heptane and toluene).
This is consistent with \eg the findings by \citet{sedghi2013} who found in
their atomistic simulations that small variations in the side groups of
model asphaltenes caused large differences in solubility.

It is notable that the two solvents in question, heptane and toluene, are not
very different in the first place: they are miscible, have the same number of
carbon atoms, and have similar Hildebrand solubility parameters\footnote{The
  Hildebrand solubility parameter is a quantity most useful for comparing
solvents, and is given by the square root of the cohesive energy density.}, 15.3
and 18.3 respectively, compared to \eg 26.2 for ethanol and 48.0 for water, all
these from \cite{barton1983} given in MPa$^{1/2}$.  The most obvious difference
between heptane and toluene is the different molecular architecture of the two
(ring-shaped and linear), which in turn enables toluene to have $\pi$-orbital
interactions with other aromatic molecules.  With the similarity in mind, an
interesting question arises: Asphaltenes are known to be a highly complex
mixture of different molecules, which originate as part of the well-stream from
crude oil reservoirs deep underground. Unlike the carefully synthesised
molecules used in man-made chemical processes, asphaltenes are created in an
essentially random fashion. How, then, do they achieve such a detailed
solubility balance?

It is likely that the explanation lies in a two-fold application of an argument analogous
to the ``anthropic principle'': the fact that asphaltenes are insoluble in
heptane comes from their penchant for $\pi$-$\pi$ ring interactions, which cause
them to cluster and subsequently fall out of solution. If the molecules had
smaller aromatic cores, they would not show this behaviour, but then they would
also not be asphaltenes in the first place, rather they would be resins. The
solubility in toluene can be understood from a similar point of view: were the
asphaltenes not soluble in toluene (or aromatic solvents in general), they would
not be soluble in crude oil at all, so they would never be found in the flow
coming up from the reservoir.

An interesting possible topic for investigation in future studies is the effect
of asphaltene polydispersity, which is an obvious next step on the
path towards better models for molecular simulations involving asphaltenes.
A very appealing idea is the construction of a coarse-grained QMR method, which
would in principle enable the construction of an arbitrary number of different
coarse-grained model asphaltene molecules.

When the model asphaltenes are introduced into the heptol-water system, they are found
to intrinsically be interfacially active. No tuning has been done of the 
cross-interactions between the aromatic cores and water, due to the lack of
experimental properties to tune this interaction to. With this in mind, it is
remarkable how well the interfacial behaviour of particularly the APCE model
asphaltenes matches what one expects based on our understanding of
the $\pi$-interactions between aromatic rings and water \cite{li2008}.

The molecular simulations presented here using the APCE model asphaltenes are the
first reported simulations of asphaltenes in oil-water systems which obtain 
the correct interfacial behaviour, as compared to experimental results using
sum frequency generation spectroscopy \cite{andrews2011}. This leads one to speculate
that perhaps a coarse-grained model asphaltene which has the correct solubility
behaviour in heptane and in toluene, such as the APCE model, somehow begins to 
take the asphaltene polydispersity into account through the coarse-grained
nature.

When it comes to the pendant drop case, simulations highlight that the
tension is highly anisotropic, and that the azimuthal tension varies
significantly along the drop profile, being zero around the neck of the drop
when volume has been removed with the needle. This is caused by the Eötvös
number being close to one, such that both gravity and the needle suction are
responsible for deforming the drop. The resulting variations in the tensions
explain the axisymmetric crumpling observed in experiments.

An important implication of the anisotropy in the tension is that when using the pendant drop
tensiometer to study asphaltene covered drops, it makes little sense to speak of a ``surface
pressure'' as a scalar number indicative of the total tension in the interface.
This is in contrast to the situation with a flat interface, as in a classical
Langmuir-Blodgett trough apparatus, where the surface pressure would remain well-defined.
This questions the findings in previous studies where conclusions are drawn on
the basis of variations in the ``surface pressure'' of asphaltene-covered 
drops measured using the pendant drop tensiometer to deform drops. It is also
important to note that the elasticity considered in these simulations is
fundamentally different from the Gibbs elasticity of surfactants.

When studying ordinary surfactants, the Marangoni effect
ensures that (at equilibrium) the distribution of surfactants on the drop
interface is uniform, and it thus makes sense to speak of a surface pressure 
in this case. For the anisotropic tensions studied here, the analogue of the Marangoni effect
serves only to remove variation of $T_s$ in the meridional direction etc. But as discussed
previously, it is not possible to have forces acting to cancel the variation in
$T_\phi$ along the meridional direction, since these forces would be in the direction
binormal to the axis of strain.

When it comes to the micropipette drop case, the Eötvös number is $Eo \ll 1$,
and thus gravity is unimportant. This means that the tension forces dominate,
and that drop deformations are entirely controlled by the needle suction. This
causes the drop to remain spherical up to the point of crumpling, such
that the crumples appear simultaneously across the drop interface. The
two-dimensional simulation at low Eötvös number demonstrates a qualitative
similarity to this, in that crumpling appears on all of the interface.

How these macroscopic phenomena relate to emulsion stability remains a topic for
future investigations. An important question which should be investigated is the
deformation of emulsion drops as they travel from the point of emulsion formation, through the
varying turbulence levels in the flow which transports them through the
pipeline, and finally when they end up in the oil-water separator, where the
flow is less turbulent and the drop shapes will be spherical. Taking this into
account, together with the adsorption time scales for asphaltenes in a highly
sheared flow, would provide important insight into the situation for real emulsion drops.
The multiscale method employed here, with future extensions \eg to fully
three-dimensional flow, is uniquely poised to help answer these and other 
important questions.

\section{Concluding remarks}
\label{sec:conclude}
To summarise, the current paper presents a multiscale approach to the simulation
of drops with complex interfaces, such as water drops in crude oil which are
covered with asphaltene molecules. The approach combines coarse-grained
molecular dynamics simulations using the SAFT-$\gamma$ Mie force field, with
detailed two-phase flow simulations using a hybrid
level-set/ghost-fluid/immersed-boundary method developed as part of this work.
At the molecular scale, the coarse-grained approach enables simulations at
unprecedented time- and length-scales, using accurate models for both the simple
fluids and the complex asphaltene molecules. The interfacial tension $\gamma$ and
elasticity $K_a$ are estimated, for use in the macroscopic simulations.

At the macroscopic scale, detailed
simulations of oil-water interfaces with both interfacial tension and elastic
properties shed new light on experimental results that  showcase one of the
peculiarities of complex fluid-fluid interfaces, namely crumpling drops. Two
categories of experimental results exist in the literature, with different
classes of crumpling behaviour. The results presented here demonstrate
that this difference is caused by the large difference in Eötvös number between
the two categories. In the pendant drop case, with an Eötvös number around
one, both gravity and the experimental setup are responsible for deforming the
drop, and this combination leads to an axisymmetric crumpling regime. In the
micropipette case, with an Eötvös number much smaller than one, the drop remains
spherical up to the point of crumpling, leading to a fully three-dimensional
behaviour.

These results present a novel hypothesis, namely anisotropic tensions in the interface,
as an explanation of the crumpling phenomenon. This hypothesis is different
from previously suggested explanations, \eg that based on glass transitions in
\cite{pauchard2014}.
The hypothesis put forward in this work makes fewer assumptions and has
greater predictive power than those presented in previous works. 
% However, which theory is the most correct must ultimately be decided by
% the scientific method, \ie making specific predictions based on the
% different hypotheses, and subsequently performing experiments to test
% these predictions.

Building on this work, there are numerous avenues open for future
investigations. The effect on interfacial properties of crude oil composition, asphaltene polydispersity,
asphaltene architecture, other crude components such as resins, etc., should all
be considered. Extending the hybrid simulation method to three dimensions would
enable not just simulations of \eg the micropipette drops, but also of drops in more
complicated (even turbulent) flows. Furthermore, the extension to drop
coalescence is a very interesting topic. Also, this study assumes that the
interfacial elasticity is of the Hookean type. While this is the simplest form
of elasticity, it could well be that asphaltene-covered interfaces can be more
accurately modelled using \eg a neo-Hookean elasticity.

All in all, the methods presented here
are well suited for increasing our understanding of dispersed two-phase flows
with complex interfaces. This is important not just for the application in
focus here, crude oil/water systems, but also for biological systems such as
the flow of blood or the transport of proteins, and for chemical processes involving
multiphase flow and macromolecules.

\section*{Acknowledgements}
{\AA}E, STM and BM acknowledge the financial support from the project
\emph{Fundamental understanding of electrocoalescence in heavy crude oils}
coordinated by SINTEF Energy Research and funded by the Petromaks programme of
the Research Council of Norway (206976), Petrobras, Statoil and W\"{a}rtsil\"{a}
Oil \& Gas Systems.

EM acknowledges the financial support for the Molecular Systems Engineering group
from the Engineering and Physical Sciences Research Council (grant numbers
EP/E016340, EP/J014958, EP/L020564).

The authors would like to thank Prof. Edo Boek for fruitful discussions on these
matters. The authors are also grateful for computational resources at the Imperial
College High Performance Computing Service, and at the Abel cluster operated 
by the University of Oslo and the Norwegian metacenter for High Performance 
Computing (NOTUR), as well as assistance from the staff at both locations.

\bibliographystyle{model1-num-names}
\bibliography{references}

\appendix

\section{Parameters for the continuum simulation cases}

\begin{table}[htbp]
  \centering
  \footnotesize
  \begin{minipage}[t]{0.48\linewidth}
  \centering
  \caption{Parameters for the drop in potential vortex.}
  \begin{tabular}{lcl}
    \toprule
    Parameter & Symbol & Value \\
    \midrule
    Drop radius              & $r$           & 0.15 \\
    Domain size              & $\Omega$      & $1 \times 1$ \\
    Stability safety factor  & $C$           & 0.5 \\
    Euler grid nodes         & $N$           & $200 \times 200$ \\
    Lagrangian point density &               & $\sfrac{5}{\Delta}$ \\
    \bottomrule
  \end{tabular}
  \label{tab_param_vortex}
  \end{minipage}
  \hfill
  \begin{minipage}[t]{0.48\linewidth}
  \centering
  \caption{Parameters for the Zalesak's disk test.}
  \begin{tabular}{lcl}
    \toprule
    Parameter & Symbol & Value \\
    \midrule
    Disk radius              & $r$           & $\sfrac{1}{3}$ \\
    Domain size              & $\Omega$      & $1 \times 1$ \\
    Time step safety factor  & $C$           & 0.5 \\
    Euler grid nodes         & $N$           & $64\times64$ \\
    Velocity field           & $\v{u}$       & $\frac{\pi}{10}[\sfrac{1}{2}-y, x-\sfrac{1}{2}]$\\
    Lagrangian point density &               & $\sfrac{5}{\Delta}$ \\
    \bottomrule
  \end{tabular}
  \label{tab_param_zalesaks}
  \end{minipage}
\end{table}

\begin{table}[htbp]
  \centering
  \footnotesize
  \begin{minipage}[t]{0.48\linewidth}
  \centering
  \caption{Parameters for the elliptical drop driven by interfacial tension.}
  \begin{tabular}{lcl}
    \toprule
    Parameter & Symbol & Value \\
    \midrule
    Drop density           & $\rho_1$      & $\unitfrac[10^3]{kg}{m^3}$ \\
    Matrix density         & $\rho_2$      & $\unitfrac[10^3]{kg}{m^3}$ \\
    Drop viscosity         & $\mu_1$       & $\unit[10^{-3}]{Pa \cdot s}$ \\
    Matrix viscosity       & $\mu_2$       & $\unit[10^{-3}]{Pa \cdot s}$ \\
    Interfacial tension        & $\gamma$      & $\unitfrac[15\times10^{-3}]{N}{m}$ \\
    Drop radius            & $r$           & $\unit[10^{-3}]{m}$ \\
    Drop axis length ratio & $\frac{a}{b}$ & 1.16 \\
    Domain size            & $\Omega$      & $\unit[0.007 \times 0.007]{m}$\\
    Time step safety factor& $C$           & 0.2\\
    Grid nodes             & $N$           & $\{ 100, 200, 400, 800 \}$\\
    \bottomrule
    \label{tab_surf_tens_parameters}
  \end{tabular}
  \end{minipage}
  \hfill
  \begin{minipage}[t]{0.48\linewidth}
  \centering
  \caption{Parameters for relaxing drop with viscosity and density jump.}
  \begin{tabular}{lcl}
    \toprule
    Parameter & Symbol & Value \\
    \midrule
    Drop density           & $\rho_1$      & $\unitfrac[10^3]{kg}{m^3}$ \\
    Matrix density         & $\rho_2$      & $\unitfrac[5 \times 10^2]{kg}{m^3}$ \\
    Drop viscosity         & $\mu_1$       & $\unit[10^{-3}]{Pa \cdot s}$ \\
    Matrix viscosity       & $\mu_2$       & $\unit[10^{-2}]{Pa \cdot s}$ \\
    Interfacial tension    & $\gamma$      & $\unitfrac[15\times10^{-3}]{N}{m}$ \\
    Drop radius            & $r$           & $\unit[10^{-3}]{m}$ \\
    Drop axis length ratio & $\frac{a}{b}$ & 1.16 \\
    Domain size            & $\Omega$      & $\unit[0.007 \times 0.007]{m}$ \\
    Time step safety factor& $C$           & 0.2 \\
    Grid nodes             & $N$           & 400 \\
    \bottomrule
    \label{tab_relaxing_jump_params}
  \end{tabular}
  \end{minipage}
\end{table}

\begin{table}[htbp]
  \centering
  \footnotesize
  \begin{minipage}[t]{0.48\linewidth}
  \centering
  \caption{Parameters for relaxing drop with an elastic membrane.}
  \begin{tabular}{lcl}
    \toprule
    Parameter & Symbol & Value \\
    \midrule
    Drop density           & $\rho_1$      & $\unitfrac[10^3]{kg}{m^3}$ \\
    Matrix density         & $\rho_2$      & $\unitfrac[10^3]{kg}{m^3}$ \\
    Drop viscosity         & $\mu_1$       & $\unit[10^{-3}]{Pa \cdot s}$ \\
    Matrix viscosity       & $\mu_2$       & $\unit[10^{-3}]{Pa \cdot s}$ \\
    Interfacial tension    & $\gamma$      & $\unitfrac[15\times10^{-3}]{N}{m}$ \\
    Elasticity             & $K_a$         & 0 and $\unitfrac[15\times10^{-2}]{N}{m}$ \\
    Drop radius            & $r$           & $\unit[10^{-3}]{m}$ \\
    Drop axis length ratio & $\frac{a}{b}$ & 3.0 \\
    Domain size            & $\Omega$      & $\unit[0.007 \times 0.007]{m}$ \\
    Time step safety factor& $C$           & 0.5 \\
    Grid nodes             & $N$           & 150 \\
    \bottomrule
  \end{tabular}
  \label{tab_relax_elastic}
  \end{minipage}
  \hfill
  \begin{minipage}[t]{0.48\linewidth}
  \centering
  \caption{Parameters for the pendant drop case.}
  \begin{tabular}{lcl}
    \toprule
    Parameter & Symbol & Value \\
    \midrule
    Drop density           & $\rho_1$      & $\unitfrac[1000]{kg}{m^3}$ \\
    Matrix density         & $\rho_2$      & $\unitfrac[830]{kg}{m^3}$ \\
    Drop viscosity         & $\mu_1$       & $\unit[1.03 \times 10^{-3}]{Pa \cdot s}$ \\
    Matrix viscosity       & $\mu_2$       & $\unit[12.4 \times 10^{-3}]{Pa \cdot s}$ \\
    Interfacial tension    & $\gamma$      & $\unitfrac[30 \times10^{-3}]{N}{m}$ \\
    Elasticity             & $K_a$         & $\unitfrac[50 \times10^{-3}]{N}{m}$ \\
    Drop radius            & $r$           & $\unit[3.3 \times 10^{-3}]{m}$ \\
    Domain size            & $\Omega$      & $\unit[(5\times 10^{-3}) \times (15\times 10^{-3})]{m}$ \\
    Time step safety factor& $C$           & 0.3 \\
    Grid nodes             & $N$           & $132\times200$ \\
    Penalisation           & $\eta$        & $3 \times 10^{-5}$\\
    \bottomrule
  \end{tabular}
  \label{tab_param_pendant}
  \end{minipage}
\end{table}

\begin{table}[htbp]
  \centering
  \footnotesize
  \caption{Parameters for the micropipette case.}
  \begin{tabular}{lcl}
    \toprule
    Parameter & Symbol & Value \\
    \midrule
    Drop density           & $\rho_1$      & $\unitfrac[1000]{kg}{m^3}$ \\
    Matrix density         & $\rho_2$      & $\unitfrac[830]{kg}{m^3}$ \\
    Drop viscosity         & $\mu_1$       & $\unit[1.03 \times 10^{-3}]{Pa \cdot s}$ \\
    Matrix viscosity       & $\mu_2$       & $\unit[12.4 \times 10^{-3}]{Pa \cdot s}$ \\
    Interfacial tension    & $\gamma$      & $\unitfrac[40 \times10^{-3}]{N}{m}$ \\
    Elasticity             & $K_a$         & $\unitfrac[50 \times10^{-3}]{N}{m}$ \\
    Drop radius            & $r$           & $\unit[1 \times 10^{-3}]{m}$ \\
    Domain size            & $\Omega$      & $\unit[(2\times 10^{-3}) \times (3\times 10^{-3})]{m}$ \\
    Time step safety factor& $C$           & 0.5 \\
    Grid nodes             & $N$           & $132\times200$ \\
    Penalisation           & $\eta$        & $5 \times 10^{-6}$\\
    \bottomrule
  \end{tabular}
  \label{tab_param_pipette}
\end{table}

\section{Parameters and scripts for the molecular simulations}
The molecular simulations reported here are performed using our raaSAFT
framework. Scripts which can be used to reproduce the molecular simulations are
available in the \texttt{replication} directory of the raaSAFT repository, at
\url{http://www.bitbucket.org/asmunder/raasaft}. The force field parameters are
mostly published elsewhere, as indicated by the DOI for each model in that repository.
However, the force field parameters for the aromatic beads used in the APCE and
APCL asphaltene models have not been published elsewhere yet. These are given in
\cref{tab:aromatic}. The parameters for the toluene model are also not published
elsewhere yet, these are given in \cref{tab:aromatic}.

\begin{table}[htbp]
  \centering
  \caption{Parameters for the aromatic beads.}
  \begin{tabular}{lll}
    \toprule
    Parameter &  Value & Unit \\
    \midrule
    & \textbf{Toluene} & \\
    \midrule
    $\epsilon/k_B$ & 269.74 & [K]  \\
    $\sigma$ & 3.6794 & [\AA]  \\
    $n$ & 11.804 & [-]\\ 
    $m$ & 6 & [-]\\ 
    \midrule
    & \textbf{APCE} & \\
    \midrule
    $\epsilon/k_B$ & 312.90& [K]  \\
    $\sigma$ & 3.975 & [\AA]\\
    $n$ & 10 & [-]\\ 
    $m$ & 6 & [-]\\ 
    \bottomrule
  \end{tabular}
  \label{tab:aromatic}
\end{table}

\end{document}